Technische Universität Berlin

Faculty VII (Economics & Management)

Workgroup for Infrastructure Policy (WIP)

# Too cheap to meter? A stochastic analysis of projected future fusion costs

Authors: Stefania Böhnlein[1], Fanny Böse[1], Christian von Hirschhausen[1,2], Claudia Kemfert[2,3], Alexander Wimmers[1,2]*

1: Workgroup for Infrastructure Policy, TU Berlin, Straße des 17. Juni 135, 10623 Berlin

2: DIW Berlin, Anton-Wilhelm-Amo-Straße 58, 10117 Berlin

3: Professur für Energiewirtschaft und Energiepolitik, Leuphana Universität Lüneburg, Universtitätsallee 1, 21335 Lüneburg

*Corresponding author and Lead contact: awi@wip.tu-berlin.de; +49 30 314 75837

# Table of contents

## List of figures

## List of tables

# Title Page Information

## Abstract/Summary

In recent years, technological developments and activities by private actors have led a reemerged discussion of the potential of nuclear fusion to meet growing global energy demands. So far, however, fusion technologies remain at comparatively low development levels and their deployment in commercial power plants is probably still decades away. Regardless, over the last decades, many cost studies have been conducted that estimate the future cost of potential fusion power plants. But to date, there is no systematic and harmonized assessment of these projections. Therefore, this study conducts a stochastic analysis of future fusion power plant costs for three distint technology lines, magnetic confinement, inertial confinement, and magneto-inertial confinement fusion, including cost assessments of different technology maturity levels. These levels are further assessed to determine projected learning rates for future fusion costs. For mature technologies, mean LCOE are determined at 114.6, 110.3, and 143.9 USD per MWh for MCF, ICF, and MIF devices, respectively. This implies learning rates of more than 30%. We find that these projected values are rather optimistic when compared to other literature or comparable technologies like fission. We therefore urge policymakers to caution when potential fusion developers refer to the potential economic competitiveness of fusion power plants.

## Highlights

- Analysis of fusion cost and technology data from 56 references
- Stochastic modeling of assumed future levelized cost for three fusion technologies
- LCOE determined to be lower than today’s fission reactors
- Analysis of assumed learning rates shows very optimistic assumptions for fusion
- Technological maturity still low; initial actual cost expected to be much higher

# 1 Introduction

In the context of necessary energy system decarbonization, nuclear fusion has increasingly been regarded as a promising electricity-generation option for future emergy systems in recent years. This arises from its potentially low environmental impact, primarily limited to low-level radioactive wastes and activated materials, as well as its high energy output and supposedly abundant fuel resources (Chatzis and Barbarino 2021; Moynihan and Bortz 2023). Given fusion's theoretical capability to provide both reliable baseload and flexible electricity generation (Segantin, Testoni, and Zucchetti 2019; Kerekeš et al. 2023; Breuning et al. 2026), the technology is increasingly considered a firm, low-carbon complement to renewables in future decarbonized power systems, especially in countries with limited capacity for large-scale renewable deployment (Schwartz et al. 2023; Spitzer et al. 2025). Due to its expected high capital costs, the economic viability of fusion would require the operation at high capacity factors, implying a primary role as baseload generation and positioning it as a competitor to other firm, low-carbon technologies such as nuclear fission (Lindley et al. 2023).

Building on this potential, some nations have already included fusion in their long-term energy strategies (IAEA 2025b), and research and development (R&D) activities have accelerated worldwide, yielding noteworthy advances in plasma physics and reactor engineering (Klinger et al. 2019; Gibney 2022; Tollefson 2024; WNA 2025). In parallel, several private companies have begun designing the first generation of fusion power plants (FPPs), with some announcing plans for initial electricity output in the late 2020s and early 2030s. However, no commercial-scale plant has yet been realized, and most large-scale facilities laying the technological foundations for industrial deployment are not expected to begin operations before 2040. The technological and economic feasibility of fusion-based electricity generation under real operating conditions therefore remains unproven to date as most concepts remain in early development stages (Dering, Wimmers, and Von Hirschhausen 2026). Their eventual commercial success will depend on achieving key technological milestones, demonstrating the feasibility of pre-concept designs, as well as establishing a global fusion industry (IAEA 2025b).

Even if technical success is achieved, fusion must also become economically competitive with other firm, low-carbon energy sources to be efficiently integrated into future power systems on a large-scale, especially beyond its initial implementation phase. Yet, the future costs of FPPs remain highly uncertain, primarily due to the lack of experience in operating FPPs and the significant technological complexity that remains mostly unresolved (Nawal et al. 2024; Dering, Wimmers, and Von Hirschhausen 2026). Cost estimates span wide uncertainty ranges and may vary considerably over time, as unforeseen developments in markets, policy frameworks, or material prices could substantially affect final project costs (Schwartz et al. 2023; Lerede et al. 2023; Lindley et al. 2023). Nevertheless, over the past decades, numerous studies have attempted to project the costs of FPPs—including generation costs and their underlying parameters such as construction and operational expenditures—across different technological concepts and design maturity levels in order to evaluate their potential economic value and competitiveness (Woodruff et al. 2017). Generation costs are commonly expressed by the levelized cost of electricity (LCOE) that enables the comparison of the economic viability of different power technologies (Kabeyi and Olanrewaju 2023). However, existing assessments have predominantly taken one of two forms: engineering-based cost studies that produce single-point LCOE estimates for specific reactor designs (e.g., Woodruff et al. (2017), Lerede et al. (2023), or Lindley et al. (2023)), or survey-based analyses that compile reported cost figures across concepts and maturity levels without statistically characterizing the resulting uncertainty (e.g, Schwartz et al. (2023) or Nawal et al. (2024)). These assessments rely on deterministic methods that treat input parameters as fixed values, thereby neglecting the inherent uncertainty of pre-commercial cost data (Z.-Y. Zhao, Chen, and Thomson 2017; Miguel and Corona 2018). This is a critical shortcoming: unlike evaluations of established technologies, which can draw on historical cost data and empirically validated assumptions, all fusion-related parameters are hypothetical rather than observed, making the explicit treatment of uncertainty even more relevant.

Assuming FPPs become commercially available, i.e., there existed a sufficient number of capable suppliers, including established supply chains (Böse et al. 2024), their economic success depends on their ability to reduce high upfront costs through learning effects that must themselves be realistically achievable (Rooney et al. 2021). Deterministic approaches are not well suited to capture this uncertainty. Stochastic methods have already demonstrated their value for assessing cost uncertainty in other energy technologies (Tran and Smith 2018; B. Wealer et al. 2021; Li et al. 2023; Steigerwald et al. 2023), but have not yet been applied to fusion in this comprehensive manner. Tang et al. (2026) conducted a survey approach to assess the validity of assumed learning rates and found that they are likely to be overestimated. Their work, however, is limited to learning rates and does not consider implied cost reductions or the spread of assumed learning rates. In addition to the ex-ante nature of all fusion cost assessments, the fragmented and heterogeneous nature of available fusion cost data, spanning several decades, multiple currencies, and varying price bases as well as heterogeneous reactor concepts, has not previously been addressed through systematic harmonization within a single analytical assessment.

Thus, so far, literature lacks a systematically applied stochastic assessment of the vast body of published fusion cost data across multiple confinement types, i.e., magnetic, inertial, and magneto-inertial, to derive probabilistic LCOE distributions and empirically implied learning rates simultaneously, a gap that this work aims to address.

## 1.1 Fusion Concepts

Nuclear fusion generates energy by combining small atoms, such as deuterium (D) and tritium (T), to larger ones. In order to ignite, a fusion reaction must pass a certain threshold which is determined by the Lawson criterion that states that for a D-T-fusion, the triple product of plasma temperature $T$, its density $n$, and energy confinement time $\tau_E$ must be greater than or equal to $6 \times 10^{21} keV m^{-3} s$ (Lawson 1957; Wurzel and Hsu 2022). Different fusion concepts have been under development since the 1950s. Since then, three main technology lines have emerged that each aim to maximize at least one of the parameters of the triple product (Wimmers, Böse, and Von Hirschhausen 2026).

Magnetic confinement fusion (MCF) devices aim to prolong the confinement time. The most promising concepts are tokamaks and stellarators for which experimental devices have demonstrated general feasibility. Yet, both concepts still face substantial challenges to be useful for commercial energy provision, such as inherent plasma instability in tokamaks or the very high geometric complexity of stellarator magnets, amongst others (Dering, Wimmers, and Von Hirschhausen 2026).

Inertial confinement fusion (ICF) devices are designed to maximize the plasma density. This is achieved by pulsed shots aimed at small fuel pellets that implode seemingly instantly. ICF devices differ regarding their pellet and shot design. The most advanced approaches in ICF aim to use high-power lasers. A substantial challenge for ICF devices is a consistently high energy level of about 100 MJ per shot and sufficient repetition rates of 10 to 20 Hz (Haefner et al. 2023; Moses 2009).

Magneto-inertial confinement (MIF) devices comprise a vast number of exotic concepts that are designed to be less complex. For example, spheromak devices do not require toroidal field coils, like tokamaks, and field reversed configurations (FRCs) are more compact and thus able to confine hot plasma in a self-generated magnetic field (Steinhauer 2011; Guo et al. 2020). However, all MIF devices also bring different drawbacks and are mostly in very early development stages (Dering, Wimmers, and Von Hirschhausen 2026).

## 1.2 Relevant Cost Metrics

Assessments of future cost developments for different energy generation technologies usually assume some form of cost reduction through economies of learning or multiples. This implies that through repetition, the costs of a first-of-a-kind (FOAK) product will be higher than those of a subsequently produced identical product. The costs $C$ of the n-th product (NOAK, n-th-of-a-kind) are determined as $C_{NOAK} = C_{FOAK} * (1-x)^d$ with $x$ as the learning rate and $n = 2^d$, i.e., $d$ as the number of doubled product volumes $n$ (Lloyd, Lyons, and Roulstone 2020). As commercial fusion power plants are currently non-existant, learning rates cannot be empirically assessed. Tang et al. (2026) determine that a range of 8 to 20% could be feasible based on their expert survey assessment. For nuclear fission plants, historical learning rates vary from 2 to 15%, depending on analyzed country and reactor-design (Stewart et al. 2021), although negative learning rates, i.e., cost increases, have also been reported (Grubler 2010).

### 1.2.1 Overnight Construction Costs

Our work provides a stochastic assessment of potential future levelized cost of electricity (LCOE) determined as $LCOE = \left(TCC + OM_{Fix} + (OM_{var} * FLH)\right)/FLH$ with $FLH$ as the full load hours calculated as the product of capacity factor $CF$ and the average number of hours per year (8760h), $OM_{Fix}$and $OM_{var}$ as the fixed and variable operation and maintenance costs, respectively, and $TCC$ as the total construction cost determined by the combination of overnight construction cost $OCC$ and the interest accured during construction $idc$ as $TCC = OCC * (1 + idc)$. $OCC$ are used to compare large-scale infrastructure projects, like fission or fusion power plants, without accounting for project-specific financing cost. The latter are accounted for by $idc$ which we approximate in our assessment as $idc \approx \frac{WACC}{2} T_{con} + \frac{WACC^2}{6} T_{con}$ with $WACC$ as the uniform weighted average cost of capital and $T_{con}$ as the construction time of the project. (Göke, Wimmers, and Von Hirschhausen 2025)

We conduct an expansive literature analysis and find 56 references providing data points on various cost parameters, such as OCC (99 data points), OM costs (136 data points), and technical parameters, such as plant lifetime (43 data points). We demonstrate the distribution of OCC for the three main technology lines and, where discussed, maturity levels. For OCC, most data points, i.e., 88, are provided for MCF, of which 86% are for tokamak devices. All data are provided in the Appendix C. All monetary values are, if not otherwise indicated, normalized to 2018-USD.

Figure 1 shows the distribution of OCC data. It is noteworthy that reported OCC values range substantially between maturity levels, i.e., a mean of 12,592 USD per kW for FOAK MCF, and 6,991 for NOAK (with unknown value for n), or 6,086 for 10th-OAK (with n = 10). The outliers in the FOAK MCF category are data points for ITER and the Chinese device CFETR. For comparison, the figure also includes OCC data points as of the year 2023 for Western fission power plants, limited to high-capacity light-water reactors, based on Göke et al. (2025).

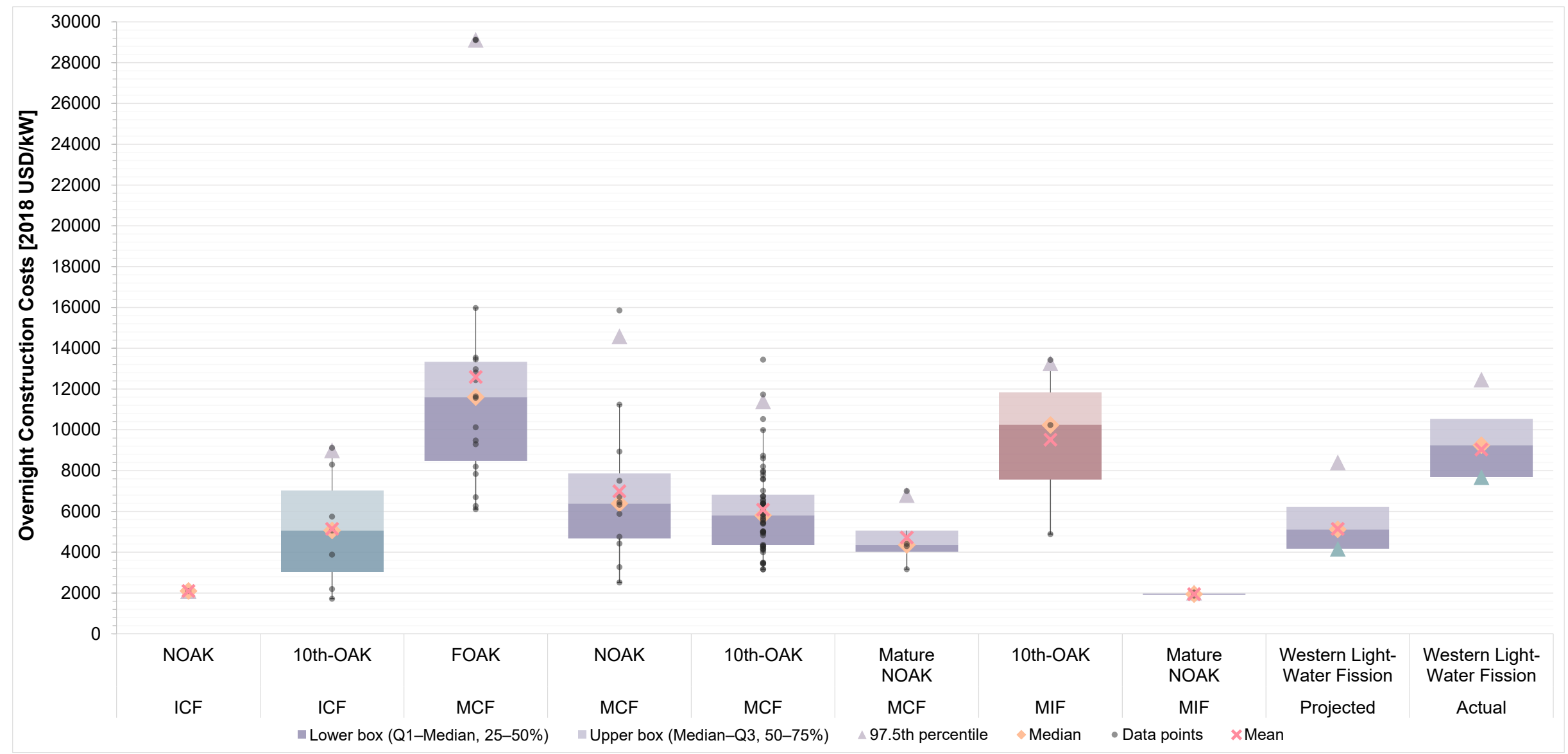

**Figure 1: Assumed overnight construction costs for FPPs by confinement type and technological maturity level as reported in literature.**

Source: Own compilation and visualization based on literature-reported data points from (Badger et al. 1975; Baker et al. 1981; Tsoulfanidis 1981; Blink et al. 1985; Maya, Schultz, and Battaglia 1985; Berwald, Whitley, and Garner 1985; J. Sheffield et al. 1986; Lee 1987; J. G. Delene et al. 1988; Logan et al. 1990; Spears 1990; J. G. Delene 1991; Miller et al. 1991; F. Najmabadi et al. 1991; Moir 1992; Waganer, Driemeyer, and Lee 1992; Krakowski 1995; M. Kikuchi 1996; Lako and Seebregts 1998; Jerry G. Delene et al. 2001; Tokimatsu et al. 2003; Dolan, Yamazaki, and Sagara 2005; Farrokh Najmabadi et al. 2006; Han and Ward 2009; Kozaki, Imagawa, and Sagara 2009; Dragojlovic et al. 2010; Bednyagin and Gnansounou 2011; Oishi et al. 2012; Helsley and Burke 2014; Chen et al. 2015; Joh Sheffield and Milora 2016; Woodruff et al. 2017; Entler et al. 2018; Hiwatari and Goto 2019; Hsu, Woodruff, and Nehl 2020; Takeda, Sakurai, and Konishi 2020; Roulstone 2021; Lindley et al. 2023).

Abbreviations used: FOAK: First-of-a-kind; ICF: Inertial confinement fusion; kW: Kilowatt; MCF: Magnetic confinement fusion; MIF: Magneto-inertial fusion; NOAK: Nth-of-a-kind; Q1: First quartile; Q3: Third quartile; 10th-OAK: Tenth-of-a-kind.

### 1.2.2 Model Parameters

We conduct a Monte Carlo simulation to calculate LCOE and implied learning rates for the fusion concept and maturity level combinations listed in Table 1. Following the distributional selection criteria outlined in Section 4.1, lognormal distributions are applied to all confinement-maturity combinations with sufficient data coverage, given the positively skewed pattern of reported construction cost estimates. For MIF 10th-OAK, a triangular distribution is applied, reflecting the narrow empirical basis of only two data points. NOAK ICF, FOAK ICF, and mature NOAK MIF are excluded from the simulation due to insufficient data coverage. Table 1 summarizes the full parameterization for the model. For fixed O&M costs, lognormal distributions are applied where data permit, with a normal distribution used for MCF 10$^{th}$-OAK given its more symmetric spread; all remaining parameters are held constant at median reported values or, where only a single estimate is available, at that reported value. Detailed analyses of the remaining parameters are provided in Appendix B.

**Table 1: Statistical and fixed input parameters for Monte Carlo-based LCOE simulations by confinement type and technological maturity level.**

| Parameter | ICF 10$^{th}$-OAK | MCF FOAK | MCF NOAK | MCF 10$^{th}$-OAK | MCF Mature NOAK | MIF 10$^{th}$-OAK |
|---|---|---|---|---|---|---|
| Overnight construction costs [USD/kW] | Lognormal (μ = 8.391; σ = 0.634); [min-max: 1,718–9,122] | Lognormal (μ = 9.339; σ = 0.443); [min-max: 6,133–29,131] | Lognormal (μ = 8.735; σ = 0.506); [min-max: 2,519–15,856] | Lognormal (μ = 8.658; σ = 0.333); [min-max: 3,161–13,440] | Lognormal (μ =8.418; σ = 0.328); [min-max: 3,164–7,004] | Triangular (a = 4,889; c = 12,391; b = 13,427) |
| Fixed O&M costs [USD/kW] | Lognormal (μ = 4.905; σ = 0.119); [min-max: 114.77–171.81] | Lognormal (μ = 4.481; σ = 0.273); [min-max: 56.48–137.84] | Lognormal (μ = 4.588; σ = 0.305); [min-max: 70.6–190.81] | Normal (μ = 100.59; σ = 28.68); [min-max: 20.47–172.69] | 95.223 | Lognormal (μ = 5.206; σ = 4.833); [min-max: 125.55–265.05] |
| Variable O&M costs [USD/MWh] | 14.858 | 9.990 | 15.012 | 14.858 | 8.604 | - |
| Fuel costs [USD/MWh] | 1.113 | 0.125 | 0.0457 | 0.115 | 0.115 | 0.0236 |
| Discount rate (%) | 7 | 7 | 7 | 7 | 7 | 7 |
| Capacity factor | 0.76 | 0.75 | 0.75 | 0.75 | 0.8 | 0.9 |
| Construction time [years] | 6 | 8 | 6 | 6 | 6.3 | 6 |
| Plant lifetime [years] | 30 | 30 | 40 | 30 | 40 | 30 |
| Capacity [MW] | 1,000 | 1,000 | 1,000 | 1,000 | 1,000 | 1,000 |
| Number of experiments N | 10,000 | 10,000 | 10,000 | 10,000 | 10,000 | 10,000 |

Note: Parameters treated stochastically in the Monte Carlo simulation (i.e., OCC and fixed O&M costs) are represented by probability distributions. Reported minimum–maximum values denote the range of literature data used to parameterize the distributions. All other parameters are held constant at the median of reported literature values, or at the single reported value where applicable. All cost values are given in constant 2018 USD. Abbreviations used: FOAK: First-of-a-kind; ICF: Inertial confinement fusion; MCF: Magnetic confinement fusion; μ: Mean; MIF: Magneto-inertial fusion; MW: Megawatt; NOAK: Nth-of-a-kind; σ: Standard deviation; 10th-OAK: Tenth-of-a-kind.

The derivation of implied learning rates is based on the reduction of OCC between FOAK and $10^{th}$-OAK configurations, capturing the most relevant early learning phase during which major design optimization, supply-chain development, and manufacturing experience are expected to occur. Given the absence of reported FOAK cost estimates for ICF and MIF, FOAK cost levels for these technologies are approximated by applying a risk premium to MCF FOAK estimates, adjusted for relative differences in technology maturity levels. ICF concepts are currently assessed at TRL 2–3 for laser-driven approaches and TRL 1 for ion-beam configurations, compared to TRL 4–5 for tokamaks and TRL 3 for stellarators, leading to a FOAK cost adjustment of approximately 30% for ICF (Wurbs et al. 2024). For MIF, which remains at an even earlier development stage with limited experimental validation, a higher adjustment of approximately 50% is applied (Wurden et al. 2016). These risk-adjusted FOAK values serve exclusively to enable consistent learning-rate estimation across confinement types and are not used in the LCOE simulations.

# 2 Results

This section introduces the stochastic modelling results for both expected LCOE and learning rates.

## 2.1 LCOE Modelling

We describe the model results before showing a conducted sensitivity analysis for WACC values.

### 2.1.1 Stochastic Results

Figure 2 summarizes the model results. We see that substantial uncertainty exists for many fusion devices; and the distribution of results is higher for lower maturity levels. Median values for FOAK, NOAK and $10^{th}$-OAK MCF devices are 206.7, 121.2 and 114.6 USD per MWh, respectively. From the literature obtained, systematically lower values for higher maturity levels are to be expected. Although it is noteworthy that NOAK costs for MCF devices are higher – on average – than $10^{th}$-OAK costs, implying that n might be less than 10 for many studies. MIF devices, although assumed to be less complex, are expected to cost more (mean value of 143.9 USD per MWh); and after “mature NOAK” MCF devices, $10^{th}$-OAK ICF are expected to be the cheapest at a mean cost of 110.3 USD per MWh. Figures 3 to 8 show the distribution of the model results that is distinctly leftward skewed, apart for MIF devices.

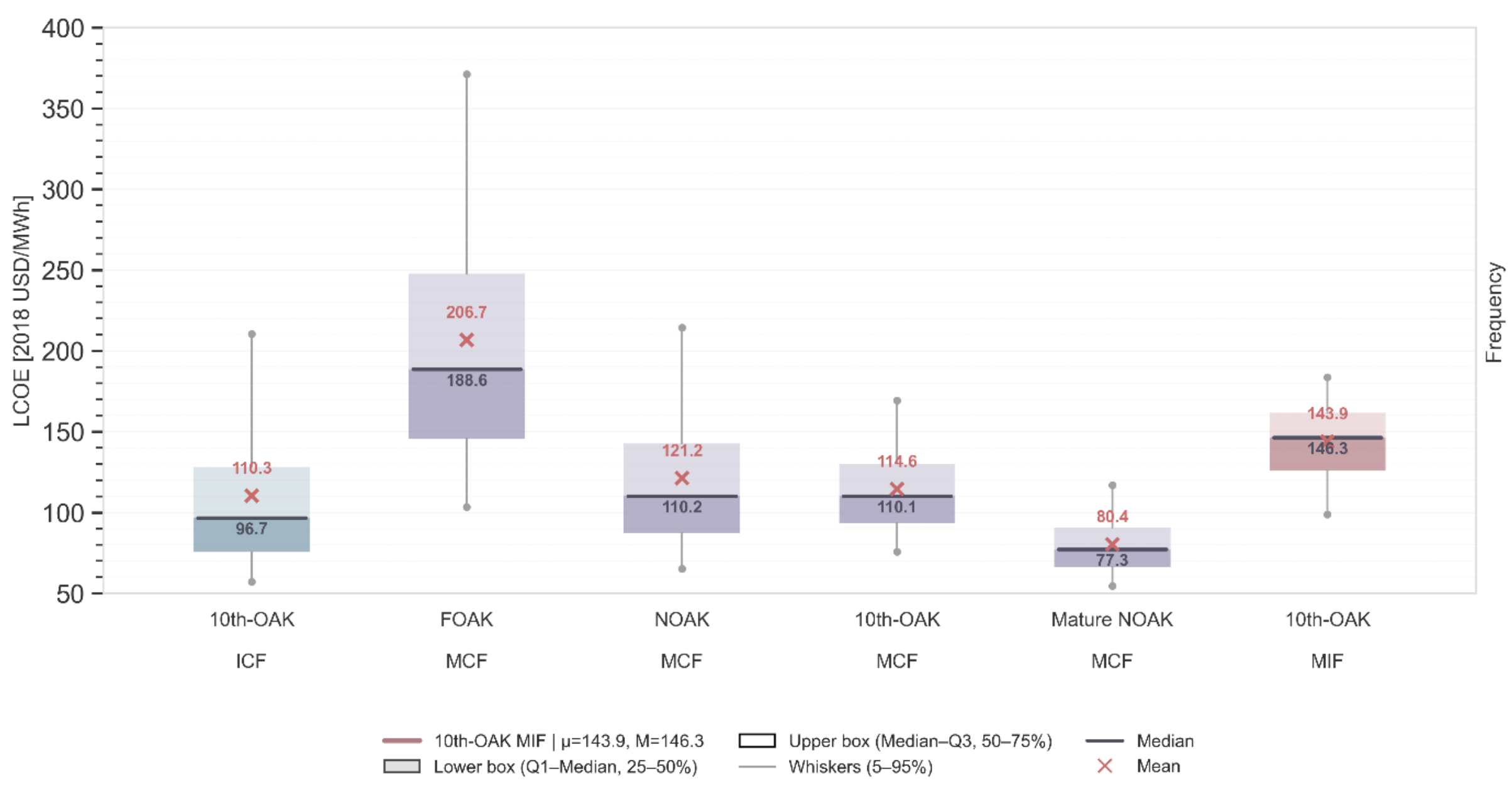

**Figure 2: Boxplots of LCOE simulations**

Source: Own calculations.

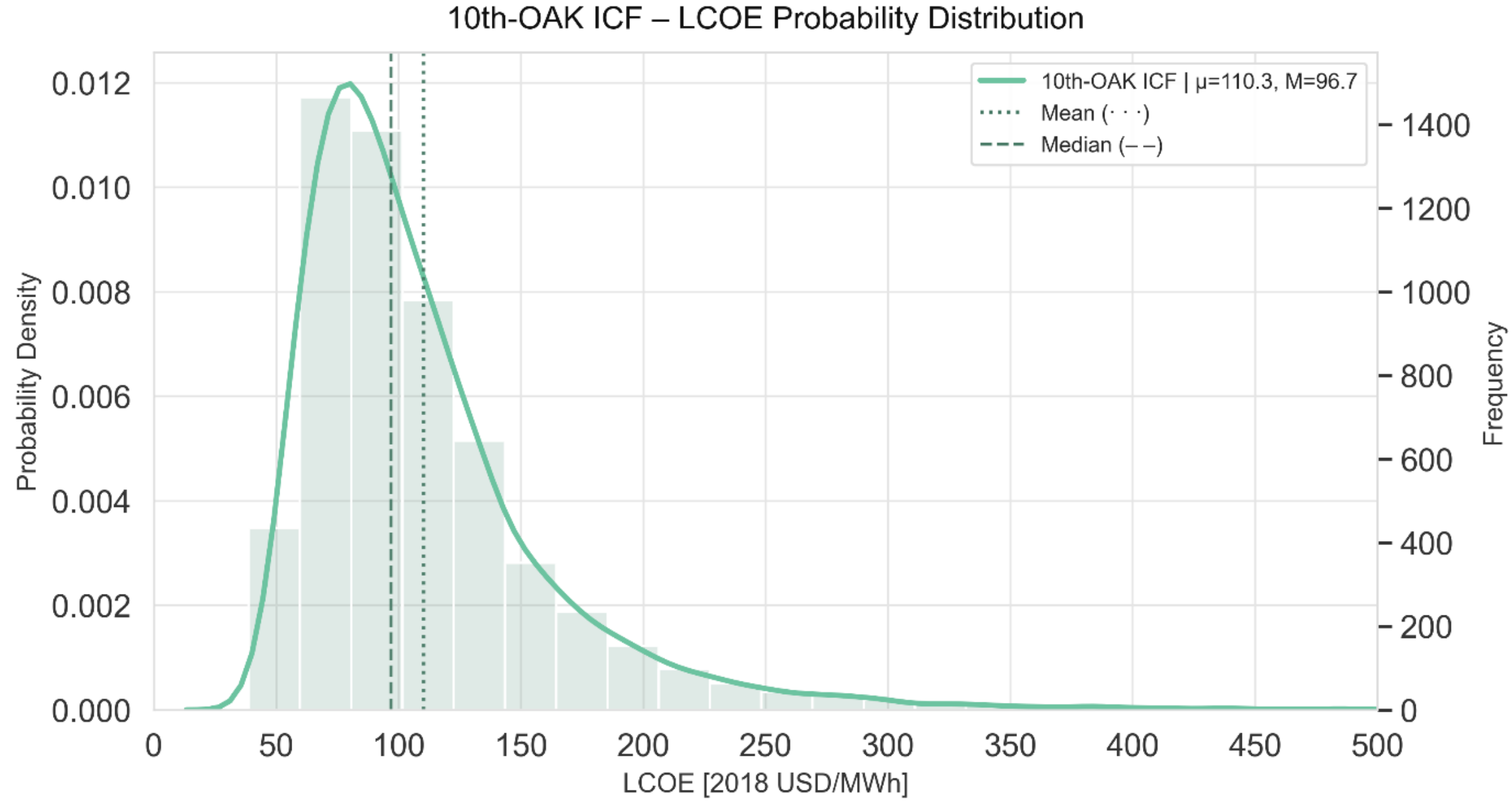

**Figure 3: Distribution of 10th-OAK LCOE for ICF devices**

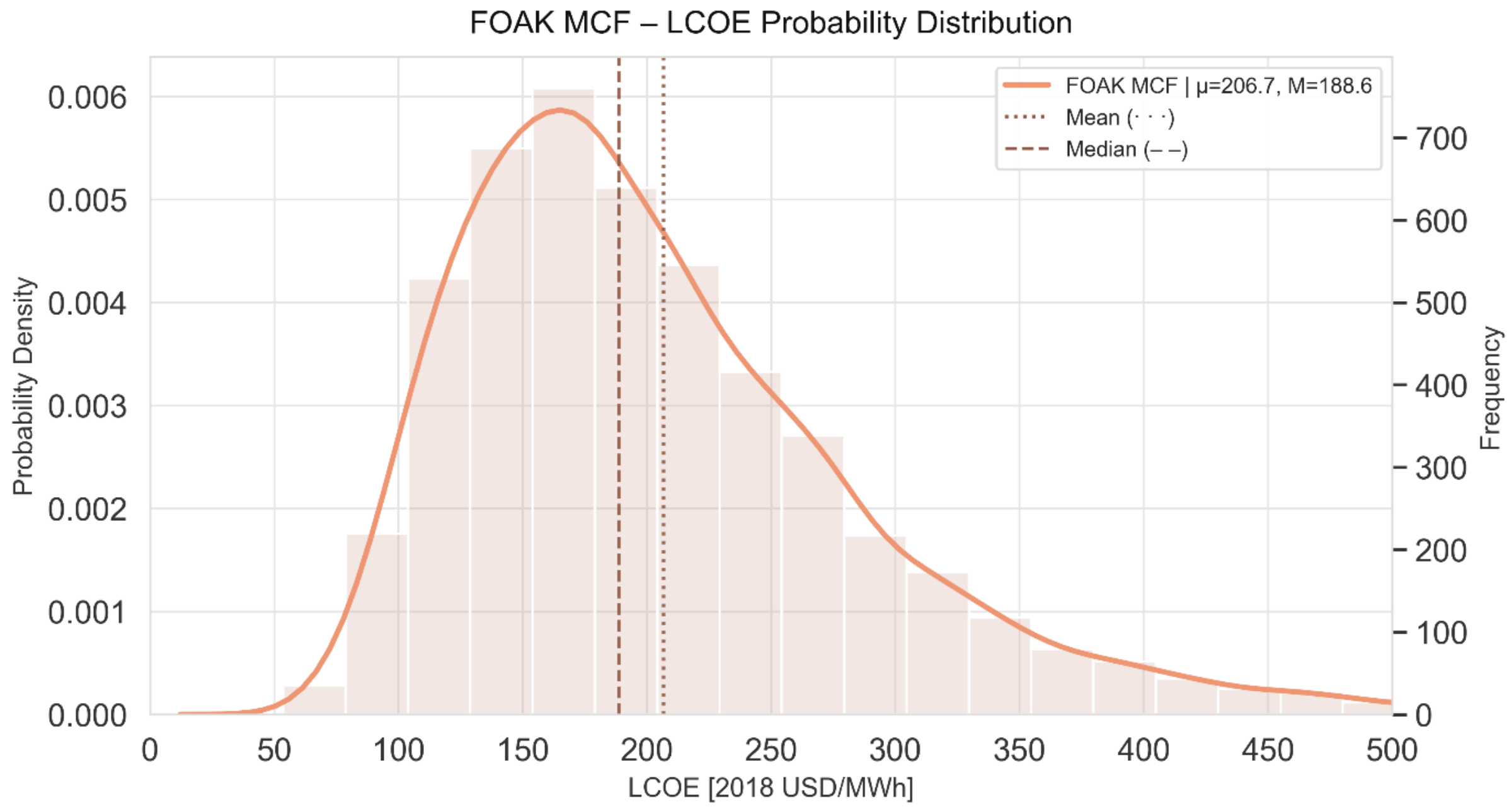

**Figure 4: Distribution of results for FOAK costs for MCF**

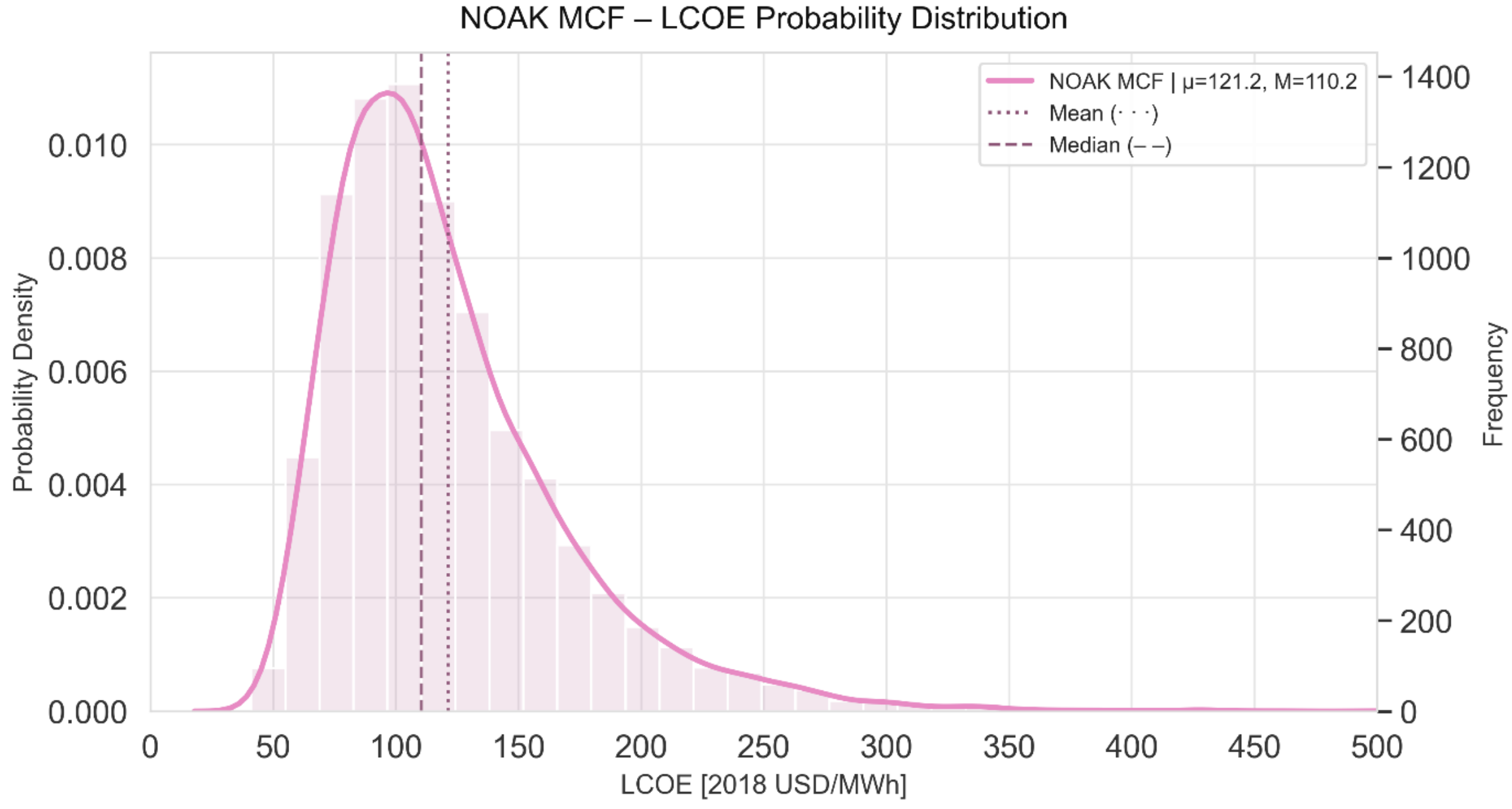

**Figure 5: Distrubtion of results for NOAK costs for MCF**

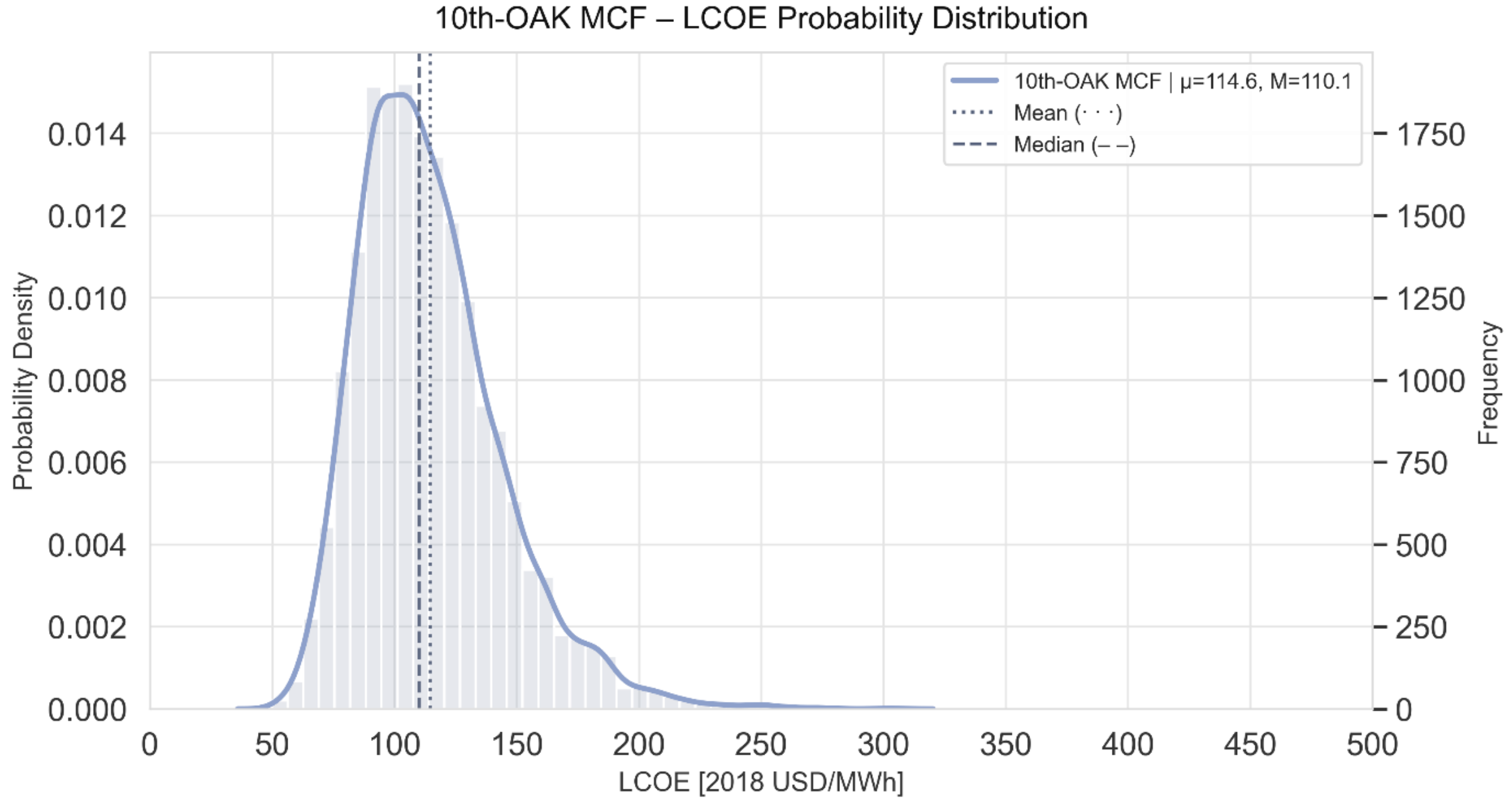

**Figure 6: Distribution of 10th-OAK results for MCF devices**

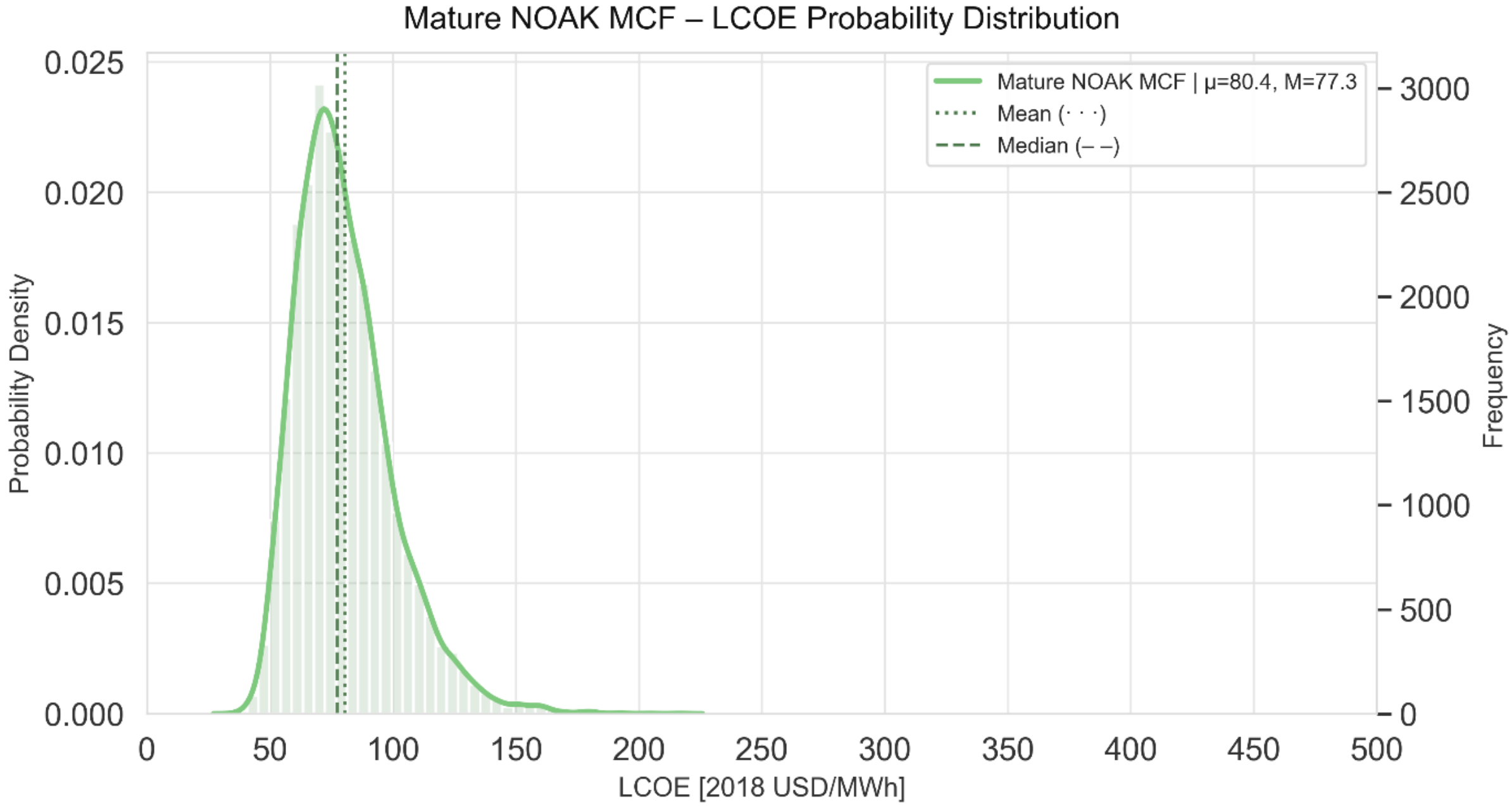

**Figure 7: Distribution of mature NOAK MCF devices**

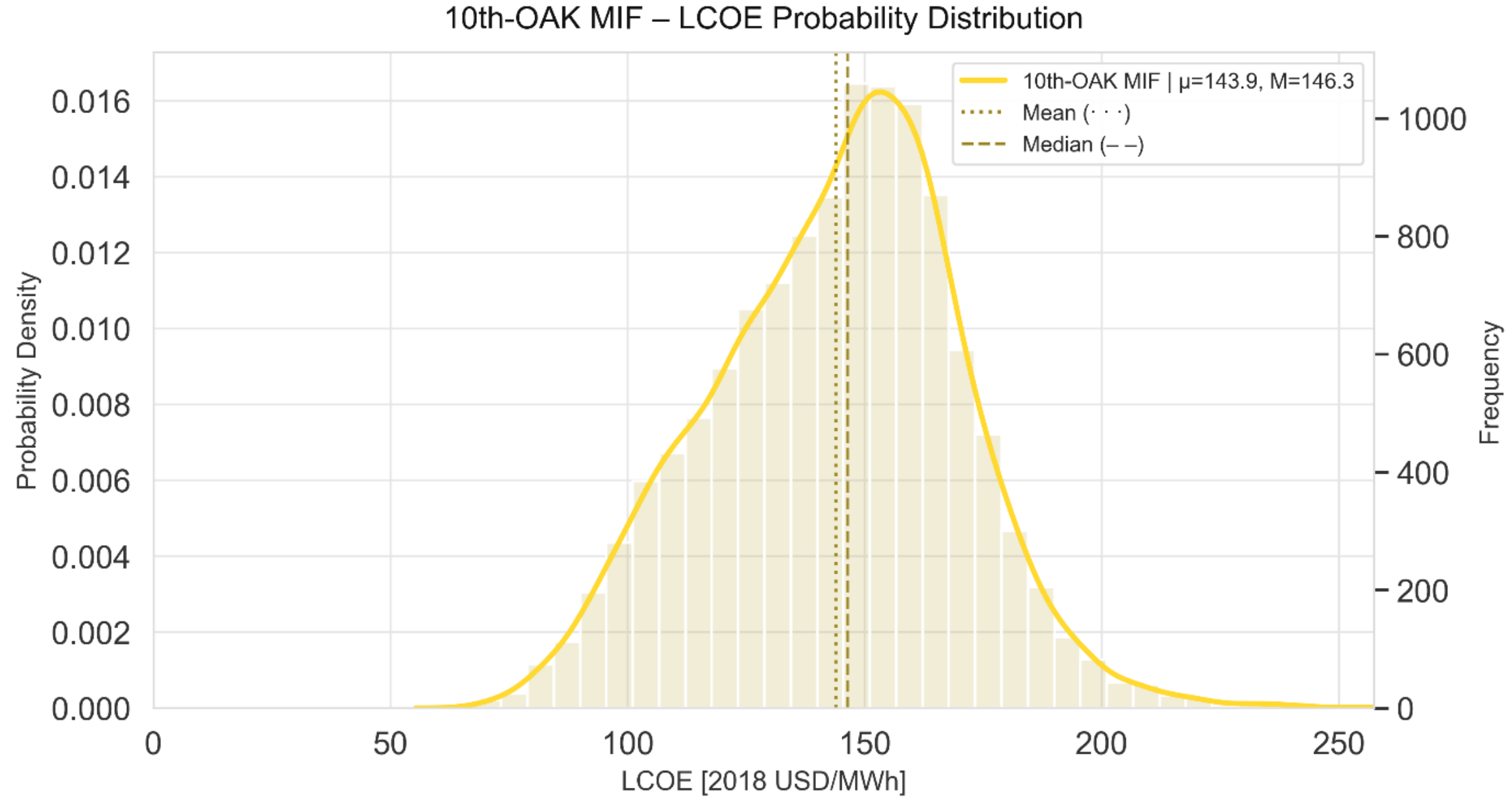

**Figure 8: Distribution of 10th-OAK costs for MIF devices**

### 2.1.2 Sensitivity

We conduct a sensitivity analysis by reducing the conservative estimate of 7% WACC to 3 and 5% in two respective scenarios. This leads to a leftward shift of the LCOE distributions to lower generation cost levels, confirming that fusion costs are dominated by upfront capital expenditures. Depending on confinement type, maturity level, and the assumed WACC, mean and median LCOE values decline by values ranging from 19 to 38%, highlighting the strong sensitivity to financing conditions. In contrast, the overall spread and skewness of the distributions remain largely unchanged. This indicates that variations in discount rates primarily affect absolute cost levels rather than the underlying cost uncertainty, which continues to be driven by assumptions on OCC. Figures 9 to 14 show the results of the sensitivity analysis.

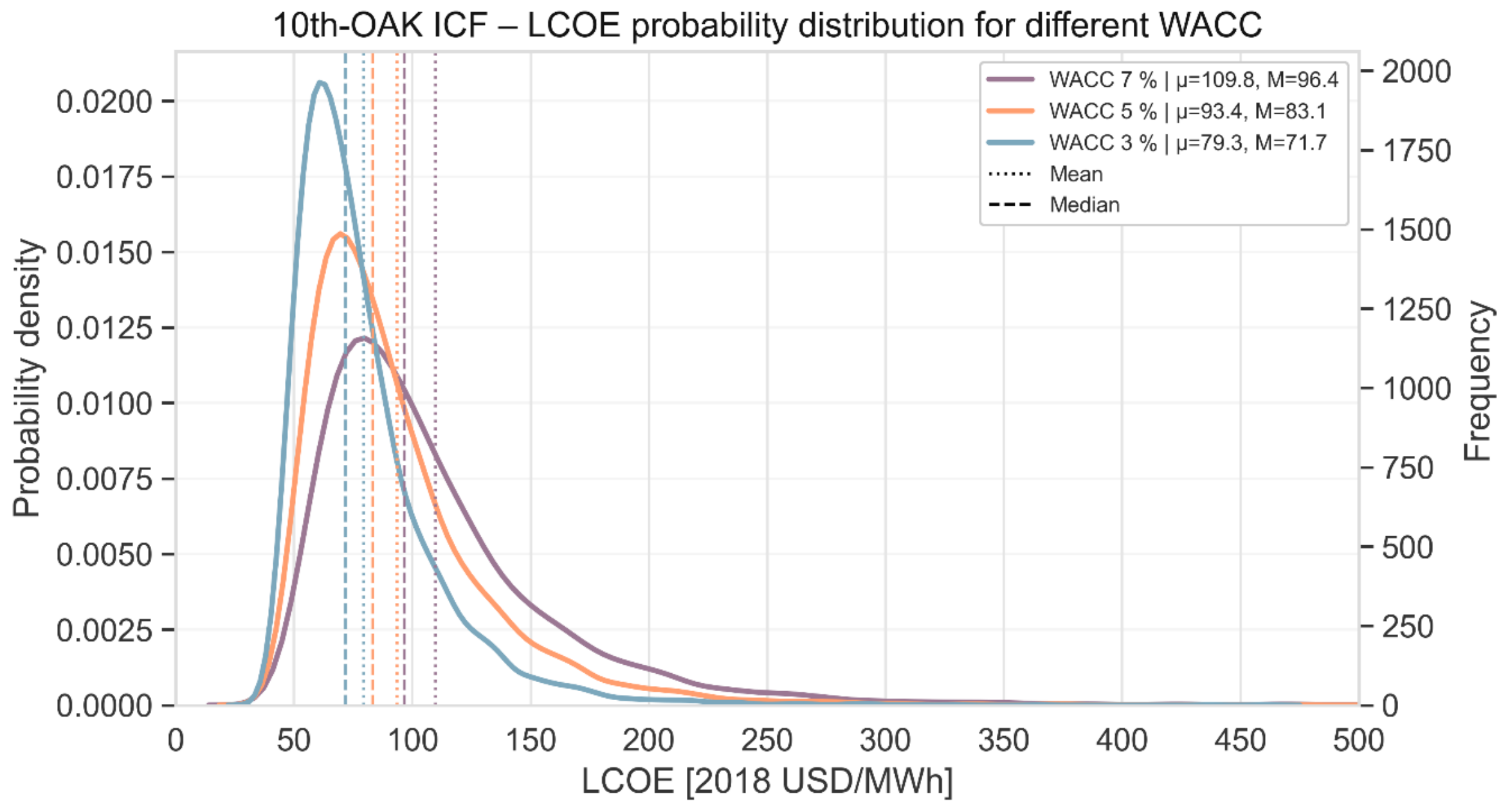

**Figure 9: Results of sensitivity analysis for 10th OAK ICF devices**

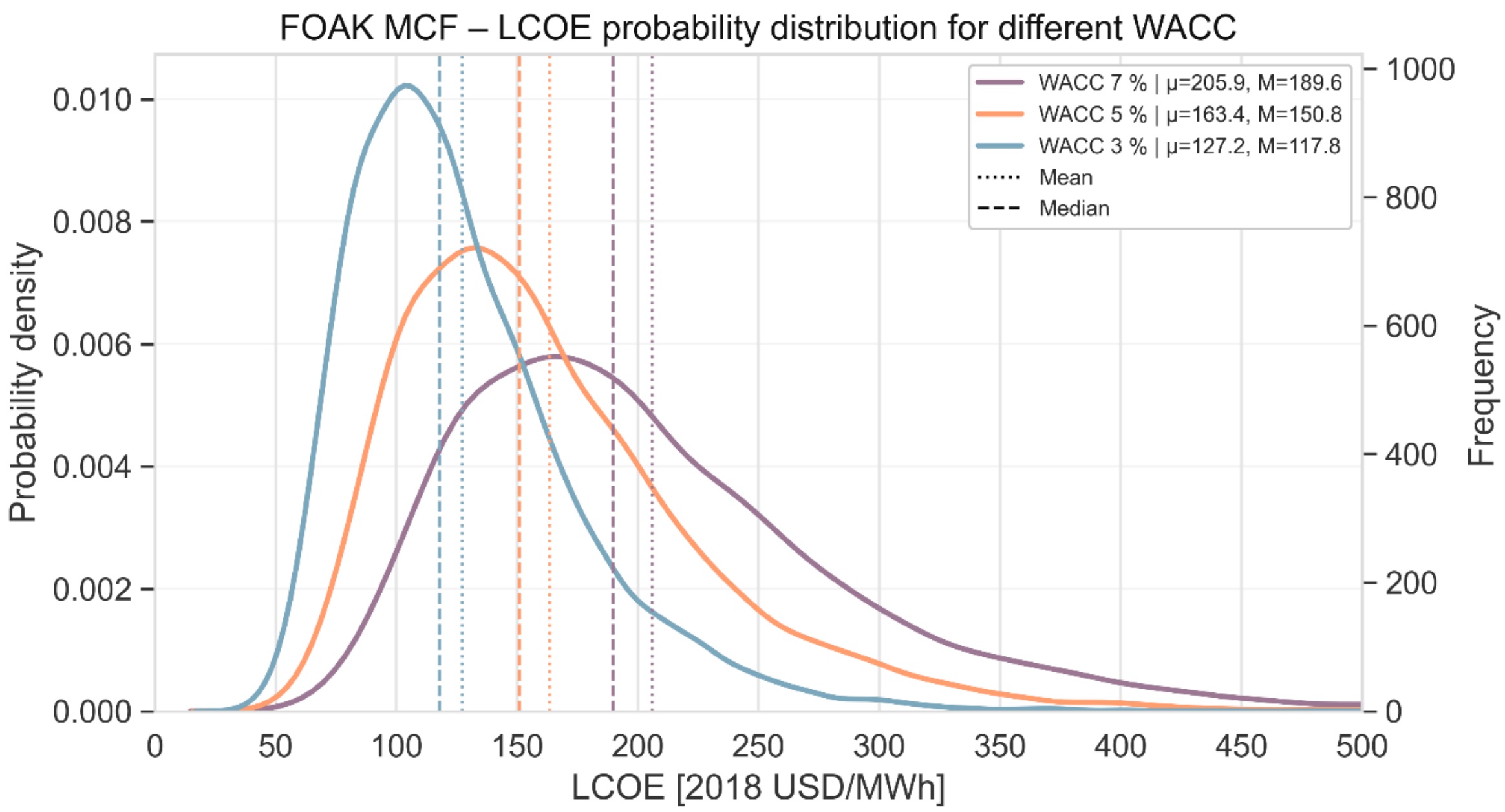

**Figure 10: results of sensifitivy analysis of FOAK MCF devices**

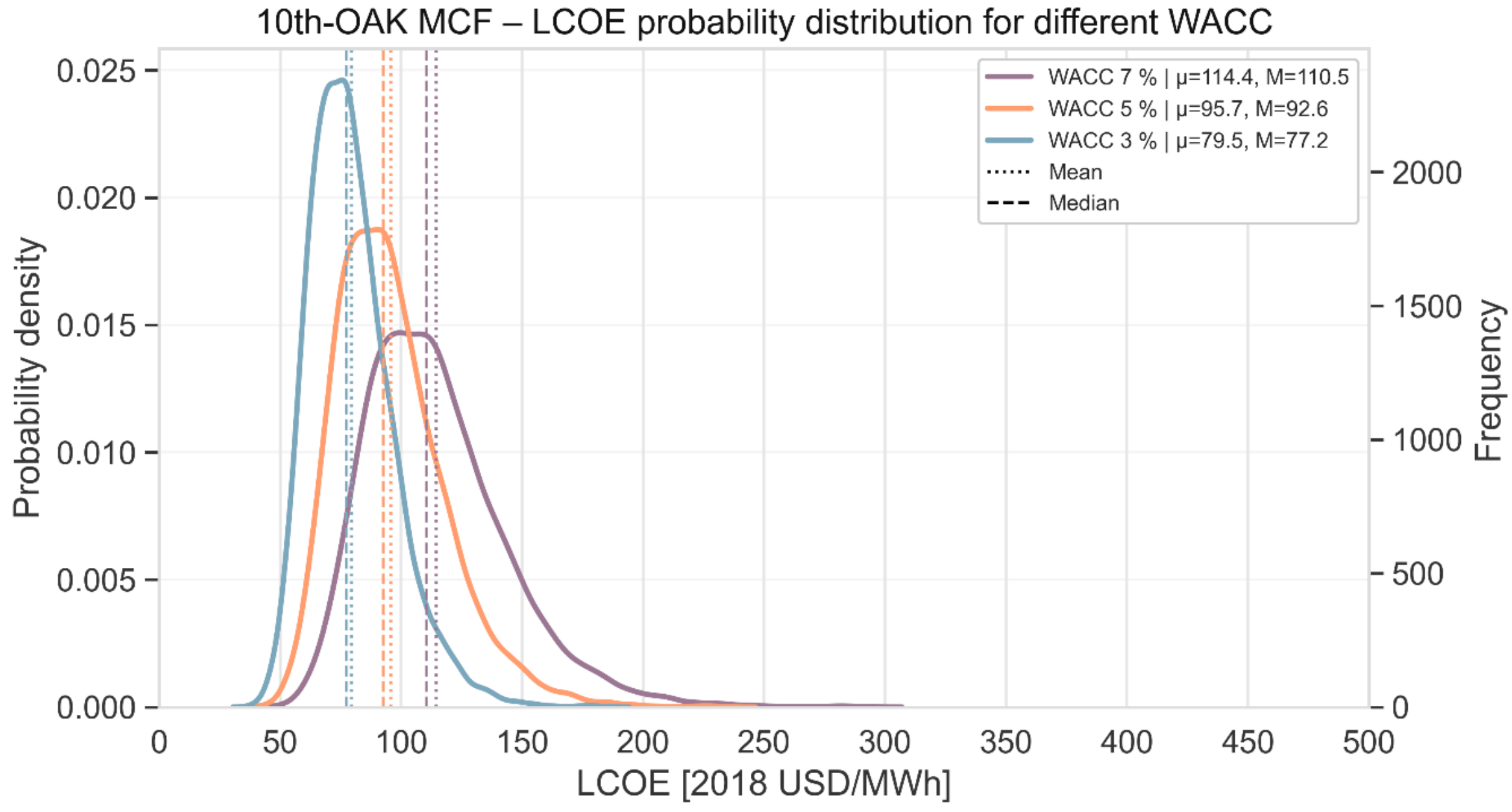

**Figure 11: Results of distribution for 10th OAK MCF**

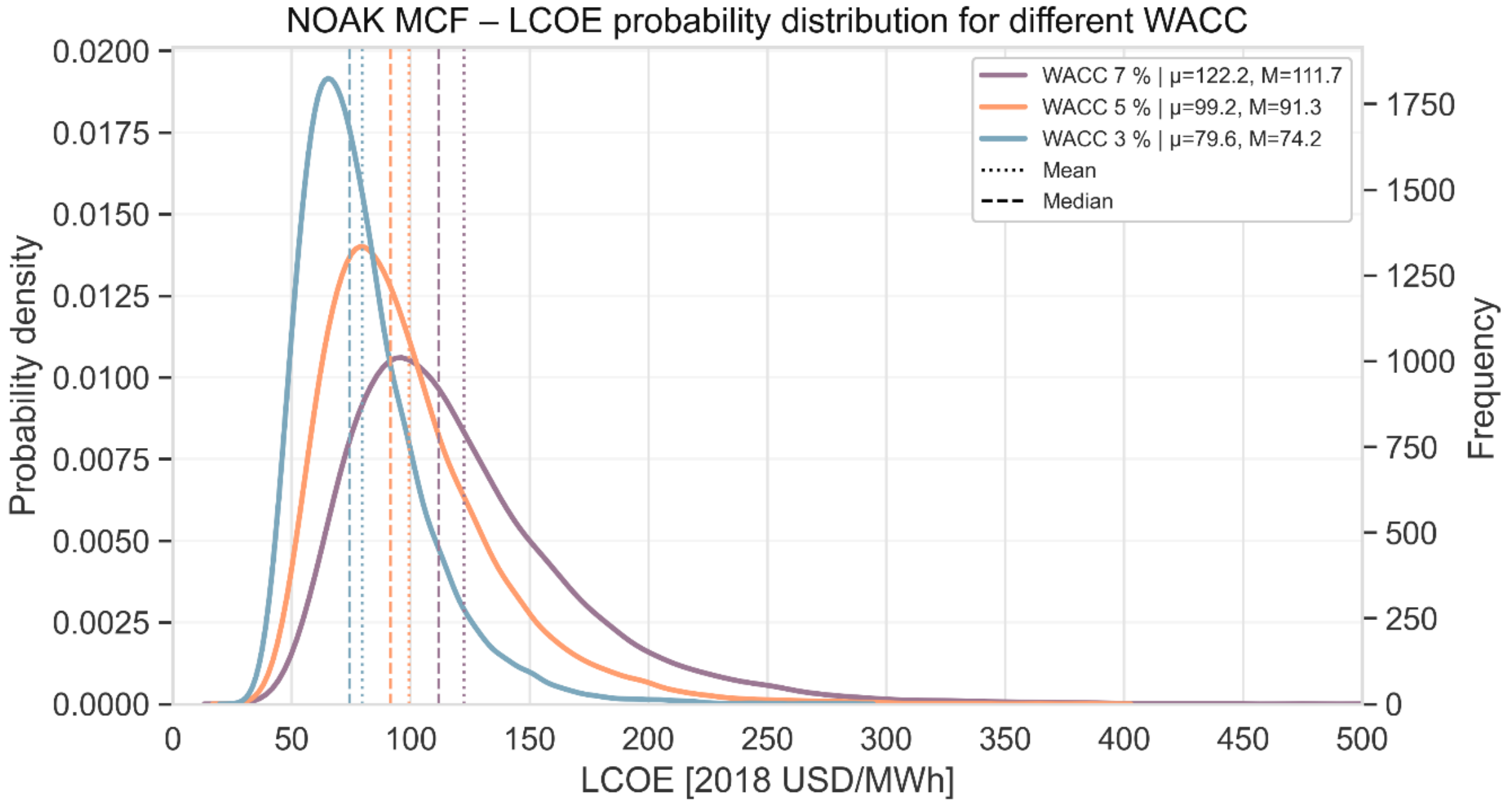

**Figure 12: Results of distribution for NOAK MCF**

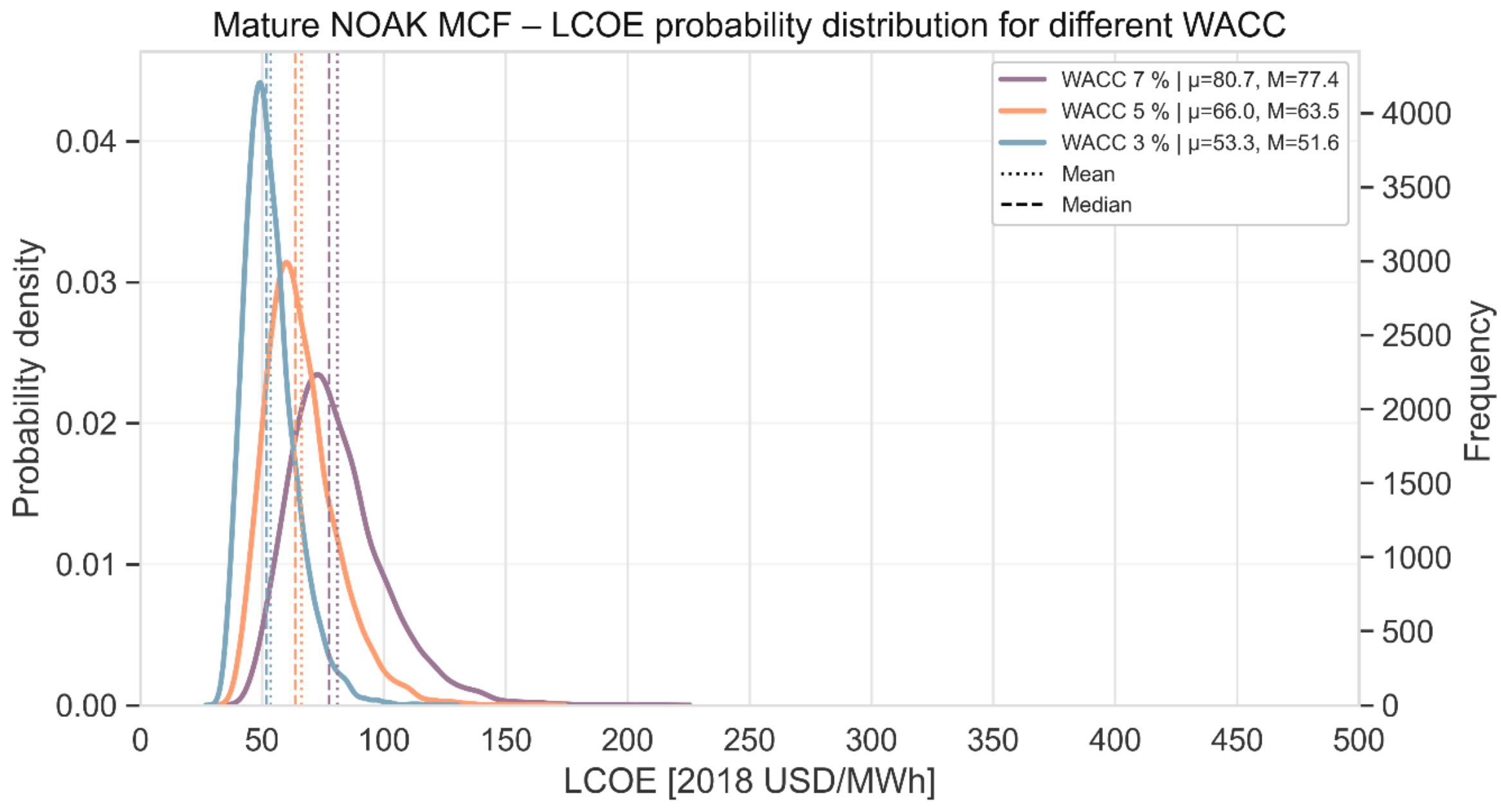

**Figure 13: Results of distribution mature NOAK**

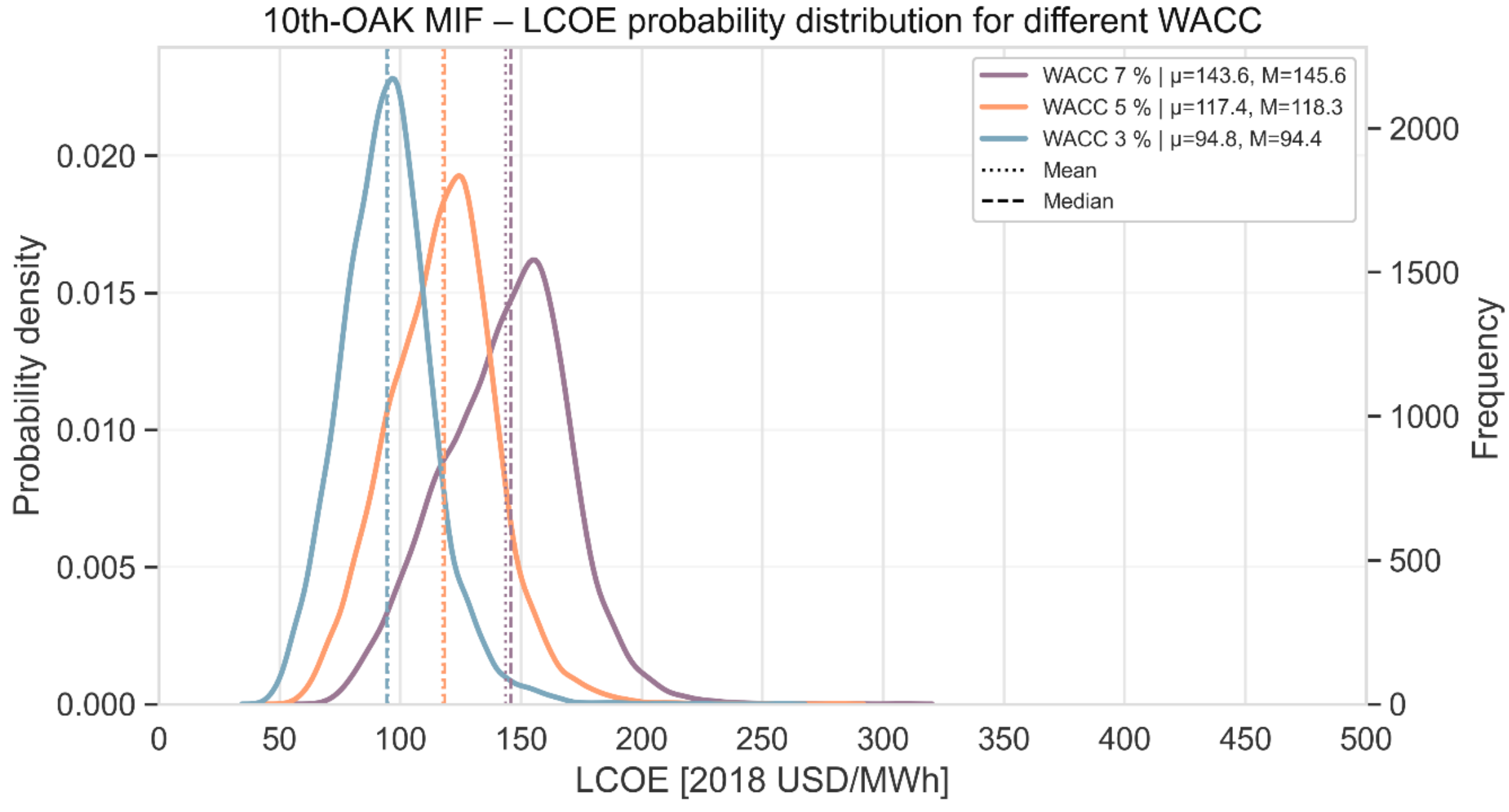

**Figure 14: Results of distribution 10th OAK MIF**

## 2.2 Learning Rates

Finally, we determine the implied learning rates as introduced in Section 1. The learning rates are calculated for FOAK to 10$^{th}$-OAK cost for MCF, ICF, and MIF devices, respectively. The results are shown in Figures 15 to 18. Mean learning rates for ICF rates are 30%, 19.5% for MCF, and 20.7% for MIF devices.

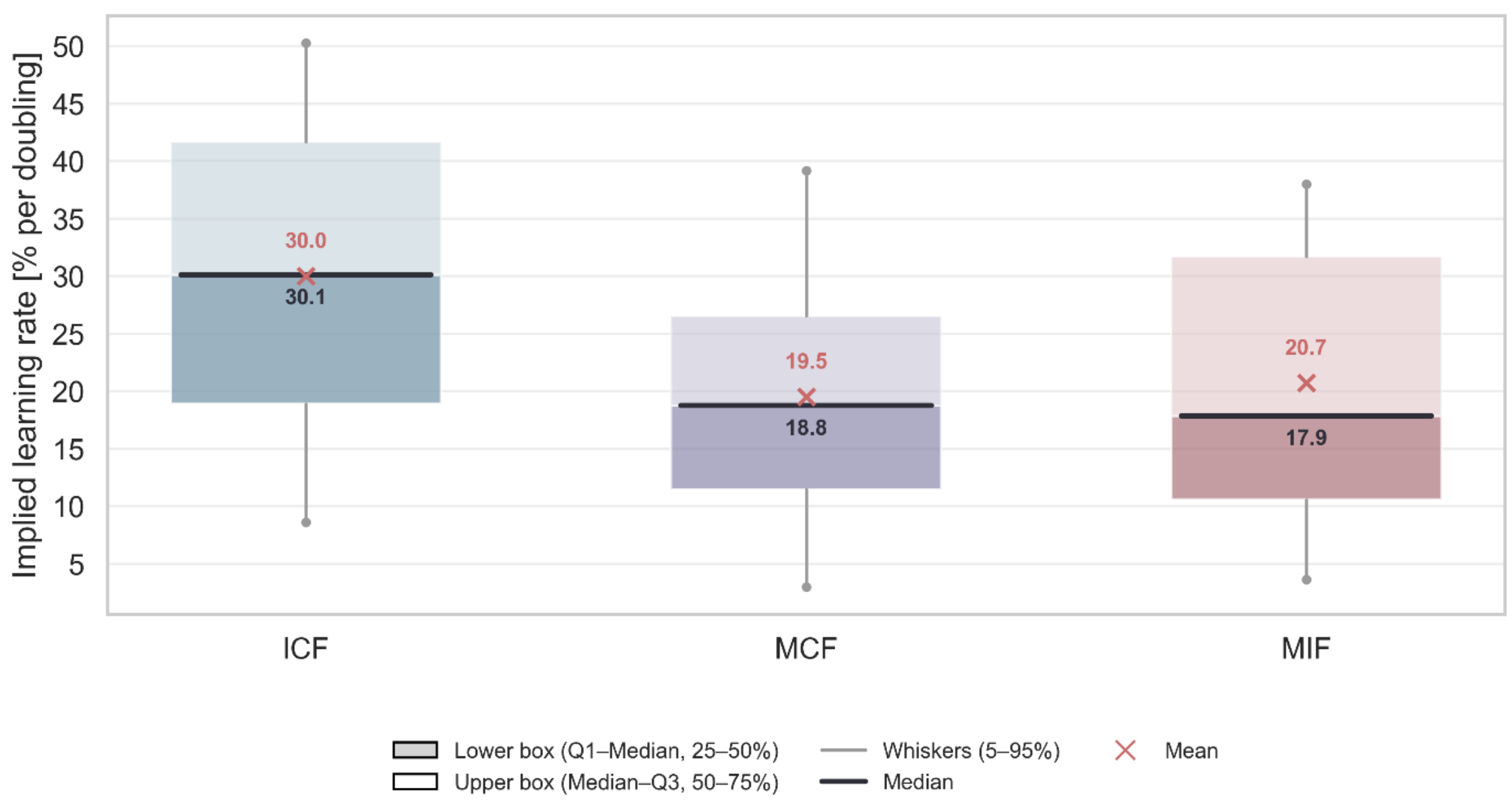

**Figure 15: Boxplots of assumed learning rates.**

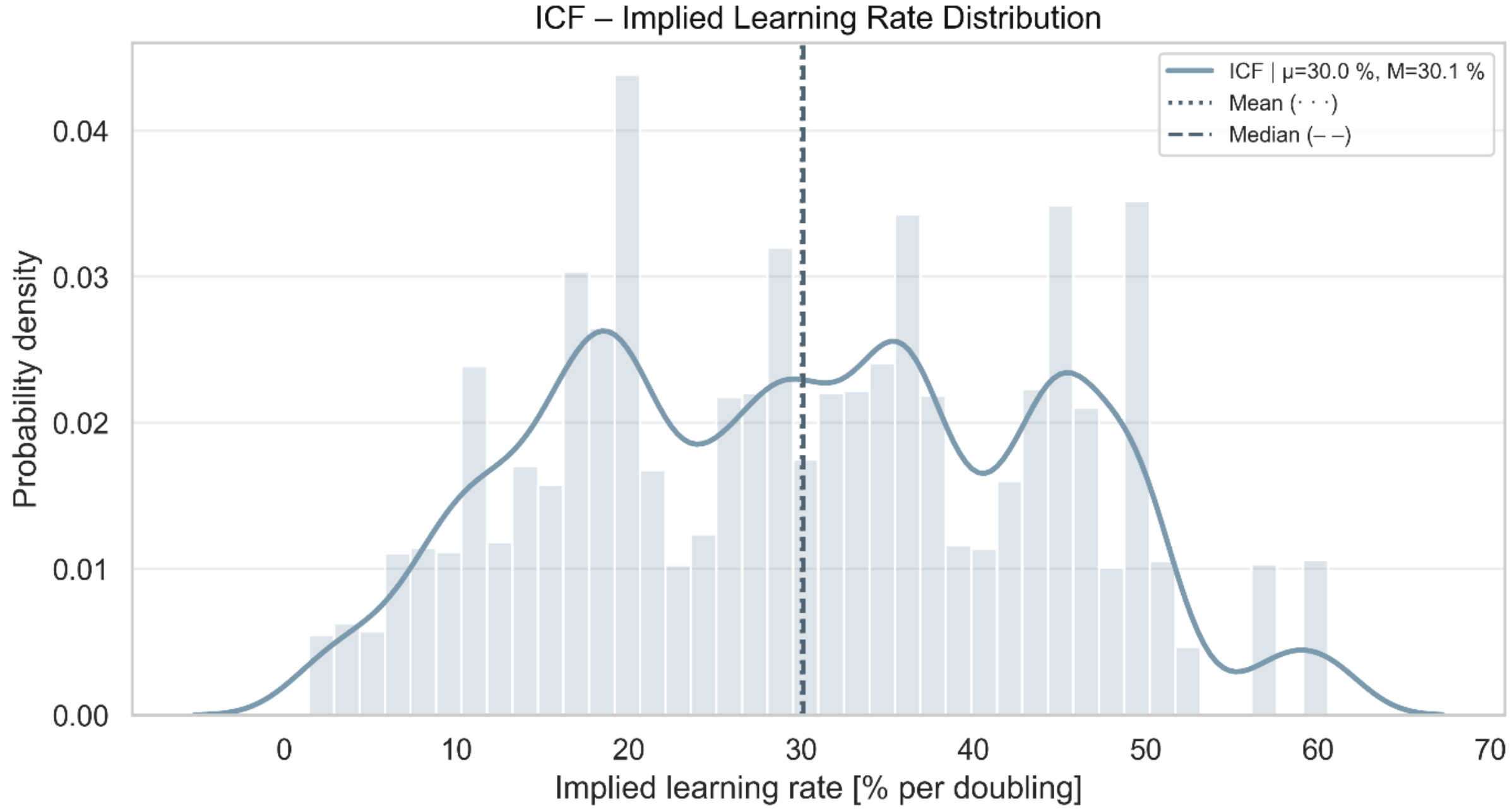

**Figure 16: Distribution of implied learning rates for ICF devices.**

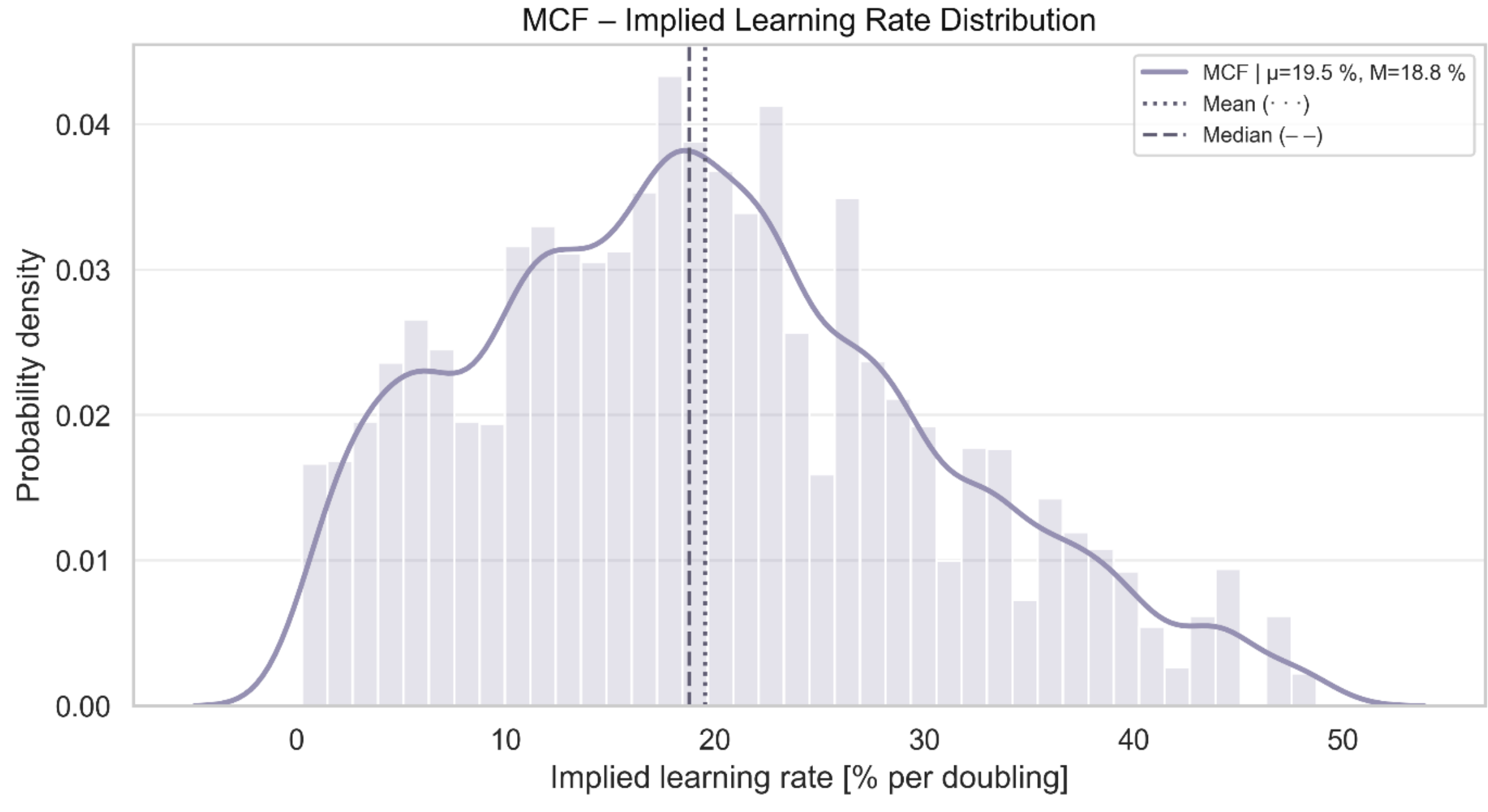

**Figure 17: Distribution of implied learning rates for MCF devices.**

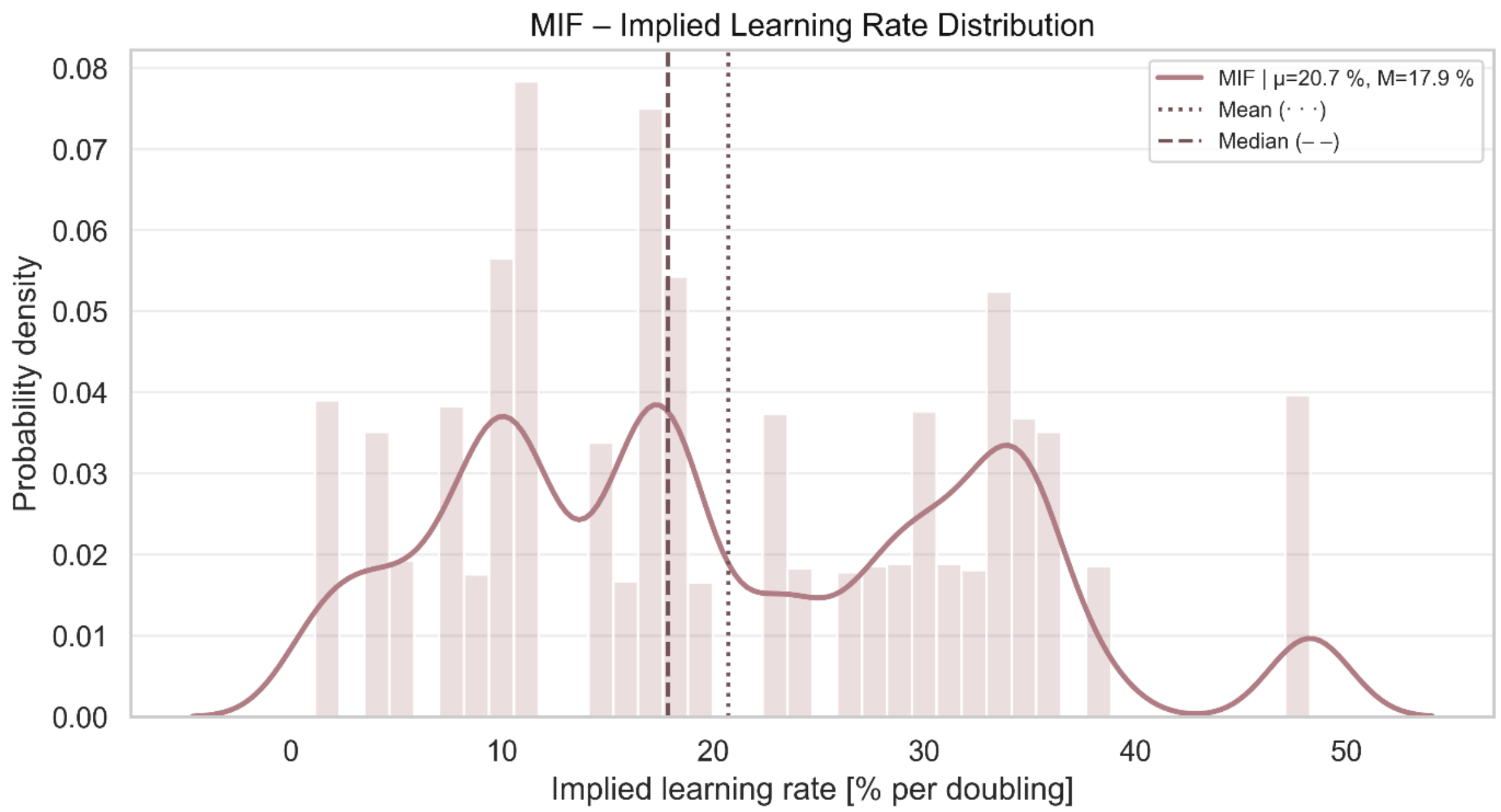

**Figure 18: Distribution of implied learning rates for ICF devices.**

# 3 Discussion

Our results show expected LCOE averaging 110.3, 114.6, and 143.9 USD per MWh for 10$^{th}$-OAK ICF, MCF, and MIF devices, respectively. These average values reflect expectations found in a broad body of literature containing subjective assumptions. The obtained expected or implied learning rates from FOAK to 10$^{th}$-OAK costs average 30% for ICF, 19.5% for MCF, and 20.7% for MIF devices.

Given the low technological development stages of fusion devices and the uncertainty of them ever becoming utilizable for commercial energy production, these results must be interpreted carefully, as the expectations are likely to have been shaped by overly optimistic experts. This is in line with findings by Böse et al. (2026) who find that, especially in the nuclear sector, that implicitly includes the fusion sector, the technological expectations are overly optimistic and might even shape global energy scenarios and, consequently, political decisions. The cost ranges are in the range of current light-water reactor fission technologies, that are already unable to economically compete against existing low-carbon alternatives like renewables in combination with batteries (Göke, Wimmers, and Von Hirschhausen 2025). These findings also coincide with Schwartz et al. (2023) who find that fusion would have to be even cheaper than fission to compete in a future US energy system; an unlikely scenario given the complexity increase of fusion compared to fission (Dering, Wimmers, and Von Hirschhausen 2026).

Regarding learning rates, our results also indicate a high degree of optimism regarding potential learning. Tang et al. (2026) find that, in general, fusion learning rates are overestimated, which is empirically validated by our results. Compared to historical learning rates for comparable technologies, like fission plants, obtained implied learning rates based on the data, are infeasible. Such an overly optimistic view is mirrored by proponents of fission reactor concepts like small modular reactors whose costs are expected to decrease via the same principle, albeit with lack of empirical evidence and unclear targets, such as a clear determination of n in n-th-of-a-kind (Steigerwald et al. 2023).

Consequently, the economic prospects of fusion power plants remain highly uncertain. Given the high number of unresolved technical issues and low development maturity, it is unlikely that fusion power plants will operate in the next decades. Therefore, fusion cost studies must be carefully scrutinized regarding their scope, design, and main assumptions. Future costs of fusion will depend on the selected technology line and therein, the selected device, its fuel, the mode of operation, and its capability to operate in a system dominated by fluctuating renewable energies. Even under these uncertainties, our determination shows that fusion is likely to be much more expensive than technologies that exist today; in times of increasingly rising global temperatures we therefore urge policymakers to focus on renewables and other options to decarbonize our global energy systems instead of waiting for a technology whose technological – and subsequentially economic – feasibility remains so uncertain.

# 4 Methods

## 4.1 Stochastic Model

The Monte Carlo method is applied to estimate the LCOE and learning-driven reductions in overnight construction costs of FPPs by repeatedly sampling input parameters from predefined probability distributions and recording the resulting output distributions as shown in Table 1. The resulting probabilistic cost estimates are subsequently compared with observed electricity costs and learning rates of other power generation technologies to assess the potential economic competitiveness of fusion and the attainability of the required cost reductions (Saltelli et al. 2007). The uncertainty associated with input parameters is represented by probability density functions derived from reported values in the fusion cost literature and available data. For parameters with sufficient data coverage exhibiting a positively skewed pattern – as is typical for capital cost data where upper-end outliers arise from design complexity or material choices – lognormal distributions are applied, as they constrain values to positive numbers and accommodate right-skewed dispersion. Where data are too sparse to reliably parameterize a lognormal fit but minimum, central, and maximum values can be identified, triangular distributions are applied, as they require only three interpretable inputs and make no assumptions about distributional symmetry. Where reported values show a more symmetric spread, a normal distribution is applied. Parameters with insufficient data variation for stochastic treatment are held constant at their median or single reported value.

For the LCOE simulation, 10,000 iterations are performed for confinement–maturity combinations with sufficient data coverage. In each iteration, uncertain techno-economic parameters – including OCC, fixed and variable O&M costs, fuel costs, discount rate, capacity factor, construction time, and plant lifetime – are sampled from their respective distributions. A subsequent sensitivity analysis varies the discount rate (WACC) across three representative scenarios (low, moderate, and high) while all other parameters remain stochastic or are held at their median values.

For the learning-rate simulation, 10,000 iterations are conducted by repeatedly pairing OCC estimates at the lowest and the most advanced maturity[1] levels – FOAK and 10th-OAK – to derive a stochastic distribution of implied learning rates. Learning beyond the 10th-OAK stage is excluded, as neither generic nor mature NOAK concepts correspond to a consistently specified cumulative deployment level across the reviewed studies, whereas the 10th-OAK stage represents a fixed deployment point with $n = 10$ in the learning-curve formulation. For ICF and MIF, where no FOAK estimates are available, risk-adjusted proxies are applied as described in Section 1.2.2.

## 4.2 Data Collection and Harmonization

Before performing the Monte Carlo simulations, a detailed examination of the cost data was conducted to identify appropriate statistical distributions and to assess the underlying assumptions of each dataset. Outliers were examined, differences across confinement types and maturity levels analyzed, and confinement–maturity combinations for which insufficient data were available were excluded from the simulations.

The analysis draws on a harmonized dataset comprising 590 entries from 56 underlying studies, of which approximately 60% are peer-reviewed journal publications, 15% are peer-reviewed technical or journal-equivalent studies, 23% are established technical and conceptual FPP design reports, and one study corresponds to an unpublished or “grey-literature” source. The parameters extracted include OCC, fixed and variable O&M costs, fuel costs, discount rates, capacity factors, construction times, and plant lifetimes, complemented by harmonization metadata such as publication year, source, confinement concept, and assumed plant maturity level. The reviewed studies cover techno-economic assessments published between 1975 and 2024, encompassing ICF, MCF, and MIF concepts, with most released between 1985 and 2015. The dataset differentiates cost assumptions according to FOAK, NOAK, and 10th-OAK maturity levels. Where not explicitly stated, NOAK maturity is considered, while a 10th-OAK assumption is applied to specific reactor design studies representing commercialized technologies that follow the NOAK stage. The mature NOAK stage encompasses approximately the 75th–100th-of-a-kind units, representing fully developed and industrialized FPPs. All monetary values are standardized to 2018 price levels using U.S. and European consumer price indices and historical exchange rates (BBk undated; BLS undated; FRED 2025). Waste management, decommissioning, and replacement costs are excluded to avoid speculative variability and potential double-counting with reported O&M assumptions. A full overview of all cost-relevant input data is provided in Appendix C.

[1] In this study, “advanced maturity” refers to deployment stages beyond FOAK and NOAK, specifically including the 10th-OAK and mature NOAK levels.

Based on the consolidated and standardized dataset, probability distributions for all techno-economic parameters required for the LCOE and learning-rate Monte Carlo simulations are derived as stochastic model inputs, as described in the following section.

## Acknowledgements

The authors thank the participants of the 20th Enerday Conference held in March 2026 in Dresden, Germany, for their helpful comments on an early version of this work. We extend our gratitude to Sophia Spitzer, Charlotte Dering, Harald Lesch, Bill Nuttall and many others for guiding us towards the exciting topic of nuclear fusion economics.

## Data Availability

All literature data can be accessed in the supplementary material (appendices) provided. The modeling results will be provided upon request due to their large sizes. Code and additional files are available at https://github.com/alexwimmers/nucFusion/tree/main/MC-Fusion.

## Author Contributions

S.B.: Writing – Original Draft; Methodology; Formal Analysis; Investigation; Data curation; Validation; Visualization; F.B.: Validation; Writing – Review & Editing; C.H.: Validation, Writing – Review & Editing; C.K.: Writing – Review & Editing; A.W.: Conceptualization; Methodology; Investigation; Writing – Original Draft, Supervision

## Declarations of Interest

The authors declare no competing interests.

## Supplementary Information

Refer to relevant data used for the modelling as well as results from the literature analysis listed in the Appendices A, B, and C.

# Appendix A: In-depth Explanation of Fusion Reactor Types

## A.1 Magnetic Confinement Fusion

Magnetic confinement fusion (MCF) systems can broadly be classified into three categories: toroidal devices, linear devices and compact toroids (Galvão and Canal 2021).

Toroidal devices include the classical and spherical tokamak (ST), reversed-field pinch (RFP) and helical systems, such as the classical and spatial-axis stellarator, heliac and heliotrion/torsatron configurations. Among these, the classical tokamak is the most extensively investigated and technologically mature concept, with research efforts dating back to the late 1950s. Its design is typically doughnut-shaped, built around a central solenoid that functions as a transformer and is surrounded by poloidal and toroidal field coils. The centrally positioned solenoid produces a time-varying axial current that induces a toroidal electric field via electromagnetic induction. This field accelerates residual free electrons, causing collisions with neutral atoms that strip off additional electrons and thereby ionize the gas to form a plasma. Once formed, the induced field drives a toroidal plasma current that generates a poloidal magnetic field, contributing to plasma confinement and stability, while the current itself provides plasma heating. Together with the externally applied toroidal and poloidal fields that are created by the surrounding coils, a helical magnetic structure is established that guides plasma motion along twisted field lines around the torus, thereby maintaining confinement (Moynihan and Bortz 2023).

Since the variation of current through the central solenoid is limited, tokamaks are inherently pulsed machines. The pulse length corresponds to the duration of the plasma discharge before confinement collapses and fusion reactions cease, once the solenoid current decays and must be recharged. To achieve continuous, steady-state operation, the plasma current must therefore be maintained by non-inductive means after its initial induction by the central solenoid (Reinders 2021). This can potentially be accomplished by maximizing the self-generated bootstrap current, which arises from strong plasma density and temperature gradients, in combination with externally driven currents supplied by high-power heating systems such as neutral beam injection (NBI) or radio-frequency (RF) waves (Maisonnier et al. 2005). Neutral beam injection (NBI) systems heat the plasma by injecting high-energy neutral atoms that penetrate the magnetic field, become ionized through collisions, and transfer their momentum and energy to the plasma ions and electrons. In contrast, radio-frequency (RF) heating systems launch electromagnetic waves at specific resonant frequencies of the plasma, where they are absorbed by charged particles, increasing their kinetic energy and thereby sustaining plasma temperature and current (Reinders 2021). In such a configuration, the central solenoid is required primarily for plasma initiation, while steady-state operation can be maintained without its continued use.

ST devices share the same fundamental confinement principle as conventional tokamaks but feature a more compact, low-aspect-ratio geometry, making them smaller and potentially cheaper to construct. The ST resembles a cored apple and is conceptually designed for steady-state operation, as its compact configuration and high bootstrap current fraction could allow for largely non-inductive current sustainment (Akers et al. 2002).
This implies that the plasma current is intended to be driven and maintained by non-inductive schemes that must achieve higher efficiencies than conventional RF or NBI systems, owing to the device's geometric constraints and the associated limited beam access and penetration. One such approach is the merging-compression technique, in which the magnetic energy of two merging plasmas is converted into thermal energy to initiate and heat the plasma (Reinders 2021). Nevertheless, early ST power-plant designs are still expected to operate in a pulsed regime, since efficient and fully validated steady-state current drive technologies have not yet been demonstrated operationally (Prajapati and Deshpande 2022).

Compared to conventional tokamaks, the ST concept shows improved plasma performance at higher power density and enhanced magnetic equilibrium, enabling operation at lower toroidal magnetic field strengths. The reduced magnetic field requires less energy for magnet operation, thereby improving the plant's overall energy efficiency, technological simplicity and cost-effectiveness. However, the compact geometry complicates construction, as limited space remains for the integration of essential subsystems (Stambaugh et al. 1998).

The RFP likewise relies on a central solenoid that provides the loop voltage required to initiate the plasma discharge and drive a toroidal plasma current, similar to a tokamak, but it operates at lower magnetic field strengths to sustain the plasma. In this configuration, the plasma becomes largely self-organized, as it is magnetized and heated almost entirely by its own current. However, the weaker magnetic field reduces overall stability and confinement performance. Therefore, RFPs are primarily used as experimental systems for studying plasma conditions rather than as power plant candidates (National Research Council 2001). Since the RFP concept lacks a strong bootstrap current, most devices operate in pulsed mode; steady-state operation would require continuous magnetic helicity injection – i.e., the controlled replenishment of linked magnetic field lines to sustain the plasma current – which remains still under investigation (Prager and Ryutov 2012).

Helical systems are characterized by a confining, helically twisted magnetic field whose three-dimensional structure is defined entirely by the geometry of their external coils. The plasma is typically created through high-frequency electromagnetic wave heating methods, such as electron cyclotron resonance heating (ECRH), a specific form of RF heating tuned to the electron cyclotron frequency corresponding to the local magnetic field strength, allowing electrons to efficiently absorb energy, accelerate, and ionize the injected neutral gas. Since helical devices do not rely on an induced plasma current and achieve confinement purely through externally generated magnetic fields, they can be operated in steady state. However, this also requires highly specific and precisely optimized coil geometries to produce the desired magnetic configuration. Variants within this device family differ mainly in plasma geometry and in the number of helical twists traced by the magnetic field lines – features that result directly from differences in coil design (Reinders 2021; Moynihan and Bortz 2023).

In (classical) stellarators, unlike tokamaks, the poloidal field required for confinement is not generated by an external plasma current but by a combination of planar toroidal coils and helically wound coils. The windings are arranged so that adjacent turns carry current in opposite directions, and this bidirectional current arrangement naturally produces the poloidal field component. Together with the toroidal field, this creates helically twisted magnetic field lines that guide and confine plasma particles within a geometry resembling a twisted ribbon coiled into a ring (Moynihan and Bortz 2023).

Modern variants of the stellarator, however, employ a set of non-planar coils that are primarily responsible for plasma confinement. Their twisted geometry simultaneously generates both toroidal and poloidal magnetic field components, supplemented by a few planar coils for fine adjustments (Reinders 2021).

Spatial-axis stellarators are characterized by a magnetic axis that follows a double-helical path – two intertwined helices wound around the torus – which, in projection, resembles a tightly twisted figure-8. This geometry enables bean-shaped plasma cross-sections and the formation of a large magnetic well, favorable for high-β[2] plasma confinement i.e. conditions in which the plasma pressure reaches a significant fraction of the magnetic pressure. The rotational transform required for plasma confinement is generated directly within the double-helix section of the axis, while coil systems conforming to this geometry provide the necessary magnetic fields and additional shaping (Schaffer and Bard 1985).

[2] The plasma β, defined as the ratio of plasma pressure to magnetic pressure, is economically desirable at high values, since higher β allows smaller and less powerful magnet coils, thereby reducing reactor size and cost. However, achieving and maintaining high β-values remains challenging due to plasma instabilities (Reinders 2021).

Heliacs are a subclass of stellarators in which the coils are shaped and arranged so that the magnetic axis – and thus the plasma particle trajectories – follow a helical path. This added twist to the magnetic field lines improves confinement by mitigating particle drifts (Reinders 2021).

Heliotrons, also known as torsatrons, employ a simplified concept that dispenses with the interwoven winding design. Instead, these devices use continuous helical field coils carrying unidirectional current, requiring the coil geometry to be defined and arranged with great precision (Beidler et al. 2012; Reinders 2021).

Linear devices such as magnetic or tandem mirrors are designed for steady-state operation, where plasma confinement is achieved purely by externally generated magnetic fields, and the plasma is typically created and heated using electromagnetic wave techniques (Post 1987; Horton et al. 2013). Although these systems are conceptually simpler and easier to construct than toroidal configurations, they face major challenges in maintaining plasma stability and preventing end losses. In magnetic mirrors, magnetic field lines are compressed at both ends of a linear chamber by magnetic coils, creating regions of high magnetic field strength that reflect charged particles back toward the center. As the particles move between regions of low and high magnetic field intensity, they are repeatedly reflected toward the midplane, where confinement is maintained and fusion reactions can occur (Moynihan and Bortz 2023).

Tandem mirrors, by contrast, are intended to counteract plasma leakage losses more effectively. They achieve this by placing multiple magnetic mirrors in series, thereby improving plasma confinement within the central region (Reinders 2021).

A compact toroid is an axisymmetric, self-organized magnetic confinement configuration that operates without external toroidal field coils. Instead, the confining magnetic fields are largely self-generated by plasma currents, which arise during rapid electromagnetic discharges, which ionize the injected gas and induce strong circulating currents, resulting in a high-β plasma equilibrium. The absence of externally applied toroidal field coils simplifies the overall design and reduces engineering complexity compared to conventional toroidal devices. Examples include the field-reversed configuration (FRC), which is characterized primarily by poloidal fields, and the spheromak, in which both poloidal and toroidal magnetic field components are generated by plasma currents (Quinn and Compact Toroid Staff 1983).

In FRCs, the plasma is typically formed by the θ-pinch method, in which a rapidly rising axial current in external θ-pinch coils generates a changing magnetic field that induces an azimuthal plasma current, ionizing the neutral gas and creating a strong poloidal magnetic field that reverses the direction of the externally applied axial field. This field reversal produces closed magnetic field lines and an axisymmetric toroidal plasma structure with self-generated circulating currents that that enables the plasma to self-confine and self-stabilize primarily through the poloidal field (B. L. Wright 2011). Owing to their dynamic versatility, FRC configurations can, in principle, be operated in both pulsed, high-density regimes and steady-state modes. In practice, however, internal resistivity causes the plasma current and magnetic field to decay over time, meaning that the FRC must be re-established, making most FRC devices inherently pulsed systems (Prager and Ryutov 2012; Zohuri 2017).
Spheromaks share similarities with FRCs in both design and physical principles, as the plasma also self-organizes into a compact toroidal structure sustained by internal currents. Unlike FRCs, however, spheromaks generate both poloidal and toroidal magnetic field components through plasma currents, resulting in a more balanced magnetic configuration and potentially greater stability. Spheromaks are commonly formed using a coaxial plasma gun, in which a high-voltage discharge between coaxial electrodes simultaneously ionizes the gas and drives axial plasma currents that generate the self-confined magnetic fields (B. L. Wright 2011). Depending on the current-drive method, such devices can operate in pulsed mode or be sustained in quasi-steady state operation through continuous current-drive schemes similar to the auxiliary heating and techniques employed in spherical tokamaks, often in combination with an externally applied poloidal magnetic field that provides more stability (Reinders 2021).

## A.2 Inertial Confinement Fusion

The most advanced inertial confinement fusion (ICF) concepts involve the compression of a fusion pellet using symmetrically converging laser beams. In the direct drive approach, the beams are focused directly onto a small pellet containing frozen fusion fuel, positioned at the chamber's center. In this configuration, laser beams from multiple directions strike the target simultaneously, ablating the pellet's outer layer and driving an implosion that compresses the fuel to fusion-relevant conditions (Moynihan and Bortz 2023). Direct drive devices enable highly efficient compression due to a direct targeting of the fuel pellet but requires exceptionally uniform laser illumination and precise beam alignment to avoid instabilities (Reinders 2021).

Indirect drive, by contrast, uses a so-called 'hohlraum' – a cylindrical cavity, typically made of gold – within which the fusion pellet is positioned. The laser beams enter the hohlraum through small openings and strike the inner walls, converting the laser light into a bath of high-energy X-rays that symmetrically irradiate and compress the pellet (Moynihan and Bortz 2023). This approach improves irradiation uniformity but results in some energy loss due to the laser-to-X-ray conversion process (Reinders 2021).

Fast ignition is a two-step process related to direct drive. In the first stage, the target is compressed using a standard long-pulse laser driver configuration to reach high density and temperature. In the second stage, an additional ultra-intense, short-pulse laser delivers a focused burst of energy to a localized region of the compressed core, heating the small spot and thereby initiating fusion reactions (Moynihan and Bortz 2023).

Alternative driver technologies include ion-beam and projectile drivers. In ion-beam ICF, the laser beams are replaced by energetic ion beams that deposit energy more deeply and uniformly into the target, offering higher repetition potential and, consequently, improved compression due to the combined energy and momentum imparted by the ions to the target material. In contrast, projectile drivers employ high-velocity impacts – using either plasma or solid projectiles launched electromagnetically – to generate shock waves that compress the fuel (Moynihan and Bortz 2023).

In the pulsed Z-pinch concept, a powerful electrical current – typically tens of megaamperes – flows through an array of fine metal wires, generating an intense magnetic field that rapidly compresses the high-Z plasma, formed from the vaporized wire array, toward the axis. The resulting hot, dense plasma emits a burst of high-intensity X-rays, which heat the surrounding hohlraum. This, in turn, causes the hohlraum to radiate a uniform flux of X-rays that heat and vaporize the outer layer of a centrally placed fusion capsule, driving its symmetric compression. While conceptually promising, this geometry is intrinsically inefficient, as only a small fraction of the X-ray power generated by the Z-pinch plasma is converted into uniform radiation from the hohlraum walls (Nuttall 2022).

## A.3 Magneto-Inertial Fusion

The MagLIF concept employs the Z-pinch principle within the magneto-inertial fusion (MIF) concept family, using an intense electric current as a driver to generate a magnetic field. The magnetized plasma, contained within a cylindrical metal liner, is preheated by a short laser pulse before being compressed by shock waves generated through the rapid current-driven implosion of the liner (Reinders 2021; Moynihan and Bortz 2023).

PJMIF, by contrast, relies on a series of high-speed plasma jets that converge to form both the target plasma and the imploding liner. The central plasma target, created inside the fusion chamber, is surrounded by a second set of plasma jets that form an imploding shell. As these jets collide and compress the target, the plasma reaches fusion conditions. The resulting fusion energy is absorbed by the chamber walls and can subsequently be converted into electricity (Moynihan and Bortz 2023). MIF studies employing FRCs are becoming increasingly popular. These concepts focus on generating high-beta plasmas within strong magnetic fields. In such experiments, FRC plasmas are created and confined within an aluminum liner, and subsequently compressed through the liner’s implosion to reach kilovolt-range temperatures and high energy densities (Wurden et al. 2016).

The sheared-flow-stabilized (SFS) Z-pinch uses a strong axial electric current to generate an azimuthal magnetic field that radially confines and compresses a plasma to fusion conditions, without external magnetic coils. Unlike a classical Z-pinch, which is rapidly destroyed by kink and sausage instabilities, the SFS Z-pinch introduces a sheared axial plasma flow along the pinch column. This velocity shear suppresses magnetohydrodynamic instabilities and allows the plasma to remain stable while the current compresses it to high density and temperatures. Experiments have demonstrated sustained, thermonuclear fusion reactions in compact, centimeter-scale plasma columns, with fusion performance increasing strongly with pinch current (Shumlak et al. 2020).

# Appendix B: Data Analysis of Techno-Economic Input Parameters

## B.1 Operations and Maintenance Costs

O&M expenditures in FPPs consist of recurring activities, including staffing, administrative tasks (e.g., insurance, general services), and operational processes including tritium handling, fuel processing, and plant operations (Badger et al. 1975; Baker et al. 1981; J. G. Delene et al. 1988; Waganer, Driemeyer, and Lee 1992). A particularly influential cost component arises from the scheduled replacement and maintenance of critical reactor systems, which require periodic shutdowns, restart procedures, and complex remote-handling operations (Badger et al. 1975; Conn et al. 1990; Tsoulfanidis 1981).

Figure 25 presents the reported fixed O&M costs in constant 2018 USD per kW by confinement type and maturity level.

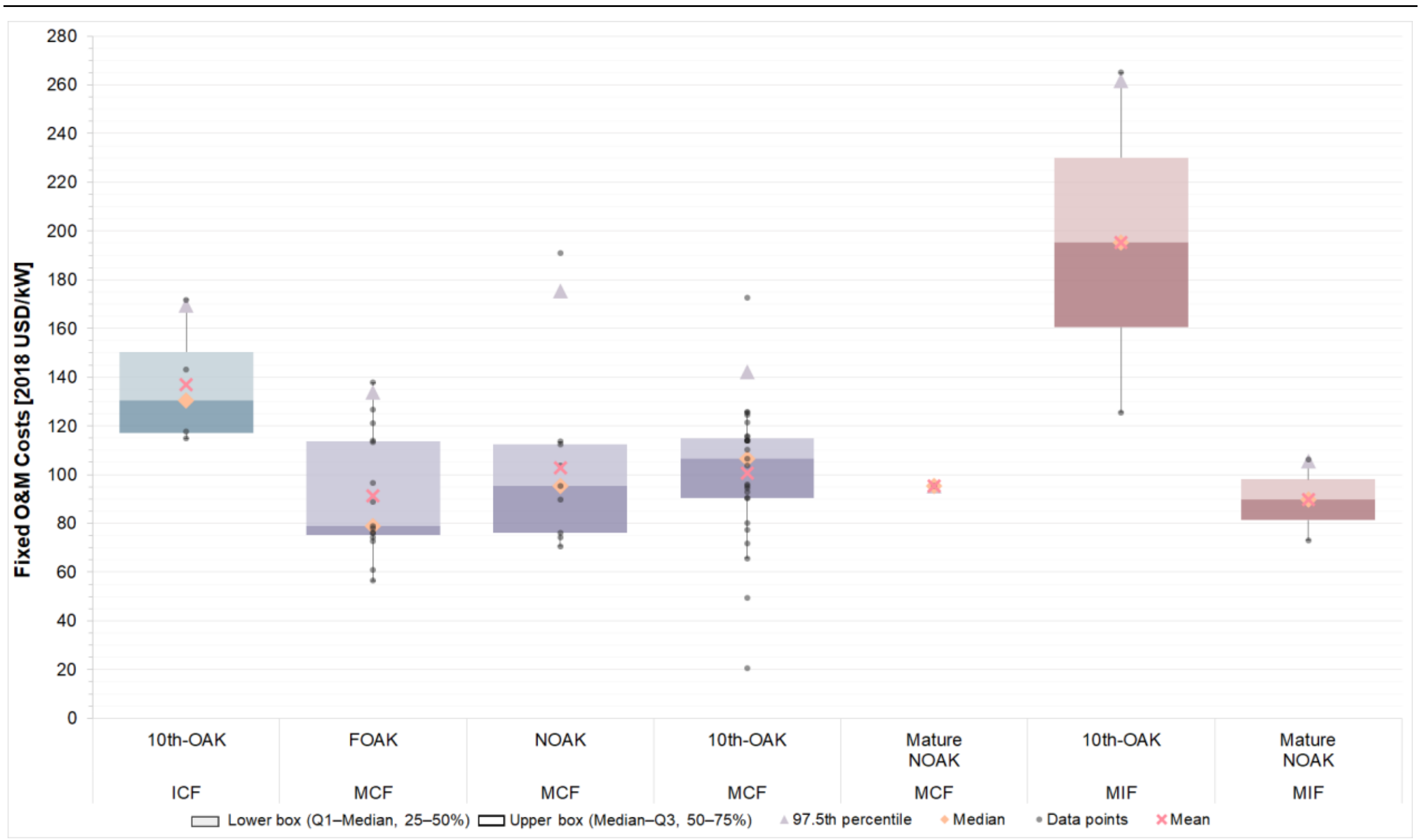

**Figure 19: Assumed fixed O&M costs [2018 USD/kW] for FPPs by confinement type and technological maturity level as reported in literature.**

Source: Own compilation and visualization based on literature-reported data points from (Baker et al. 1981; Tsoulfanidis 1981; Maya, Schultz, and Battaglia 1985; Lee 1987; J. G. Delene et al. 1988; Spears 1990; Waganer, Driemeyer, and Lee 1992; Jerry G. Delene et al. 2001; Dolan, Yamazaki, and Sagara 2005; Vaillancourt et al. 2008; Han and Ward 2009; Joh Sheffield and Milora 2016; Woodruff et al. 2017; Hsu, Woodruff, and Nehl 2020; Takeda, Sakurai, and Konishi 2020; Jo et al. 2021; Famà et al. 2024; Lerede et al. 2023).

Note: Boxplots summarize the distribution of reported overnight construction cost estimates from the literature. Boxes represent the central 50% of values (interquartile range) and are split at the median; darker shading denotes

the lower quartile (Q1–median), lighter shading the upper quartile (median–Q3). Whiskers extend to 1.5 × IQR, beyond the first and third quartiles, with values outside this range representing outliers.

Abbreviations used: FOAK: First-of-a-kind; ICF: Inertial confinement fusion; kW: Kilowatt; MCF: Magnetic confinement fusion; MIF: Magneto-inertial fusion; NOAK: $N^{th}$-of-a-kind; O&M: Operations and maintenance; Q1: First quartile; Q3: Third quartile; USD: United States dollar $10^{th}$-OAK: Tenth-of-a-kind.

In ICF plants, the operation of on-site target fabrication facilities is expected to contribute significantly to the overall fixed O&M burden, resulting in considerably higher costs than in MCF or fission power plants. For comparison, fission systems show a median of about 65 $USD_{2018}$/kW (NEA/IEA 2020), whereas reported ICF values span a full range of approximately 114 to 171 $USD_{2018}$/kW, with a median of around 130 $USD_{2018}$/kW.

For MCF systems, no consistent cost-reduction trend can be observed from early to mature designs. The median fixed O&M cost increases from around 78 $USD_{2018}$/kW in FOAK reactors to approximately 95 $USD_{2018}$/kW in NOAK and mature systems, and further to about 106 $USD_{2018}$/kW in $10^{th}$-OAK plants. The IQR for FOAK designs extends from roughly 75 to 113 $USD_{2018}$/kW, with a full uncertainty band of 56-137 $USD_{2018}$/kW, indicating substantial cost variability among early estimates. The relatively broad upper quartile suggests that several studies have assumed high-cost maintenance or operational risk factors for initial reactor generations. In contrast, NOAK systems exhibit a similar IQR but a slightly narrower overall spread, indicating greater convergence in cost expectations as experience accumulates. The $10^{th}$-OAK data are more tightly clustered, with most values ranging between 90 and 114 $USD_{2018}$/kW. The distribution shows a mild left skew, where a few low-cost assumptions pull the mean slightly below the median. This moderate variance supports the use of a normal distribution for simulation purposes, suggesting that by the $10^{th}$-OAK stage, MCF O&M cost expectations had begun to stabilize around a central estimate.

MIF systems exhibit the highest fixed O&M costs at $10^{th}$-OAK maturity stage, ranging between approximately 125 and 265 $USD_{2018}$/kW, as captured by the full range of reported values. This wide range reflects both the immaturity of technology and the substantial uncertainty associated with extrapolating experimental prototypes. As the designs mature, the reported data cluster more tightly between 71 and 106 $USD_{2018}$/kW, consistent with expectations of cost reductions through system optimization and operational learning. Notably, MIF is the only confinement type that shows a clear downward trend in fixed O&M costs with increasing technological maturity, suggesting that the experience gained from early prototype development could translate into measurable improvements in operational efficiency and maintainability.

Figure 20 illustrates the reported variable O&M costs expressed in constant 2018 USD per MWh.

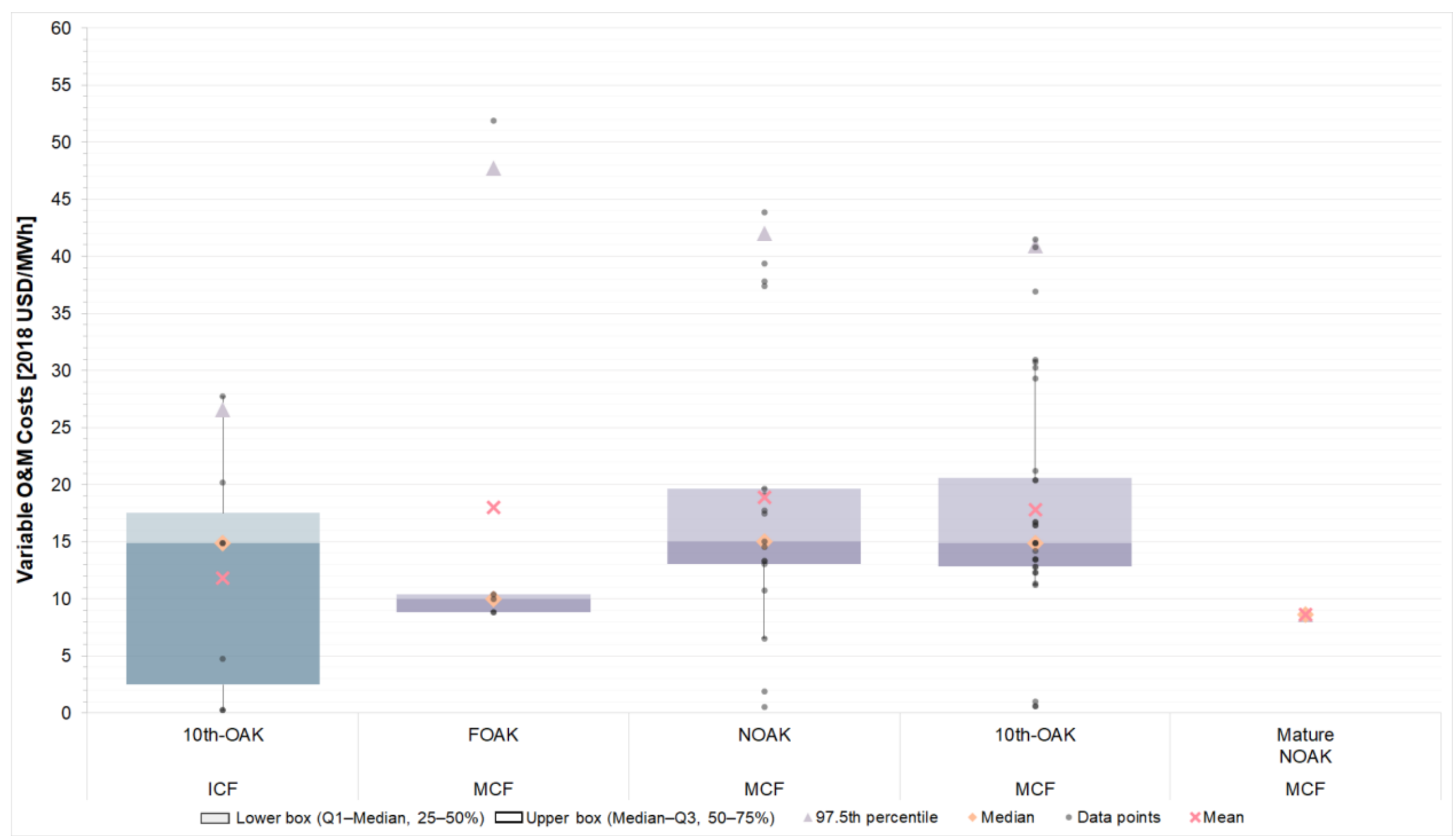

**Figure 20: Assumed variable O&M costs [2018 USD/MWh] for FPPs by confinement type and technological maturity level as reported in literature.**

Source: Own compilation and visualization based on literature-reported data points from (Blink et al. 1985; Maya, Schultz, and Battaglia 1985; J. G. Delene et al. 1988; Conn et al. 1990; Logan et al. 1990; J. G. Delene 1991; Miller et al. 1991; F. Najmabadi et al. 1991; Moir 1992; Krakowski 1995; Jerry G. Delene et al. 2001; Tokimatsu et al. 2003; Maisonnier et al. 2005; Farrokh Najmabadi et al. 2006; Vaillancourt et al. 2008; Kozaki, Imagawa, and Sagara 2009; Yamazaki et al. 2009; Han and Ward 2009; Dragojlovic et al. 2010; Yamazaki, Oishi, and Mori 2011; Entler et al. 2018; Lerede et al. 2023).

Note: Boxplots summarize the distribution of reported overnight construction cost estimates from the literature. Boxes represent the central 50% of values (interquartile range) and are split at the median; darker shading denotes the lower quartile (Q1–median), lighter shading the upper quartile (median–Q3). Whiskers extend to 1.5 × IQR, beyond the first and third quartiles, with values outside this range representing outliers.

Abbreviations used: FOAK: First-of-a-kind; ICF: Inertial confinement fusion; MCF: Magnetic confinement fusion; MIF: Magneto-inertial fusion; MWh: Megawatt hour; NOAK: N$^{th}$-of-a-kind; O&M: Operations and maintenance; Q1: First quartile; Q3: Third quartile; 10$^{th}$-OAK: Tenth-of-a-kind; USD: United States dollar.

Among all confinement types, 10$^{th}$-OAK ICF designs exhibit the broadest variation, with reported variable O&M estimates ranging from 0.23 to 27 USD$_{2018}$/MWh and a median of approximately 14.8 USD$_{2018}$/MWh. This extraordinary spread largely arises from methodological inconsistencies in cost accounting, since several analyses classify all O&M expenses as variable, without distinguishing fixed component (Blink et al. 1985; Conn et al. 1990; Moir 1992). Such treatment inflates variability and complicates cross-study comparability.

For FOAK MCF plants, the IQR is relatively narrow at 8.8–10.4 USD$_{2018}$/MWh, implying consistent accounting practices across studies. However, the mean lies noticeably above the median of about 10 USD$_{2018}$/MWh, at roughly 18 USD$_{2018}$/MWh, due to conservative or high-risk assumptions that skewed the distribution upward (Tokimatsu et al. 2003). Both NOAK and 10$^{th}$-OAK MCF datasets display similar IQRs of around 12–20 USD$_{2018}$/MWh, with median values of 15 and 14.8 USD$_{2018}$/MWh, respectively.

This close alignment indicates that no consistent reduction in variable O&M costs can be observed with increasing technological maturity. The wide scatter within these groups reflects differences in methodological scope – for instance, whether auxiliary services or secondary operational costs are included (Logan et al. 1990; Kozaki, Imagawa, and Sagara 2009). For mature MCF systems, only two values of approximately 8.6 $USD_{2018}$/MWh is available, and although lower than earlier-stage estimates, no explicit rationale for this reduction is provided (Han and Ward 2009).

Overall, the wide spread in both fixed and variable O&M costs across confinement types stems primarily from differences in modeling approaches and cost-accounting assumptions, although technological characteristics also play a secondary role. Some studies include replacement work (Lee 1987; Logan et al. 1990; Takeda, Sakurai, and Konishi 2020), decommissioning, or waste-handling within O&M (Han and Ward 2009), while others omit these elements entirely (Famà et al. 2024), or apply proportionally higher mark-ups than those used for comparable PWR systems (Maya, Schultz, and Battaglia 1985). Additional discrepancies arise from differences in the treatment of O&M expenditures, particularly the separation into fixed and variable components, with most studies allocating the majority of these costs to fixed rather than variable categories (Maya, Schultz, and Battaglia 1985; J. G. Delene et al. 1988; Vaillancourt et al. 2008). This methodological inconsistency explains much of the observed variance, the presence of outliers, and the lack of a clear cost-reduction trend across maturity stages. Consequently, no robust conclusions regarding true cost-reduction potential or dominant O&M cost drivers can yet be drawn from the available data.

For the LCOE simulations, lognormal distributions are applied to fixed O&M cost data of the ICF and MCF systems to reflect their continuous and positively skewed nature, while normal distributions are used for fixed O&M costs of $10^{th}$-OAK MCF plants, consistent with the slightly left-skewed yet nearly symmetric spread observed in the underlying data. Triangular distributions are assigned to fixed O&M values of $10^{th}$-OAK MIF systems due to the explicitly reported minimum–maximum uncertainty range, which captures the technological uncertainty inherent in these early-stage confinement designs. Where reported values are tightly clustered or only a single estimate is available – such as for fixed O&M costs of mature NOAK MCF plants – the median value, or the explicitly reported single value, is adopted as the simulation input. For variable O&M costs, the median is applied across all confinement types and maturity levels as a robust measure of central tendency, since the observed variations and outliers primarily reflect differences in cost-accounting conventions rather than pure technological assumptions. For $10^{th}$-OAK MIF systems, no separate variable O&M data are included in the simulation, as all reported cost components for this confinement type originate from a single study in which operational expenditures are fully subsumed within the fixed O&M cost category.

## B.2 Fuel Costs

Fusion fuel based on the D-T cycle represents a negligible share of total generation costs, with values several orders of magnitude below those of fission or fossil-based fuels. Across all confinement types, reported estimates range from 0.015 to 2.91 $USD_{2018}$/MWh, with most clustering between 0.046 and 0.115 $USD_{2018}$/MWh. For comparison, uranium-fueled fission reactors average around 10

USD$_{2018}$/MWh, while coal and combined-cycle gas turbine (CCGT) power plants range between 20–56 USD$_{2018}$/MWh (NEA/IEA 2020).

The overall low fusion fuel costs primarily result from the assumption that only deuterium must be externally supplied, while tritium is generated in lithium-based breeding blankets (Pearson, Antoniazzi, and Nuttall 2018; U.S. GAO 2023). Most studies assume that, once plant operation begins, a surplus of tritium will be produced, making tritium effectively available at zero cost and eliminating the need for external resupply (Badger et al. 1975). This assumption requires the tritium breeding ratio (TBR) – defined as the ratio of produced to consumed tritium – to exceed unity (TBR > 1) to ensure self-sufficiency, including start-up requirements and reserve inventories (Abdou et al. 2020). While deuterium benefits from being abundant and inexpensive, tritium is extremely scarce and currently obtained almost exclusively from decommissioned CANDU fission reactors, with estimated costs reaching up to 2 billion USD$_{2018}$/kg. As the International Thermonuclear Experimental Reactor (ITER) – the first large-scale experimental fusion device – is expected to consume most of the remaining recoverable tritium inventory during its operation, future FPPs will necessarily rely on self-sufficient tritium breeding to remain economically viable (Kovari et al. 2017). However, no tritium breeding system has yet been constructed or demonstrated under plant-operational conditions (Wurbs et al. 2024).

Differences in reported fuel costs primarily stem from variations in cost accounting methodologies, although most studies report values within a similar and consistently low range. Some analyses include miscellaneous replacement costs (J. G. Delene et al. 1988), blanket and spent-fuel material disposal (Gi et al. 2020), or – in the case of ICF – target fabrication and processing costs within the fuel cost term (Waganer, Driemeyer, and Lee 1992). These additions account for the upper tail of the reported fuel cost distributions.

Table 2 summarizes descriptive statistics based on 49 normalized fuel cost data points, categorized by confinement type and maturity level.

| Confinement Type | Maturity | N | Median [USD/MWh] | Mean [USD/MWh] | Min-max range [USD/MWh] | Source |
|---|---|---|---|---|---|---|
| ICF | 10$^{th}$-OAK | 3 | 1.113 | 0.887 | 0.021–1.527 | (Waganer, Driemeyer, and |
| MCF | FOAK | 10 | 0.125 | 0.159 | 0.069–0.466 | (Miller et al. 1991; Entler et al. 2018; Famà et al. 2024) |
| MCF | NOAK | 11 | 0.0457 | 0.0423 | 0.029–0.055 | (Yamazaki et al. 2009; Yamazaki, Oishi, and Mori 2011; Kozaki, |
| MCF | 10th-OAK | 23 | 0.115 | 0.583 | 0.048–2.916 | (Baker et al. 1981; J. Sheffield et al. 1986; J. G. Delene et al. 1988; Miller et al. 1991; F. Najmabadi et al. 1991; Krakowski 1995; Farrokh Najmabadi et al. 2006; Dragojlovic et al. 2010; Joh Sheffield and |
| MIF | Mature NOAK | 2 | 0.0236 | 0.0236 | 0.015–0.032 | (Hsu, Woodruff, and Nehl 2020) |

**Table 2: Descriptive statistics of fusion fuel cost estimates derived from literature-based assumptions, categorized by confinement type and technological maturity level.**

Note: N represents the number of data points. All cost values are given in constant 2018 USD.

Abbreviations used: FOAK: First-of-a-kind; ICF: Inertial confinement fusion; MCF: Magnetic confinement fusion; MWh: Megawatt hour; MIF: Magneto-inertial fusion; NOAK: Nth-of-a-kind; USD: United States dollar; 10th-OAK: Tenth-of-a-kind.

Among the reviewed literature, only one study explicitly quantifies the tritium inventory required for plant start-up, estimating roughly 200 gram (g) per reactor (Woodruff et al. 2017), whereas other studies omit considerations of start-up inventory requirements (Badger et al. 1975). In principle, FPPs could initiate operations using D-D reactions that produce tritium as a byproduct. However, this process takes several months and is considered economically infeasible. Alternatively, reactors within a future fusion fleet could exchange surplus tritium bred during operation through an internal market, although achieving sufficient inter-plant tritium self-sufficiency remains uncertain (Kovari et al. 2017). Given the limited data availability, the common assumption of tritium self-sufficiency during operation across all studies, the theoretical but impractical option of tritium-free start-up, and the unpredictable development of a tritium market, start-up inventory costs are excluded from this analysis.

Beyond the D-T cycle, several studies investigate D-$^3$He as an alternative fuel. This option exhibits significantly higher and highly uncertain costs (4.59–31.32 $USD_{2018}$/MWh) due to the extreme scarcity of $^3$He, which is found mainly in lunar rigoleth and only in trace quantities on Earth. However, the intrinsically high fuel cost and speculative supply chain render this cycle economically uncompetitive in the foreseeable future (Niechciał et al. 2020). Accordingly, this analysis focuses on D-T-based fuel cycles, representing the most realistic near- to mid-term conditions for FPPs. Once a self-sufficient tritium breeding cycle is established, operational fuel costs become almost negligible compared to capital and O&M expenditures.

For the Monte Carlo-based LCOE simulations, the median fuel cost is applied for ICF and MIF systems to account for variability within the reported data and to reduce the influence of extreme values. Since no explicit fuel cost data are reported for 10$^{th}$-OAK MIF plants, the values for mature NOAK MIF are adopted, as only limited variation between the two maturity levels is expected. Similarly, due to the absence of data for mature NOAK MCF systems, the fuel cost assumptions from 10$^{th}$-OAK MCF are applied.

## B.3 Discount Rate

The discount rate is not specific to fusion but is a generic economic parameter; nevertheless, it has a substantial influence on fusion LCOE estimates because FPPs entail very high upfront capital expenditures that are discounted over long construction and operating periods, making the LCOE highly sensitive to financing assumptions (Maisonnier et al. 2005; Entler et al. 2018; Lindley et al. 2023). Reported discount rate assumptions vary widely across the literature, ranging from approximately 3% to 9%. This divergence is primarily driven by differences in assumed financing structures –public versus private capital—and by the perceived risk profile of future fusion plants. Across all studies, discount rates are implicitly assumed to remain constant across technological maturity stages; therefore, learning-related risk reduction is typically not reflected in capital costs (Tokimatsu et al. 2002; Maisonnier et al. 2005; Kozaki, Imagawa, and Sagara 2009; Anklam et al. 2011; Entler et al. 2018; Gi et al. 2020; Roulstone 2021; Lindley et al. 2023; Famà et al. 2024). Lower discount rates (roughly 3–7%) are used when analogies to fission-reactor cost assessments are drawn (Kozaki, Imagawa, and Sagara 2009), when fusion is evaluated within energy-system models competing with other capital-intensive technologies (Tokimatsu et al. 2002; Entler et al. 2018; Gi et al. 2020; Famà et al. 2024), or when future fusion plants are assumed to receive partial or full government-backed financing analogous to subsidized renewable energy sources. The latter assumptions reflect reduced construction costs through governmental risk-sharing and the view that large-scale fusion deployment without state support is unlikely to succeed in competitive markets (Roulstone 2021; Lindley et al. 2023). Discount rates between approximately 3.5% and 7% are further justified by their position between public-sector, risk-free investment benchmarks (Famà et al. 2024) and the higher return requirements typically applied in private-sector energy investments (Maisonnier et al. 2005). Higher discount rates (7–9%) appear in studies assuming fully private financing and high risk premiums for novel, unproven technologies, with 7% often interpreted as the standard return for mature private investments and 9% as representative of early-stage, high-risk technologies (Roulstone 2021; Lindley et al. 2023).

Given this variability and the uncertainty regarding the degree of government involvement and future financing structures, this study adopts a 7% discount rate for the LCOE simulation across all confinement types and maturity levels. This value balances the extremes: lower rates presuppose strong public support that may not be provided, whereas extremely high rates would imply risk levels under which fusion is unlikely to achieve cost competitiveness. Since this study aims to evaluate fusion's best-case competitiveness potential, a moderate and still plausible value is therefore preferable. Furthermore, for being able to compare it with other generation technologies where 7% is typically applied (NEA/IEA 2020).

## B.4 Capacity Factor and Operational Mode

The capacity factor strongly influences the LCOE, as it determines how capital and operating costs are distributed over the total electricity output. A higher capacity factor implies more continuous and reliable operation, thereby lowering unit generation costs, whereas lower factors – resulting from maintenance requirements, replacement operations, or unplanned outages – lead to higher LCOEs (WNA 2021). Given their high upfront costs, FPPs are therefore predominantly designed and assumed to operate as baseload sources with a high utilization rate and minimal downtime to minimize total generation costs (J. G. Delene et al. 1988; Nicholas et al. 2021; Lindley et al. 2023; Famà et al. 2024).

In fusion systems, the achievable capacity factor is primarily constrained by neutron wall loading and irradiation damage to first-wall, fusion-chamber, and blanket components, which limit their operational lifetimes and necessitate periodic replacements that require plant shutdowns (J. G. Delene et al. 1988; Farrokh Najmabadi et al. 2006; Kozaki, Imagawa, and Sagara 2009; Foster, Knight, and Muldrew 2025). However, replacement frequencies and maintenance cycles remain unpredictable, as the long-term performance of these components has not yet been verified under operating conditions. Current structural materials cannot yet withstand the extreme conditions of continuous operation in the fusion environment and have so far been tested only through computer-based simulations or small-scale experimental setups (Meschini et al. 2023).

To mitigate the availability losses associated with such replacement intervals, modular reactor architectures are proposed, in which several smaller fusion modules operate in parallel. This allows individual modules to undergo maintenance while others remain online, maintaining steady electricity output (Anklam et al. 2011; Chuyanov and Gryaznevich 2017) and achieving availability levels comparable to those of nuclear fission baseload plants (Kabeyi and Olanrewaju 2023).

Several studies equate the capacity factor with the availability factor, which is defined as the fraction of total time a power plant is technically capable of operating, excluding major maintenance or replacement outages. Such simplification assumes an optimistic operational scenario characterized by minimal startup periods, short maintenance interruptions, and low failure rates during operation (Hender, Knight, and Cook 1996; Mulder, Melese, and Lopes Cardozo 2021). Under these assumptions, the reported availability values are therefore treated as equivalent to the capacity factors in this study, consistent with the conventions in the underlying literature.

Figure 21 presents the reported capacity factors by confinement type and technological maturity level.

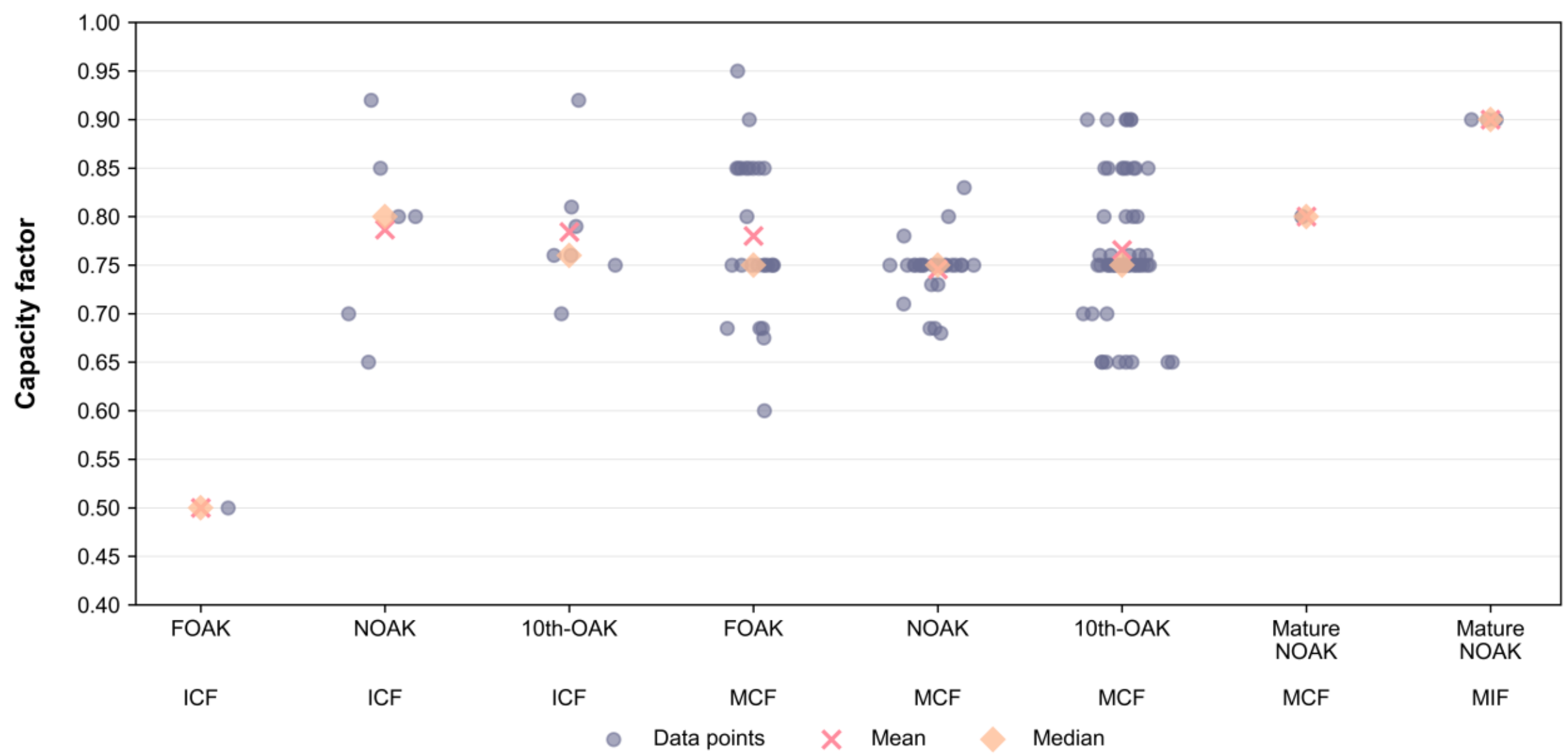

**Figure 21: Assumed capacity factors for FPPs by confinement type and technological maturity level as reported in literature.**

Source: Own compilation and visualization based on literature-reported data points from (Tsoulfanidis 1981; Blink et al. 1985; Maya, Schultz, and Battaglia 1985; J. Sheffield et al. 1986; Lee 1987; Berwald, Whitley, and Garner 1985; J. G. Delene et al. 1988; Conn et al. 1990; Spears 1990; F. Najmabadi et al. 1991; Miller et al. 1991; J. G. Delene 1991; Waganer, Driemeyer, and Lee 1992; Krakowski 1995; Hender, Knight, and Cook 1996; Lako and Seebregts 1998; Jerry G. Delene et al. 2001; Tokimatsu et al. 2002; Dolan, Yamazaki, and Sagara 2005; Maisonnier et al. 2005; Farrokh Najmabadi et al. 2006; Yamazaki and Dolan 2006; Maisonnier et al. 2007; F. Najmabadi et al. 2008; Kozaki, Imagawa, and Sagara 2009; Yamazaki et al. 2009; Dragojlovic et al. 2010; Yamazaki, Oishi, and Mori 2011; Anklam et al. 2011; Bednyagin and Gnansounou 2011; Oishi et al. 2012; Helsley and Burke 2014; Joh Sheffield and Milora 2016; Chuyanov and Gryaznevich 2017; Entler et al. 2018; Hiwatari and Goto 2019; Gi et al. 2020; Hsu, Woodruff, and Nehl 2020; Jo et al. 2021; Lindley et al. 2023; Lerede et al. 2023; Famà et al. 2024).

Abbreviations used: FOAK: First-of-a-kind; ICF: Inertial confinement fusion; MCF: Magnetic confinement fusion; MIF: Magneto-inertial fusion; NOAK: $N^{th}$-of-a-kind; $10^{th}$-OAK: Tenth-of-a-kind.

Across all confinement types, most reported values range between 0.65 and 0.90, reflecting assumptions of steady-state or quasi-continuous baseload operation, while mean and median values are broadly stable across most confinement types and maturity levels. The dataset is dominated by MCF values, where reported capacity factors cluster between 0.70 and 0.85 across all maturity levels. This reflects the assumption of near-continuous operation with limited downtime for maintenance and blanket replacement from the first deployment stage onward. Several scattered data points reach up to 0.95; these outliers mainly represent assumptions for configurations such as spherical or advanced tokamaks (Farrokh Najmabadi et al. 2006; Joh Sheffield and Milora 2016; Chuyanov and Gryaznevich 2017), and stellarators (Famà et al. 2024), which are expected to achieve higher availability due to their inherent steady-state capabilities. The limited spread and stable averages across maturity levels suggest that most cost studies apply standardized operational availability assumptions rather than empirically derived performance data, as a steady-state regime has not yet been demonstrated under power-plant-relevant conditions (Xu et al. 2024).

For ICF power plants, regardless of maturity level, reported capacity factors span a broad range – from 0.50 to 0.92 – reflecting divergent technological assumptions ranging from early, low-repetition laser systems to advanced, high-repetition-rate designs capable of quasi-steady-state operation, illustrating innovation-driven improvements in driver technology that increase operational availability and achievable capacity factors (Blink et al. 1985; Maya, Schultz, and Battaglia 1985; Conn et al. 1990; Waganer, Driemeyer, and Lee 1992; Anklam et al. 2011; Helsley and Burke 2014).

Data points for MIF report a capacity factor of approximately 0.90, indicating an assumed high level of availability, although this estimate is provided without detailed technical justification (Hsu, Woodruff, and Nehl 2020).

For comparison, other firm or baseload low-carbon technologies operate at similar availability levels: modern nuclear fission reactors typically reach capacity factors of 0.80–0.90, while coal and CCGT plants achieve 0.50–0.80, depending on dispatch strategy and maintenance schedules (Kabeyi and Olanrewaju 2023).

Four data points correspond to pulsed MCF tokamak systems designed for non-continuous operation (Badger et al. 1975; Spears 1990; Lindley et al. 2023). As future FPPs are intended to provide continuous baseload generation, and their economic competitiveness depends critically on achieving high values for capacity factor, such configurations are not representative of commercial fusion designs and would bias the overall results downward.
Recent analyses indicate that the pulsed operation of FPPs is unlikely to be economically feasible (Lindley et al. 2023). Tokamak systems – such as EU-DEMO, the planned successor to ITER and one of the first planned FOAK MCF plants – are currently constrained to pulsed operation but are being developed with the objective of achieving steady-state operation (or at least extended pulse durations of up to two hours) through technological advances. However, current research remains far from demonstrating such performance, as existing technologies required for steady-state operation remain inefficient for commercial applications and therefore constrain most early tokamak designs to inherently pulsed operation (Bachmann et al. 2024; Siccinio 2024). While pulsed reactors could, in principle, operate as quasi-baseload plants if coupled with TES that buffers heat output during dwell periods, which is necessary when the central solenoid recharges and plasma burning pauses (Kerekeš et al. 2023; Breuning et al. 2026), such integration would introduce additional system complexity and cost (Badger et al. 1975). Moreover, current TES technologies cannot store sufficient thermal energy to fully bridge these dwell times, rendering such configurations both technically and economically uncompetitive at present (Schwartz et al. 2023). For this reason, pulsed reactor designs are excluded from the capacity factor dataset.

Because FPPs have not yet demonstrated operational availability, the reported capacity factors represent design targets rather than empirically validated performance. Steady-state operation and sufficiently high laser repetition rates have not yet been achieved under reactor-relevant conditions, and current experimental results remain far below the levels required for commercial FPP operation (Wurbs et al. 2024). The resulting dataset therefore shows substantial variability across confinement types and maturity levels, primarily driven by differing design assumptions and steady-state expectations. To avoid overemphasizing optimistic or concept-specific values, the median is applied for all confinement types and maturity levels within the Monte Carlo LCOE simulation setup.

## B.5 Construction Time

Construction time is a critical determinant of LCOE, as longer build periods delay revenue generation and increase IDC, thereby raising total capital-related costs (WNA 2023). The construction phase for FPPs typically comprises site excavation, infrastructure and plant development, as well as the assembly and installation of major reactor and supporting systems (IAEA 2001; Böhnlein, Böse, and Wimmers 2026).

**Fehler! Verweisquelle konnte nicht gefunden werden.** presents the expected construction times of FPPs by confinement type and technological maturity level as reported in the literature. Across all confinement types, the assumed durations mostly range between 5 and 10 years, consistent with empirical evidence from nuclear fission power plant projects. Historical data for pressurized water reactors (PWRs) indicate average global construction times of approximately 7.7 years, with FOAK or large-scale units typically requiring longer periods. In some cases, project durations have extended beyond 10 years due to regulatory complexity, project-specific design modifications, and the loss of industrial standardization and construction experience (Shykinov, Rulko, and Mroz 2016; Ritchie 2023). Furthermore, current fusion projects show that actual construction schedules can extend well beyond the initially planned timelines. For instance, the ITER project was originally scheduled with a construction duration of around 10 years, yet the facility has remained under construction since 2010, with completion dates repeatedly revised, while full operation is now projected for 2039 (Gibney 2024).

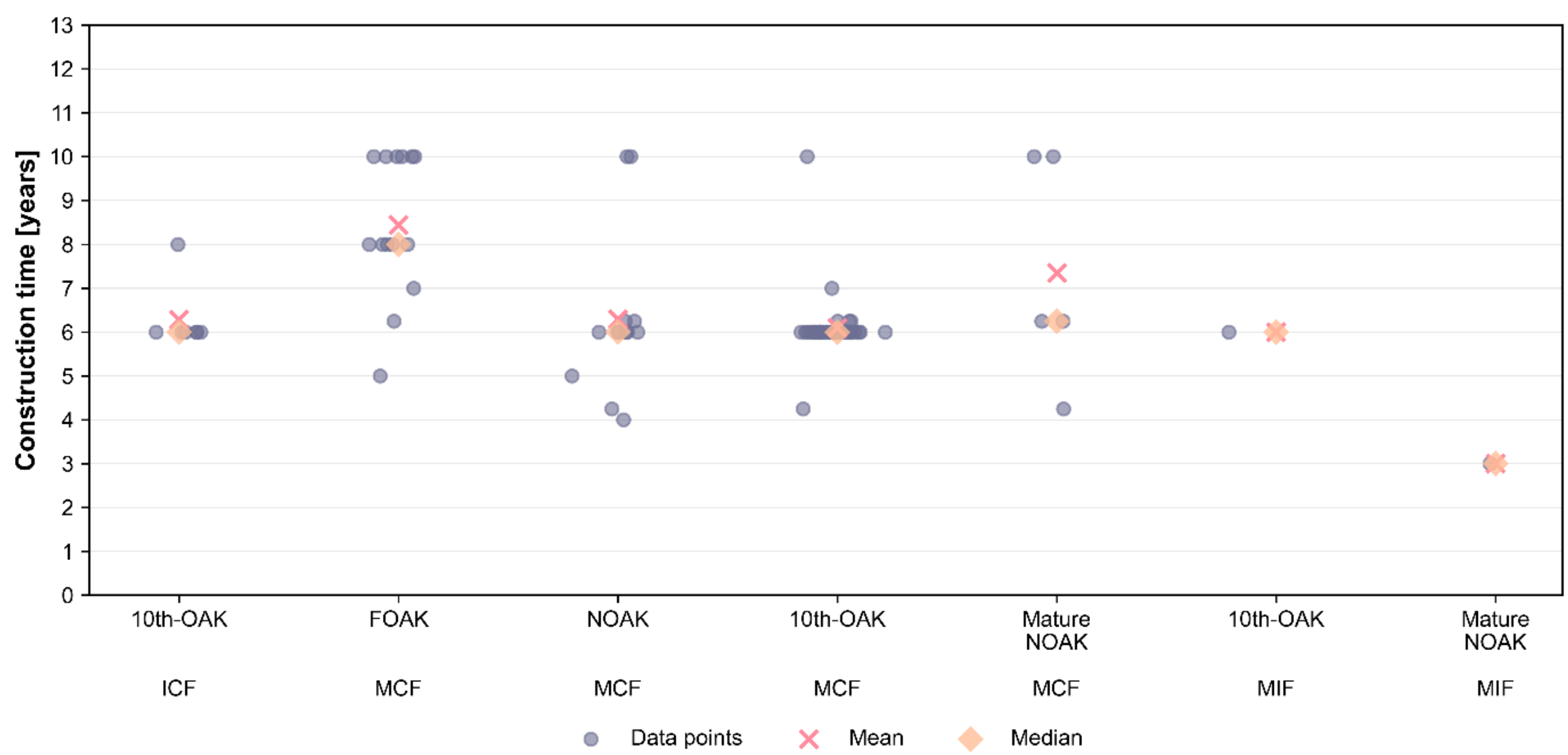

**Figure 22: Assumed construction time [years] for FPPs by confinement type and technological maturity level as reported in the literature.**

Source: Own compilation and visualization based on literature-reported data points from (Baker et al. 1981; Berwald, Whitley, and Garner 1985; Maya, Schultz, and Battaglia 1985; Blink et al. 1985; J. G. Delene et al. 1988; Conn et al. 1990; Logan et al. 1990; Spears 1990; J. G. Delene 1991; Miller et al. 1991; Moir 1992; M. Kikuchi 1996; Lako and Seebregts 1998; Jerry G. Delene et al. 2001; IAEA 2001; Dolan, Yamazaki, and Sagara 2005; Maisonnier et al. 2005; Dragojlovic et al. 2010; Joh Sheffield and Milora 2016; Woodruff et al. 2017; Entler et al. 2018; Hsu, Woodruff, and Nehl 2020; Jo et al. 2021; Lindley et al. 2023; Lerede et al. 2023).

Abbreviations used: FOAK: First-of-a-kind; ICF: Inertial confinement fusion; MCF: Magnetic confinement fusion; MIF: Magneto-inertial fusion; NOAK: $N^{th}$-of-a-kind; Q1: First quartile; Q3: Third quartile; $10^{th}$-OAK: Tenth-of-a-kind.

For MCF systems, assumed construction times are expected to decrease modestly with technological maturity – from a median and mean of approximately 8 years for FOAK plants to around 6 years for NOAK and mature NOAK designs – indicating moderate efficiency gains achieved through accumulated project management and construction experience, as similarly observed in serial nuclear fission plant deployments (Ritchie 2023). Shorter build times are mainly associated with modular plant designs, which enable manufacturing and installation processes in parallel (Baker et al. 1981; Dolan, Yamazaki, and Sagara 2005; Lindley et al. 2023). Extreme values extend up to 10 years, reflecting more pessimistic assumptions in individual studies.

Comparable construction times are reported for ICF systems at $10^{th}$-OAK maturity, with most estimates clustering around 6 years, indicating similar construction requirements.

In contrast, MIF plants are projected to reduce construction times substantially – from about 6 to 3 years – through centralized manufacturing of modular components and transport of pre-fabricated subsystems to the site for integration during construction (Woodruff et al. 2017; Hsu, Woodruff, and Nehl 2020).

For the Monte Carlo-based LCOE simulations, the median construction time is applied across all confinement types and maturity levels to avoid bias from extreme outliers lacking technical justification, as reported values generally cluster around a central range with only minor deviations. Furthermore, these assumed construction periods are broadly consistent with those observed in large-scale nuclear fission projects, which show comparable project complexity and construction requirements (Ben Wealer and von Hirschhausen 2020; Böhnlein, Böse, and Wimmers 2026).

## B.6 Plant Lifetime

The plant lifetime of an electricity generation technology has a significant influence on the LCOE, as the total cost of building and operating a power plant is discounted over its operational period and divided by the total amount of electricity generated during that time (WNA 2023).

The overall lifetime of an FPP is primarily determined by the durability of its structural and outer components, which must withstand the harsh environmental conditions of fusion operation, including intense neutron irradiation, high thermal loads, and ion-induced material damage (Baker et al. 1981; Maisonnier et al. 2005; Dolan, Yamazaki, and Sagara 2005; Dongiovanni et al. 2025). In addition, plant lifetime also depends on the sustained performance and reliability of its components and supporting infrastructure, which can be affected by technological obsolescence – such as the loss of spare part availability, supplier capabilities, and technical support within the industrial supply chain (IAEA 2018). The expected operational lifetime of FPPs remains subject to substantial uncertainty, as long-term material resistance under continuous fusion-relevant conditions has not yet been experimentally validated, and the industrial supply chain and market environment required to ensure component replacement and sustained technical support have not yet been established to the extent necessary for commercial deployment (Meschini et al. 2023; FIA 2024), thus limiting the reliability of lifetime projections under commercial reactor conditions.

**Fehler! Verweisquelle konnte nicht gefunden werden.** presents the expected plant lifetimes of FPPs by confinement type and technological maturity level. Across all confinement types, assumed design lifetimes mostly cluster between 30 and 40 years – broadly consistent with expectations for large nuclear fission plants, which generally operate for 30 to 40 years and may be extended up to 60 years depending on technical reliability and regulatory approvals (Hrinchenko et al. 2025). Given that reported values show similar lifetime assumptions across all maturity levels, with median values of approximately 30–40 years, this indicates that no substantial changes are reflected in the underlying lifetime assumptions and that major improvements in component durability, maintainability, or material resilience under high neutron fluence are not yet anticipated. Consequently, the reported lifetimes represent rather conservative estimates that implicitly account for the potential effects of both physical aging and technological component obsolescence.

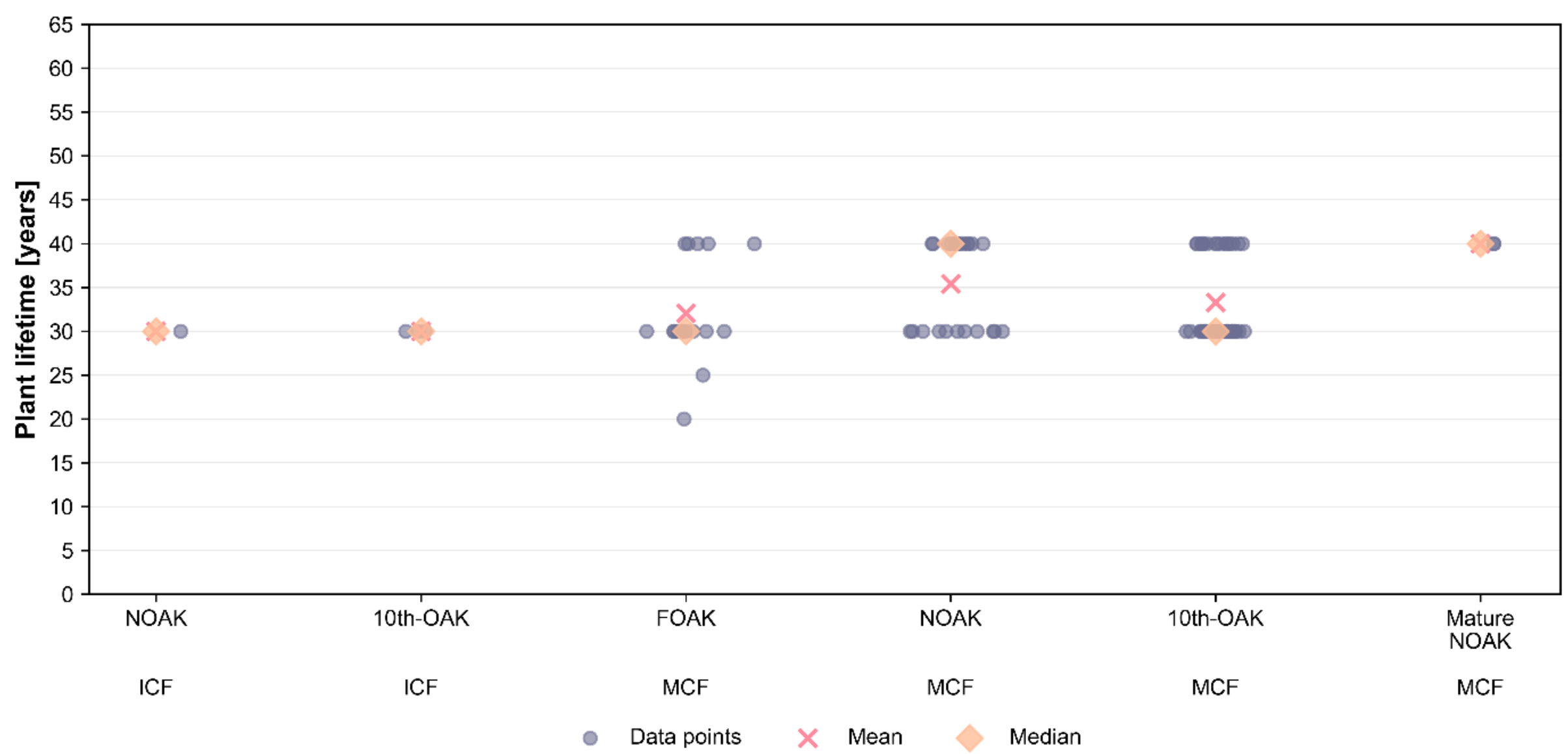

**Figure 23: Assumed plant lifetime [years] for FPPs by confinement type and technological maturity level as reported in the literature.**

Source: Own compilation and visualization based on literature-reported data points from (Baker et al. 1981; Tsoulfanidis 1981; Maya, Schultz, and Battaglia 1985; Blink et al. 1985; J. Sheffield et al. 1986; J. G. Delene et al. 1988; Logan et al. 1990; Spears 1990; J. G. Delene 1991; Miller et al. 1991; Moir 1992; Lako and Seebregts 1998; Jerry G. Delene et al. 2001; IAEA 2001; Maisonnier et al. 2005; Vaillancourt et al. 2008; Farrokh Najmabadi et al. 2006; Kozaki, Imagawa, and Sagara 2009; Yamazaki et al. 2009; Anklam et al. 2011; Bednyagin and Gnansounou 2011; Oishi et al. 2012; Kessel et al. 2015; Joh Sheffield and Milora 2016; Entler et al. 2018; Gi et al. 2020; Jo et al. 2021; Lerede et al. 2023).

Abbreviations used: FOAK: First-of-a-kind; ICF: Inertial confinement fusion; MCF: Magnetic confinement fusion; MIF: Magneto-inertial fusion; NOAK: N$^{th}$-of-a-kind; 10$^{th}$-OAK: Tenth-of-a-kind.

For the LCOE simulations, the median plant lifetime is applied for all confinement types and maturity levels to minimize bias from speculative lifetime assumptions regarding long-term component performance – assumptions that have not yet been empirically validated and are primarily included for calculation purposes. The reported values generally cluster around a central range with only minor deviations, supporting the use of the median as a robust and representative input. Since no lifetime data are reported for MIF plants, the same values as for ICF systems of comparable maturity are applied, reflecting the similar technological uncertainty inherent in both concepts.

# Appendix C: Literature-Based Cost and Technology Data

| Reactor | Confinement Type | Reactor Type | Capacity [MW] | Readiness | Value | Unit | Cost Reference Year | Comment | Reference |
|---|---|---|---|---|---|---|---|---|---|
| UWMAK-III | Magnetic | Tokamak | 1985 | 10th-OAK | 3270000000 | USD | 1975 | | Badger et al. (1975) |
| MARS | Magnetic | Magnetic Mirror | 1198 | 10th-OAK | 4969000000 | USD | 1985 | 6 years construction time | Berwald et al. (1985) |
| HYLIFE-I (single unit) | Inertia | Direct Drive | 1010 | 10th-OAK | 552000000 | USD | 1983 | 30 years lifetime; 8 years construction time | Blink et al. (1985) |
| HYLIFE-I (dual unit) | Inertia | Direct Drive | 2000 | 10th-OAK | 855000000 | USD | 1983 | Dual-unit | Blink et al. (1985) |
| STARFIRE | Magnetic | Tokamak | 1200 | 10th-OAK | 1935530000 | USD | 1980 | 30 years lifetime; 6 years construction time | Baker et al. (1981) |
| CFETR (Full Superconducting Tokamak) | Magnetic | Tokamak | 121.10 | FOAK | 2579400000 | USD | 2009 | Experimental; Without net electricity for output and needs external power supply | Chen et al. (2015) |
| CFETR (Water-cooling Cu magnets) | Magnetic | Tokamak | 121.1 | FOAK | 1414500000 | USD | 2009 | Experimental; Without net electricity for output and needs external power supply | Chen et al. (2015) |
| V-Li/TOK (level safety assurance = 4) | Magnetic | Tokamak | 1200 | 10th-OAK | 2853000000 | USD | 1990 | Plant startup: 2025; 6 years construction time; 30 years lifetime | Delene (1991) |
| V-Li/TOK (level safety assurance = 3) | Magnetic | Tokamak | 1200 | 10th-OAK | 2547000000 | USD | 1990 | Plant startup: 2025; 6 years construction time; 30 years lifetime | Delene (1991) |
| ARIES-I | Magnetic | Tokamak | 1200 | 10th-OAK | 4092000000 | USD | 1990 | Plant startup: 2025; 6 years construction time; 30 years lifetime | Delene (1991) |
| Generomak - ESECOM - RAF-LiPb/RFP | Magnetic | Reversed-Field Pinch | 1200 | 10th-OAK | 1566300000 | USD | 1986 | 6 years construction time; 30 years lifetime | Delene et al. (1988) |
| Generomak - ESECOM - V-Li/RFP | Magnetic | Reversed-Field Pinch | 1200 | 10th-OAK | 1589200000 | USD | 1986 | 6 years construction time; 30 years lifetime | Delene et al. (1988) |
| Generomak - ESECOM - V-LI/TOK | Magnetic | Tokamak | 1200 | 10th-OAK | 2273100000 | USD | 1986 | 6 years construction time; 30 years lifetime | Delene et al. (1988) |
| Generomak - ESECOM - RAF-He/TOK | Magnetic | Tokamak | 1200 | 10th-OAK | 2288500000 | USD | 1986 | 6 years construction time; 30 years lifetime | Delene et al. (1988) |
| Generomak - ESECOM - SiC-He/TOK | Magnetic | Tokamak | 1200 | 10th-OAK | 2675000000 | USD | 1986 | 6 years construction time; 30 years lifetime | Delene et al. (1988) |
| Generomak - ESECOM - V-FLiBe/TOK | Magnetic | Tokamak | 1200 | 10th-OAK | 1953700000 | USD | 1986 | 6 years construction time; 30 years lifetime | Delene et al. (1988) |
| Generomak - ESECOM - V-MHD/TOK | Magnetic | Tokamak | 1200 | 10th-OAK | 1439800000 | USD | 1986 | 6 years construction time; 30 years lifetime | Delene et al. (1988) |
| Generomak - ESECOM - V-DHe3/TOK | Magnetic | Tokamak | 1200 | 10th-OAK | 2907700000 | USD | 1986 | 6 years construction time; 30 years lifetime | Delene et al. (1988) |
| ARIES-RS | Magnetic | Tokamak | 1000 | 10th-OAK | 3521500000 | USD | 1999 | 6 years construction time; 30 years lifetime | Delene et al. (2001) |
| ARIES-ST | Magnetic | Tokamak | 1000 | 10th-OAK | 3880800000 | USD | 1999 | 6 years construction time; 30 years lifetime | Delene et al. (2001) |
| Base Case Heliotron | Magnetic | Heliotron | 1940 | NOAK | 5767800000 | USD | 2003 | | Dolan et al. (2005) |
| New ARIES | Magnetic | Tokamak | 1000 | 10th-OAK | 2105150000 | USD | 1992 | | Dragojlovic et al. (2010) |
| DEMO2 | Magnetic | Tokamak | 953 | FOAK | 7516000000 | USD | 2015 | 40 years lifetime; discount rate: 7%; 10 years construction time | Entler et al. (2018) |
| ARIES-AT | Magnetic | Tokamak | 1000 | 10th-OAK | 2849 | USD/kW | 2000 | | Han and Ward (2009) |
| Fusion basic early | Magnetic | Tokamak | 1000 | 10th-OAK | 3940 | USD/kW | 2000 | in 2050 | Han and Ward (2009) |
| Fusion basic early mature | Magnetic | Tokamak | 1000 | Highly Advanced NOAK | 2950 | USD/kW | 2000 | in 2060 | Han and Ward (2009) |
| Fusion Advanced early | Magnetic | Tokamak | 1000 | 10th-OAK | 2820 | USD/ kW | 2000 | in 2070 | Han and Ward (2009) |
| Fusion Advanced mature | Magnetic | Tokamak | 1000 | Highly Advanced NOAK | 2170 | USD/ kW | 2000 | in 2080 | Han and Ward (2009) |
| PPCS-A | Magnetic | Tokamak | 1000 | 10th-OAK | 5625 | USD/ kW | 2000 | | Han and Ward (2009) |
| PPCS-B | Magnetic | Tokamak | 1000 | 10th-OAK | 5215 | USD/ kW | 2000 | | Han and Ward (2009) |
| PPCS-C | Magnetic | Tokamak | 1000 | 10th-OAK | 4155 | USD/ kW | 2000 | | Han and Ward (2009) |
| PPCS-D | Magnetic | Tokamak | 1000 | 10th-OAK | 2868 | USD/ kW | 2000 | | Han and Ward (2009) |
| Fusion Power Corporation Reactor | Inertia | Ion Beam Driver | 38000 | NOAK | 75244000000 | USD | 2014 | Including hydrogen production | Helsley and Burke (2014) |
| CREST | Magnetic | Tokamak | 1160 | FOAK | 7980000 | USD/ MW | 2019 | | Hiwatari and Goto (2019) |
| CREST | Magnetic | Tokamak | 1160 | 10th-OAK | 6870000 | USD/ MW | 2019 | | Hiwatari and Goto (2019) |
| FPP_low | Magneto-Inertial | SLC / PJMIF / Staged Z-Pinch / SFS Z-Pinch | 383.1 | Advanced NOAK | 799800000 | USD | 2020 | Four different fusion technologies; 3 years construction time | Hsu et al. (2020) |
| FPP_high | Magneto-Inertial | SLC / PJMIF / Staged Z-Pinch / SFS Z-Pinch | 814.4 | Advanced NOAK | 1572000000 | USD | 2020 | Four different fusion technologies; 3 years construction time | Hsu et al. (2020) |
| SSTR | Magnetic | Tokamak | 1000 | FOAK | 6326914000 | USD | 1996 | | Kikuchi et al. (1996) |
| A-SSTR | Magnetic | Tokamak | 1700 | NOAK | 4697100000 | USD | 1996 | 4 years construction time | Kikuchi et al. (1996) |
| Large Helical Device | Magnetic | Heliotron | 1604 | NOAK | 7270000000 | USD | 2002 | 40 years lifetime; discount rate: 3% | Kozaki et al. (2009) |
| Large Helical Device | Magnetic | Heliotron | 1601 | NOAK | 7393000000 | USD | 2002 | 40 years lifetime; discount rate: 3% | Kozaki et al. (2009) |
| Large Helical Device | Magnetic | Heliotron | 1598 | NOAK | 7705000000 | USD | 2002 | 40 years lifetime; discount rate: 3% | Kozaki et al. (2009) |
| Large Helical Device | Magnetic | Heliotron | 1598 | NOAK | 6735000000 | USD | 2002 | 40 years lifetime; discount rate: 3% | Kozaki et al. (2009) |
| ARIES-I | Magnetic | Tokamak | 1000 | $10^{th}$-OAK | 3289800000 | USD | 1992 | 40 years lifetime | Krakowski (1995) |
| ARIES-II | Magnetic | Tokamak | 1000 | $10^{th}$-OAK | 3060900000 | USD | 1992 | 40 years lifetime | Krakowski (1995) |
| ARIES-III | Magnetic | Tokamak | 1000 | $10^{th}$-OAK | 3114900000 | USD | 1992 | 40 years lifetime | Krakowski (1995) |
| ARIES-IV | Magnetic | Tokamak | 1000 | $10^{th}$-OAK | 2746300000 | USD | 1992 | 40 years lifetime | Krakowski (1995) |
| Commercial Twin Reactor | Magnetic | Tokamak | 1500 | Advanced NOAK | 2576 | EUR/kW | 1998 | in 2100; 6.25 years construction time | Lako & Seebregts (1998) |
| Commercial Twin Reactor | Magnetic | Tokamak | 1500 | $10^{th}$-OAK | 2926 | EUR/kW | 1998 | in 2070; 6.25 years construction time | Lako & Seebregts (1998) |
| Commercial Single Reactor | Magnetic | Tokamak | 1000 | NOAK | 5220 | EUR/kW | 1998 | in 2050; 6.25 years construction time | Lako & Seebregts (1998) |
| DEMO | Magnetic | Tokamak | 1000 | FOAK | 7906 | EUR/kW | 1998 | in 2020; 7 years construction time; ITER-type | Lako & Seebregts (1998) |
| Commercial breeder | Magnetic | Tandem Mirror | 1100 | $10^{th}$-OAK | 3637000000 | USD | 1982 | Selling excess electricity and fissile material to commercial markets | Lee (1987) |
| | | | | | | | | while producing tritium and plutonium | |
| 10th-OAK Reactor | Magnetic | Tokamak | 2000 | 10th-OAK | 9000 | USD/kW | 2020 | in 2050; 6.25 years construction time | Lindley et al. (2023); Roulstone (2021) |
| Early large | Magnetic | Tokamak | 1500 | NOAK | 11580 | USD/kW | 2020 | | Lindley et al. (2023) |
| Early large | Magnetic | Tokamak | 1500 | 10th-OAK | 8168 | USD/kW | 2020 | | Lindley et al. (2023) |
| Advanced Large | Magnetic | Tokamak | 1500 | NOAK | 7733 | USD/kW | 2020 | | Lindley et al. (2023) |
| Advanced Large | Magnetic | Tokamak | 1500 | 10th-OAK | 5747 | USD/kW | 2020 | | Lindley et al. (2023) |
| Small | Magnetic | Tokamak | 200 | NOAK | 16343 | USD/kW | 2020 | 4-5 years construction time | Lindley et al. (2023) |
| Small | Magnetic | Tokamak | 200 | 10th-OAK | 12095 | USD/kW | 2020 | 4-5 years construction time | Lindley et al. (2023) |
| Small | Magnetic | Tokamak | 200 | Advanced NOAK | 7219 | USD/kW | 2020 | 4-5 years construction time | Lindley et al. (2023) |
| Generomak - ESECOM - V-Li/TOK | Magnetic | Tokamak | 1200 | 10th-OAK | 2178 | USD/kW | 1986 | 6 years construction time; 30 years lifetime | Logan et al. (1990) |
| Generomak - ESECOM - RAF-He/TOK | Magnetic | Tokamak | 1200 | 10th-OAK | 2193 | USD/kW | 1986 | 6 years construction time; 30 years lifetime | Logan et al. (1990) |
| Generomak - ESECOM - RAF-LiPb/RFP | Magnetic | Reversed-Field Pinch | 970 | 10th-OAK | 1501 | USD/kW | 1986 | 6 years construction time; 30 years lifetime | Logan et al. (1990) |
| Generomak - ESECOM - V-Li/RFP | Magnetic | Reversed-Field Pinch | 970 | 10th-OAK | 1523 | USD/kW | 1986 | 6 years construction time; 30 years lifetime | Logan et al. (1990) |
| Generomak - ESECOM - SiC-He/TOK | Magnetic | Tokamak | 1200 | $10^{th}$-OAK | 2563 | USD/kW | 1986 | 6 years construction time; 30 years lifetime | Logan et al. (1990) |
| Generomak - ESECOM - V-FLiBe/TOK | Magnetic | Tokamak | 1200 | $10^{th}$-OAK | 1873 | USD/kW | 1986 | 6 years construction time; 30 years lifetime | Logan et al. (1990) |
| Generomak - ESECOM - V-MHD/TOK | Magnetic | Tokamak | 1200 | $10^{th}$-OAK | 1380 | USD/kW | 1986 | 6 years construction time; 30 years lifetime | Logan et al. (1990) |
| Generomak - ESECOM - V-D3He/TOK | Magnetic | Tokamak | 1200 | $10^{th}$-OAK | 2787 | USD/kW | 1986 | 6 years construction time; 30 years lifetime | Logan et al. (1990) |
| CASCADE (all Conventional) | Inertia | Direct Drive | 815 | $10^{th}$-OAK | 1233000000 | USD | 1985 | 6 years construction time; 30 years lifetime | Maya et al. (1985) |
| CASCADE | Inertia | Direct Drive | 815 | $10^{th}$-OAK | 1559800000 | USD | 1985 | 6 years construction time; 30 years lifetime | Maya et al. (1985) |
| Reactor based on SCAN Code | Magnetic | Tokamak | 1200 | FOAK | 7227000000 | USD | 1990 | | Miller et al. (1991) |
| Reactor based on ARIES Code | Magnetic | Tokamak | 1200 | $10^{th}$-OAK | 3646000000 | USD | 1990 | 6 years construction time; 30 years lifetime | Miller et al. (1991) |
| HYLIFE-II | Inertia | Ion Beam Driver | 1083 | $10^{th}$-OAK | 4231500000 | USD | 1988 | 30 years lifetime | Moir (1992) |
| ARIES-I | Magnetic | Tokamak | 1000 | $10^{th}$-OAK | 2890000000 | USD | 1988 | | Najmabadi et al. (1991) |
| ARIES-AT | Magnetic | Tokamak | 1000 | $10^{th}$-OAK | 2129000000 | USD | 1992 | 40 years lifetime | Najmabadi et al. (2006) |
| SSTR-like DEMO | Magnetic | Tokamak | 1080 | FOAK | 4228700000 | USD | 2003 | 30 years lifetime | Oishi et al. (2012) |
| PPCS-A | Magnetic | Tokamak | 500 | FOAK | 30000 | USD/kW | 2020 | | Roulstone (2021) |
| Larger FOAK | Magnetic | Tokamak | 2000 | FOAK | 12000 | USD/kW | 2020 | | Roulstone (2021) |
| European Model C Reactor | Magnetic | Tokamak | 1450 | 10th-OAK | 6093 | USD/kW | 2010 | 6 years construction time; 40 years lifetime | Sheffield and Milora (2016) |
| Generic Toroidal Reactor | Magnetic | Tokamak | 1000 | 10th-OAK | 6252 | USD/kW | 2010 | 6 years construction time; 30 years lifetime | Sheffield and Milora (2016) |
| ARIES-RS | Magnetic | Tokamak | 1000 | 10th-OAK | 4897 | USD/kW | 2010 | 7 years Construction time | Sheffield and Milora (2016) |
| ARIES-AT | Magnetic | Tokamak | 1000 | 10th-OAK | 3045 | USD/kW | 2010 | 6 years Construction time | Sheffield and Milora (2016) |
| ARIES-AT Advanced | Magnetic | Tokamak | 1000 | 10th-OAK | 3949 | USD/kW | 2010 | 6 years Construction time | Sheffield and Milora (2016) |
| ITER-like | Magnetic | Tokamak | 1000 | FOAK | 5324 | USD/kW | 2010 | 10 years Construction time | Sheffield and Milora (2016) |
| STARFIRE | Magnetic | Tokamak | 1200 | 10th-OAK | 2307000000 | USD | 1983 | 30 years lifetime | Sheffield et al. (1986) |
| PCSR-E | Magnetic | Tokamak | 1247 | FOAK | 6121062000 | USD | 1984 | 8 years construction time ; 25 years lifetime | Spears (1990) |
| Reactor-1 | Magnetic | Tokamak | 1203 | FOAK | 5801517000 | USD | 1984 | 8 years construction time ; 25 years lifetime | Spears (1990) |
| Reactor-2 | Magnetic | Tokamak | 1209 | FOAK | 4251921000 | USD | 1984 | 8 years construction time ; 25 years lifetime | Spears (1990) |

| | | | | | | | | | | |
|---|---|---|---|---|---|---|---|---|---|---|
| ESECOM | Magnetic | Tokamak | 1164 | FOAK | 3948156000 | USD | 1984 | | 8 years construction time ; 25 years lifetime | Spears (1990) |
| Fusion Power Plant | Magnetic | Tokamak | 1200 | NOAK | 3116000000 | USD | 2020 | | installation in 2060s | Takeda et al. (2020) |
| n.a. | Magnetic | Tokamak | 1000 | NOAK | 2400 | USD/kW | 2003 | | High; experimental reactor in 2010, succeeded by a demonstration reactor in 2030, in order to begin using a power reactor in 2050 | Tokimatsu et al. (2003) |
| n.a. | Magnetic | Tokamak | 1000 | FOAK | 4480 | USD/kW | 2003 | | High; experimental reactor in 2010, succeeded by a demonstration reactor in 2030, in order to begin using a power reactor in 2050 | Tokimatsu et al. (2003) |
| UWMAK-III | Magnetic | Tokamak | 1185 | 10th-OAK | 3609000000 | USD | 1976 | | 30 years lifetime | Tsoulfanidis (1981) |
| Prometheus-L | Inertia | Direct Drive | 972 | 10th-OAK | 3929900000 | USD | 1991 | | 6 years construction time | (Waganer et al. 1992) |
| Prometheus-H | Inertia | Indirect Drive | 999 | 10th-OAK | 2716020000 | USD | 1991 | | 6 years construction time | (Waganer et al. 1992) |
| "FPP"_min | Magneto-Inertial | SLC / PJMIF / Staged Z-Pinch / SFS Z-Pinch | 150 | 10th-OAK | 603000000 | USD | 2016 | | Four different fusion technologies ; fusion power core is 10th-of-a-kind and the BOP is nth- | (Woodruff et al. 2017) |

| | | | | | | | | | |
|---|---|---|---|---|---|---|---|---|---|
| | | | | | | | | of-a-kind; 6 years construction time; highly uncertain and pre-conceptual level | |
| "FPP"_ average | Magneto-Inertial | SLC / PJMIF / Staged Z-Pinch / SFS Z-Pinch | 150 | $10^{th}$-OAK | 1313000000 | USD | 2016 | Four different fusion technologies; fusion power core is 10th-of-a-kind and the BOP is nth-of-a-kind; 6 years construction time | (Woodruff et al. 2017) |
| "FPP"_ max | Magneto-Inertial | SLC / PJMIF / Staged Z-Pinch / SFS Z-Pinch | 150 | $10^{th}$-OAK | 1714000000 | USD | 2016 | Four different fusion technologies; fusion power core is 10th-of-a-kind and the BOP is nth-of-a-kind; 6 years construction time | (Woodruff et al. 2017) |

**Table 3: Literature-reported overnight construction cost assumptions for FPPs by confinement type and technological maturity.**

Note: In addition to overnight construction costs, supplementary techno-economic assumptions reported in the source studies – such as construction time, plant lifetime, start of operation, and configuration-specific characteristics – are extracted where available and documented in the table. Construction time and plant lifetime values reported here are used as input parameters for the LCOE simulations.

Abbreviations used: ARIES: Advanced Reactor Innovation and Evaluation Study; BOP: Balance of plant; CASCADE: Conceptual Advanced Stellarator/Direct-Drive Energy; CFETR: China Fusion Engineering Test Reactor; DEMO: Demonstration Fusion Power Plant; ESECOM: Energy Systems and Economic Modeling; FLiBe: Lithium fluoride-beryllium fluoride molten salt; FOAK: First-of-a-kind; FPP: Fusion power plant; GENEROMAK: Generic Reactor Model for Magnetic Fusion Power Plants; HYLIFE: High Yield Lithium Injection Fusion Energy; ICF: Inertial confinement fusion; ITER: International Thermonuclear Experimental Reactor; MIF: Magneto-inertial fusion; MW: Megawatt; NOAK: $N^{th}$-of-a-kind; PJMIF: Plasma Jet Magneto-Inertial Fusion; PPCS: Power Plant Conceptual Study; RFP: Reversed-field pinch; SCAN: Systems Code for Advanced Nuclear Reactors; SFS: Sheared-flow stabilized; SLC: Stabilized liner compressor; SSTR: Steady-State Tokamak Reactor; $10^{th}$-OAK: Tenth-of-a-kind; TOK: TokamakUSD: United States dollar; UWMAK: University of Wisconsin Magnetic Fusion Reactor Design Study.

| Reactor | Confinement Type | Reactor Type | Capacity [MW] | Cost Type | Readiness | Value | Unit | Cost Reference Year | Comment | Reference |
|---|---|---|---|---|---|---|---|---|---|---|
| HYLIFE-I (single unit) | Inertia | Direct Drive | 1010 | O&M Cost | 10th-OAK | 11 | mills/ kWh | 1983 | | Blink et al. (1985) |
| HYLIFE-I (dual unit) | Inertia | Direct Drive | 2000 | O&M Cost | 10th-OAK | 8 | mills/ kWh | 1983 | | Blink et al. (1985) |
| STARFIRE | Magnetic | Tokamak | 1200 | O&M Cost | 10th-OAK | 19410 000 | USD | 1980 | | Baker et al. (1981) |
| CASCADE | Inertia | Direct Drive | 815 | O&M Cost | 10th-OAK | 7 | mills/ kWh | 1988 | 6 years constructio n time | Conn et al. (1990) |
| HYLIFE | Inertia | Direct Drive | 1010 | O&M Cost | 10th-OAK | 7 | mills/ kWh | 1988 | 6 years constructio n time | Conn et al. (1990) |
| Generomak - ESECOM - LI/LI/V | Magnetic | Reversed-Field Pinch | 1200 | O&M Cost | 10th-OAK | 10 | mills/ kWh | 1988 | 6 years constructio n time | Conn et al. (1990) |
| TITAN-I | Magnetic | Reversed-Field Pinch | 970 | O&M Cost | 10th-OAK | 7 | mills/ kWh | 1988 | 6 years constructio n time | Conn et al. (1990) |
| ARIES-II | Magnetic | Tokamak | 1000 | O&M Cost | 10th-OAK | 7 | mills/ kWh | 1988 | 6 years constructio n time | Conn et al. (1990) |
| ARIES-I | Magnetic | Tokamak | 1000 | O&M Cost | 10th-OAK | 7 | mills/ kWh | 1988 | 6 years constructio n time | Conn et al. (1990) |
| V-Li/TOK Reactor | Magnetic | Tokamak | 1200 | O&M Cost | 10th-OAK | 8.7 | mills/ kWh | 1990 | | Delene (1991) |
| ARIES-I | Magnetic | Tokamak | 1200 | O&M Cost | 10th-OAK | 8.7 | mills/ kWh | 1990 | | Delene (1991) |
| Generomak - ESECOM - V-Li/TOK | Magnetic | Tokamak | 1200 | O&M Cost | 10th-OAK | 65910 000 | USD | 1986 | | Delene et al. (1988) |
| Generomak - ESECOM - RAF-He/TOK | Magnetic | Tokamak | 1200 | O&M Cost | 10th-OAK | 60560 000 | USD | 1986 | | Delene et al. (1988) |
| Generomak - ESECOM - RAF-LiPb/RFP | Magnetic | Reversed-Field Pinch | 1200 | O&M Cost | 10th-OAK | 65260 000 | USD | 1986 | | Delene et al. (1988) |
| Generomak - ESECOM - V-Li/RFP | Magnetic | Reversed-Field Pinch | 1200 | O&M Cost | 10th-OAK | 63640 000 | USD | 1986 | | Delene et al. (1988) |
| Generomak - ESECOM - SiC-He/TOK | Magnetic | Tokamak | 1200 | O&M Cost | 10$^{th}$-OAK | 59750000 | USD | 1986 | | Delene et al. (1988) |
| Generomak - ESECOM - V-FLiBe/TOK | Magnetic | Tokamak | 1200 | O&M Cost | 10$^{th}$-OAK | 65700000 | USD | 1986 | | Delene et al. (1988) |
| Generomak - ESECOM - V-MHD/TOK | Magnetic | Tokamak | 1200 | O&M Cost | 10$^{th}$-OAK | 60700000 | USD | 1986 | | Delene et al. (1988) |
| Generomak - ESECOM - V-DHe3/TOK | Magnetic | Tokamak | 1200 | O&M Cost | 10$^{th}$-OAK | 40500000 | USD | 1986 | | Delene et al. (1988) |
| Generomak - ESECOM - V-LI/TOK | Magnetic | Tokamak | 1200 | O&M Fixed | 10$^{th}$-OAK | 57700000 | USD | 1986 | | Delene et al. (1988) |
| Generomak - ESECOM - V-LI/TOK | Magnetic | Tokamak | 1200 | O&M Variable | 10$^{th}$-OAK | 0.44 | mills/kWh | 1986 | | Delene et al. (1988) |
| Generomak - ESECOM | Magnetic | Tokamak | 1200 | O&M Cost | 10$^{th}$-OAK | 8.9 | mills/kWh | 1986 | | Delene et al. (1988) as cited by Woodruff et al. (2017) |
| ARIES-ST | Magnetic | Tokamak | 1000 | O&M Fixed | 10$^{th}$-OAK | 60000000 | USD | 1999 | | Delene et al. (2001) |
| ARIES-RS | Magnetic | Tokamak | 1000 | O&M Fixed | 10$^{th}$-OAK | 60000000 | USD | 1999 | | Delene et al. (2001) |
| ARIES-ST | Magnetic | Tokamak | 1000 | O&M Variable | 10$^{th}$-OAK | 0.4 | mills/kWh | 1999 | | Delene et al. (2001) |
| ARIES-RS | Magnetic | Tokamak | 1000 | O&M Variable | 10$^{th}$-OAK | 0.4 | mills/kWh | 1999 | | Delene et al. (2001) |
| Base Case Heliotron | Magnetic | Heliotron | 1940 | O&M Cost | NOAK | 100364000 | USD | 2003 | | Dolan et al. (2005) |
| New ARIES | Magnetic | Tokamak | 1000 | O&M Cost | 10$^{th}$-OAK | 6.87 | mills/kWh | 1992 | | Dragojlovic et al. (2010) |
| DEMO2 | Magnetic | Tokamak | 953 | O&M Cost | FOAK | 9.81 | mills/kWh | 2015 | | Entler et al. (2018) |
| Option 1-Reactor - Greenfield | Magnetic | Stellarator | 1204 | O&M Cost | FOAK | 113800000 | USD | 2023 | | Famà et al. (2024) |
| Option 2-Reactor - Greenfield | Magnetic | Stellarator | 1313 | O&M Cost | FOAK | 118100000 | USD | 2023 | | Famà et al. (2024) |
| Option 3-Reactor - Greenfield | Magnetic | Stellarator | 2272 | O&M Cost | FOAK | 155700000 | USD | 2023 | | Famà et al. (2024) |
| Option 4-Reactor - Greenfield | Magnetic | Stellarator | 1368 | O&M Cost | FOAK | 120650000 | USD | 2023 | | Famà et al. (2024) |
| Option 1-Reactor - restructured | Magnetic | Stellarator | 914 | O&M Cost | FOAK | 98610000 | USD | 2023 | | Famà et al. (2024) |
| Option 2-Reactor - restructured | Magnetic | Stellarator | 774 | O&M Cost | FOAK | 90750000 | USD | 2023 | | Famà et al. (2024) |
| Option 3-Reactor - restructured | Magnetic | Stellarator | 1945 | O&M Cost | FOAK | 143850000 | USD | 2023 | | Famà et al. (2024) |
| Option 4-Reactor - restructured | Magnetic | Stellarator | 1165 | O&M Cost | FOAK | 111340000 | USD | 2023 | | Famà et al. (2024) |
| Fusion basic early | Magnetic | Tokamak | 1000 | O&M Fixed | $10^{th}$-OAK | 65800000 | USD | 2000 | | Han and Ward (2009) |
| Fusion Advanced early | Magnetic | Tokamak | 1000 | O&M Fixed | $10^{th}$-OAK | 65300000 | USD | 2000 | | Han and Ward (2009) |
| Fusion Advanced mature | Magnetic | Tokamak | 1000 | O&M Fixed | Highly Advanced NOAK | 65300000 | USD | 2000 | | Han and Ward (2009) |
| Fusion basic early | Magnetic | Tokamak | 1000 | O&M Variable | $10^{th}$-OAK | 7.78 | mills/kWh | 2000 | | Han and Ward (2009) |
| Fusion basic early mature | Magnetic | Tokamak | 1000 | O&M Variable | Highly Advanced NOAK | 5.9 | mills/kWh | 2000 | | Han and Ward (2009) |
| Fusion Advanced early | Magnetic | Tokamak | 1000 | O&M Variable | $10^{th}$-OAK | 7.7 | mills/kWh | 2000 | | Han and Ward (2009) |
| Fusion Advanced mature | Magnetic | Tokamak | 1000 | O&M Variable | Highly Advanced NOAK | 5.9 | mills/kWh | 2000 | | Han and Ward (2009) |
| FPP_low | Magneto-Inertial | SLC / PJMIF / Staged Z-Pinch / SFS Z-Pinch | 383.1 | O&M Cost | Advanced NOAK | 42000000 | USD | 2020 | | Hsu et al. (2020) |
| FPP_high | Magneto-Inertial | SLC / PJMIF / Staged Z-Pinch / SFS Z-Pinch | 814.4 | O&M Cost | Advanced NOAK | 61300000 | USD | 2020 | | Hsu et al. (2020) |
| With Central Solenoid min | Magnetic | Tokamak | 1000 | O&M Cost | NOAK | 98590060 | USD | 2010 | | Jo et al. (2021) |
| With Central Solenoid max | Magnetic | Tokamak | 1600 | O&M Cost | NOAK | 124707658 | USD | 2010 | | Jo et al. (2021) |
| No Central Solenoid min | Magnetic | Tokamak | 1200 | O&M Cost | NOAK | 108000000 | USD | 2010 | | Jo et al. (2021) |
| No Central Solenoid max | Magnetic | Tokamak | 1420 | O&M Cost | NOAK | 117483616 | USD | 2010 | | Jo et al. (2021) |
| Large Helical Device | Magnetic | Heliotron | 1604 | O&M Cost | NOAK | 26.8 | mills/kWh | 2002 | | Kozaki et al. (2009) |
| Large Helical Device | Magnetic | Heliotron | 1601 | O&M Cost | NOAK | 27.1 | mills/kWh | 2002 | | Kozaki et al. (2009) |
| Large Helical Device | Magnetic | Heliotron | 1598 | O&M Cost | NOAK | 28.2 | mills/kWh | 2002 | | Kozaki et al. (2009) |
| Large Helical Device | Magnetic | Heliotron | 1598 | O&M Cost | NOAK | 31.4 | mills/kWh | 2002 | | Kozaki et al. (2009) |
| ARIES-III | Magnetic | Tokamak | 1000 | O&M Variable | $10^{th}$-OAK | 9.2 | mills/kWh | 1992 | | Krakowski (1995) |
| ARIES-IV | Magnetic | Tokamak | 1000 | O&M Variable | $10^{th}$-OAK | 7.5 | mills/kWh | 1992 | | Krakowski (1995) |
| ARIES-I | Magnetic | Tokamak | 1000 | O&M Variable | $10^{th}$-OAK | 7.5 | mills/kWh | 1992 | | Krakowski (1995) |
| ARIES-II | Magnetic | Tokamak | 1000 | O&M Variable | $10^{th}$-OAK | 9.2 | mills/kWh | 1992 | | Krakowski (1995) |
| Commercial breeder | Magnetic | Tandem Mirror | 1100 | O&M Cost | $10^{th}$-OAK | 73000000 | USD | 1982 | | Lee (1987) |
| ARC-2035 | Magnetic | Tokamak | 190 | O&M Fixed | FOAK | 20000000 | USD | 2010 | 40 years lifetime; 5 years construction time; start of operation: 2035; ARC-like commercial reactor; modular | Lerede et al. (2023) |
| EU-DEMO | Magnetic | Tokamak | 500 | O&M Fixed | FOAK | 33000000 | USD | 2010 | 40 years lifetime; 10 years construction time; start of operation: 2060 | Lerede et al. (2023) |
| EU-DEMO Advanced | Magnetic | Tokamak | 500 | O&M Fixed | NOAK | 33000000 | USD | 2010 | 40 years lifetime; 10 years construction time; start of operation: 2080 | Lerede et al. (2023) |
| Asian-DEMO | Magnetic | Tokamak | 500 | O&M Fixed | FOAK | 33000000 | USD | 2010 | 40 years lifetime; 10 years construction time; start of operation: 2060 | Lerede et al. (2023) |
| ARC-2050 | Magnetic | Tokamak | 250 | O&M Variable | NOAK | 9.33 | mills/kWh | 2010 | 40 years lifetime; 5 years construction time; start of operation: 2050; ARC-like commercial reactor; modular | Lerede et al. (2023) |
| EU-DEMO Advanced | Magnetic | Tokamak | 500 | O&M Variable | NOAK | 5.67 | mills/kWh | 2010 | 40 years lifetime; 10 years construction time; start of operation: 2080 | Lerede et al. (2023) |
| EU-DEMO | Magnetic | Tokamak | 500 | O&M Variable | FOAK | 7.67 | mills/kWh | 2010 | 40 years lifetime; 10 years construction time; start of operation: 2060 | Lerede et al. (2023) |
| Asian-DEMO | Magnetic | Tokamak | 500 | O&M Variable | FOAK | 7.67 | mills/ kWh | 2010 | 40 years lifetime; 10 years construction time; start of operation: 2060 | Lerede et al. (2023) |
| Generomak - ESECOM - V-Li/TOK | Magnetic | Tokamak | 1200 | O&M Cost | 10th-OAK | 18.1 | mills/ kWh | 1986 | 6 years construction time; 30 years lifetime | Logan et al. (1990) |
| Generomak - ESECOM - RAF-He/TOK | Magnetic | Tokamak | 1200 | O&M Cost | 10th-OAK | 13.2 | mills/ kWh | 1986 | 6 years construction time; 30 years lifetime | Logan et al. (1990) |
| Generomak - ESECOM - RAF-LiPb/RFP | Magnetic | Reversed-Field Pinch | 970 | O&M Cost | 10th-OAK | 13.5 | mills/ kWh | 1986 | 6 years construction time; 30 years lifetime | Logan et al. (1990) |
| Generomak - ESECOM - V-Li/RFP | Magnetic | Reversed-Field Pinch | 970 | O&M Cost | 10th-OAK | 12.8 | mills/ kWh | 1986 | 6 years construction time; 30 years lifetime | Logan et al. (1990) |
| Generomak - ESECOM - SiC-He/TOK | Magnetic | Tokamak | 1200 | O&M Cost | 10th-OAK | 13.4 | mills/ kWh | 1986 | 6 years construction time; 30 years lifetime | Logan et al. (1990) |
| Generomak - ESECOM - V-FLiBe/TOK | Magnetic | Tokamak | 1200 | O&M Cost | 10th-OAK | 17.8 | mills/ kWh | 1986 | 6 years construction time; 30 years lifetime | Logan et al. (1990) |
| Generomak - ESECOM - V-MHD/TOK | Magnetic | Tokamak | 1200 | O&M Cost | 10th-OAK | 16.1 | mills/ kWh | 1986 | 6 years construction time; 30 years lifetime | Logan et al. (1990) |
| Generomak - ESECOM - V-D3He/TOK | Magnetic | Tokamak | 1200 | O&M Cost | 10th-OAK | 8.9 | mills/ kWh | 1986 | 6 years construction time; 30 years lifetime | Logan et al. (1990) |
| PPCS-A | Magnetic | Tokamak | 1546 | O&M Cost | 10$^{th}$-OAK | 0.00855 | EUR/ kWh | 2005 | 6% discount rate; 40 years lifetime | Maisonnier et al. (2005) |
| PPCS-B | Magnetic | Tokamak | 1332 | O&M Cost | 10$^{th}$-OAK | 0.008 | EUR/ kWh | 2005 | 6% discount rate; 40 years lifetime | Maisonnier et al. (2005) |
| PPCS-C | Magnetic | Tokamak | 1449 | O&M Cost | 10$^{th}$-OAK | 0.0084 | EUR/ kWh | 2005 | 6% discount rate; 40 years lifetime | Maisonnier et al. (2005) |
| PPCS-D | Magnetic | Tokamak | 1527 | O&M Cost | 10$^{th}$-OAK | 0.0085 | EUR/ kWh | 2005 | 6% discount rate; 40 years lifetime | Maisonnier et al. (2005) |
| CASCADE (all Conventional) | Inertia | Direct Drive | 815 | O&M Variable | 10$^{th}$-OAK | 0.1 | mills/ kWh | 1985 | | Maya et al. (1985) |
| CASCADE | Inertia | Direct Drive | 815 | O&M Variable | 10$^{th}$-OAK | 0.1 | mills/ kWh | 1985 | | Maya et al. (1985) |
| CASCADE (all Conventional) | Inertia | Direct Drive | 815 | O&M Fixed | 10$^{th}$-OAK | 50000000 | USD | 1985 | | Maya et al. (1985) |
| CASCADE | Inertia | Direct Drive | 815 | O&M Fixed | 10$^{th}$-OAK | 60000000 | USD | 1985 | | Maya et al. (1985) |
| Reactor based on ARIES Code | Magnetic | Tokamak | 1200 | O&M Cost | 10$^{th}$-OAK | 6.7 | mills/ kWh | 1990 | | Miller et al. (1991) |
| Reactor based on SCAN Code | Magnetic | Tokamak | 1200 | O&M Cost | FOAK | 5.2 | mills/ kWh | 1990 | 8 years construction time; 30 years lifetime | Miller et al. (1991) |
| Generic Magnetic Fusion Reactor | Magnetic | Tokamak | 1200 | O&M Cost | 10$^{th}$-OAK | 49100000 | USD | 1982 | | Moir (1986) as cited by Woodruff et al. (2017) |
| HYLIFE-II | Inertia | Ion Beam Driver | 1083 | O&M Cost | 10$^{th}$-OAK | 2.24 | mills/ kWh | 1988 | | Moir (1992) |
| ARIES-I | Magnetic | Tokamak | 1000 | O&M Cost | 10$^{th}$-OAK | 6.7 | mills/ kWh | 1988 | | Najmabadi et al. (1991) |
| ARIES-AT | Magnetic | Tokamak | 1000 | O&M Cost | 10th-OAK | 6.87 | mills/ kWh | 1992 | | Najmabadi et al. (2006) |
| Generic Toroidal Reactor | Magnetic | Tokamak | 1200 | O&M Cost | 10th-OAK | 108000000 | USD | 2010 | | Sheffield and Milora (2016) |
| Generic Toroidal Reactor | Magnetic | Tokamak | 1500 | O&M Cost | 10th-OAK | 121000000 | USD | 2010 | | Sheffield and Milora (2016) |
| Generic Toroidal Reactor | Magnetic | Tokamak | 2000 | O&M Cost | 10th-OAK | 139000000 | USD | 2010 | | Sheffield and Milora (2016) |
| Generic Toroidal Reactor | Magnetic | Tokamak | 2500 | O&M Cost | 10th-OAK | 156000000 | USD | 2010 | | Sheffield and Milora (2016) |
| Generic Toroidal Reactor | Magnetic | Tokamak | 3000 | O&M Cost | 10th-OAK | 171000000 | USD | 2010 | | Sheffield and Milora (2016) |
| Generic Toroidal Reactor | Magnetic | Tokamak | 1000 | O&M Cost | 10th-OAK | 99000000 | USD | 2010 | | Sheffield and Milora (2016) |
| ARIES-RS | Magnetic | Tokamak | 1000 | O&M Cost | 10th-OAK | 99000000 | USD | 2010 | | Sheffield and Milora (2016) |
| ARIES-AT | Magnetic | Tokamak | 1000 | O&M Cost | 10th-OAK | 99000000 | USD | 2010 | | Sheffield and Milora (2016) |
| ARIES-AT Advanced | Magnetic | Tokamak | 1000 | O&M Cost | 10th-OAK | 99000000 | USD | 2010 | | Sheffield and Milora (2016) |
| European Model C Reactor | Magnetic | Tokamak | 1450 | O&M Cost | 10th-OAK | 119000000 | USD | 2010 | | Sheffield and Milora (2016) |
| ESECOM | Magnetic | Tokamak | 1164 | O&M Cost | FOAK | 54598800 | USD | 1984 | 8 years construction time ; 25 years lifetime | Spears (1990) |
| PCSR-E | Magnetic | Tokamak | 1247 | O&M Cost | FOAK | 65360760 | USD | 1984 | | Spears (1990) |
| Reactor-1 | Magnetic | Tokamak | 1203 | O&M Cost | FOAK | 68611440 | USD | 1984 | | Spears (1990) |
| Reactor-2 | Magnetic | Tokamak | 1209 | O&M Cost | FOAK | 56997360 | USD | 1984 | | Spears (1990) |
| Fusion power plant | Magnetic | Tokamak | 1200 | O&M Cost | NOAK | 236000000 | USD | 2020 | | Takeda et al. (2020) |
| n.a. | Magnetic | Tokamak | 1000 | O&M Cost | NOAK | 14 | mills/ kWh | 2003 | Low | Tokimatsu et al. (2003) |
| n.a. | Magnetic | Tokamak | 1000 | O&M Cost | FOAK | 38 | mills/ kWh | 2003 | High | Tokimatsu et al. (2003) |
| UWMAK-III | Magnetic | Tokamak | 1185 | O&M Cost | $10^{th}$-OAK | 3226667 | USD | 1967 | 30 years lifetime | Tsoulfanidis (1981) |
| Fusion power reactor | Magnetic | Tokamak | n.a. | O&M Fixed | NOAK | 51 | USD/ kW | 2000 | Minimum; availability of this technology for commercial uses is planned for 2048; 40 year lifetime; differences in costs due to regional variations | Vaillancourt et al. (2008) |
| Fusion power reactor | Magnetic | Tokamak | n.a. | O&M Fixed | NOAK | 77 | USD/ kW | 2000 | Maximum; availability of this technology for commercial uses is planned for 2048; 40 year lifetime; differences in costs due to regional variations | Vaillancourt et al. (2008) |
| Fusion power reactor | Magnetic | Tokamak | n.a. | O&M Variable | NOAK | 0.1 | USD/ GJ | 2000 | Minimum; availability of this technology for commercial uses is | Vaillancourt et al. (2008) |

| | | | | | | | | | planned for 2048; 40 year lifetime; differences in costs due to regional variations | |
|---|---|---|---|---|---|---|---|---|---|---|
| Fusion power reactor | Magnetic | Tokamak | n.a. | O&M Variable | NOAK | 0.36 | USD/ GJ | 2000 | Maximum; differences in costs due to regional variations | Vaillancourt et al. (2008) |
| Prometheus-L | Inertia | Direct Drive | 972 | O&M Cost | $10^{th}$-OAK | 62030 000 | USD | 1991 | | Waganer et al. (1992) |
| Prometheus-H | Inertia | Indirect Drive | 999 | O&M Cost | $10^{th}$-OAK | 62190 000 | USD | 1991 | | Waganer et al. (1992) |
| "FPP"_min | Magneto-Inertial | SLC / PJMIF / Staged Z-Pinch / SFS Z-Pinch | 150 | O&M Cost | $10^{th}$-OAK | 18000 000 | USD | 2016 | Four different fusion MIF technologies; fusion power core is 10th-of-a-kind and the BOP is nth-of-a-kind; 6 years construction time | Woodruff et al. (2017) |
| "FPP"_max | Magneto-Inertial | SLC / PJMIF / Staged Z-Pinch / SFS Z-Pinch | 150 | O&M Cost | $10^{th}$-OAK | 38000 000 | USD | 2016 | Four different fusion MIF technologies; fusion power core is 10th-of-a-kind and the BOP is nth-of-a-kind; 6 years construction time | Woodruff et al. (2017) |
| Tokamak Reactor-1 | Magnetic | Tokamak | 1000 | O&M Cost | NOAK | 13 | mills/ kWh | 2003 | | Yamazaki et al. (2009) |
| Spherical Reactor-1 | Magnetic | Spherical Tokamak | 1000 | O&M Cost | NOAK | 14.4 | mills/ kWh | 2003 | | Yamazaki et al. (2009) |
| Helical Reactor-1 | Magnetic | Heliotron | 1000 | O&M Cost | NOAK | 11 | mills/ kWh | 2003 | | Yamazaki et al. (2009) |
| Tokamak (Fusion-Fission Hybrid) | Magnetic | Tokamak | 1000 | O&M Cost | NOAK | 11.7 | mills/ kWh | 2011 | | Yamazaki et al. (2011) |
| Tokamak (D-3He) | Magnetic | Tokamak | 1000 | O&M Cost | NOAK | 11.9 | mills/ kWh | 2011 | | Yamazaki et al. (2011) |
| Tokamak Reactor (DT) | Magnetic | Tokamak | 1000 | O&M Cost | NOAK | 13 | mills/ kWh | 2011 | | Yamazaki et al. (2011) |
| Spherical Tokamak (DT) | Magnetic | Spherical Tokamak | 1000 | O&M Cost | NOAK | 14.4 | mills/ kWh | 2007 | | Yamazaki et al. (2011) |
| Helical Reactor (DT) | Magnetic | Heliotron | 1000 | O&M Cost | NOAK | 11 | mills/ kWh | 2007 | | Yamazaki et al. (2011) |

**Table 4: Literature-reported fixed and variable O&M cost assumptions for FPPs by confinement type and technological maturity.**

Note: In addition to fixed and variabl O&M costs, supplementary techno-economic assumptions reported in the source studies – such as construction time, plant lifetime, start of operation, and configuration-specific characteristics – are extracted where available and documented in the table. Construction time and plant lifetime values reported here are used as input parameters for the LCOE simulations.

Abbreviations used: ARIES: Advanced Reactor Innovation and Evaluation Study; ARC: Advanced Reactor Concept; BOP: Balance of plant; CASCADE: Conceptual Advanced Stellarator/Direct-Drive Energy; DEMO: Demonstration Fusion Power Plant; DEMO2: Second-generation demonstration fusion power plant; DHe3: Deuterium-helium-3; DT: Deuterium-tritium; ESECOM: Energy Systems and Economic Modeling; FLiBe: Lithium fluoride-beryllium fluoride molten salt; FOAK: First-of-a-kind; FPP: Fusion power plant; GENEROMAK: Generic Reactor Model for Magnetic Fusion Power Plants; HYLIFE: High Yield Lithium Injection Fusion Energy; MHD: Magnetohydrodynamic; MIF: Magneto-inertial fusion; MW: Megawatt; NOAK: N$^{th}$-of-a-kind; O&M: Operations and maintenance; PCSR: Power Core Support Reactor; PPCS: Power Plant Conceptual Study; PJMIF: Plasma Jet Magneto-Inertial Fusion; RAF: Reduced activation ferritic; RFP: Reversed-field pinch; RS: Reversed shear; SFS: Sheared-flow stabilized; SiC: Silicon carbide; SLC: Stabilized liner compressor; ST: Spherical tokamak; 10$^{th}$-OAK: Tenth-of-a-kind; TOK: Tokamak; TITAN: Tokamak Ignition/Thermonuclear Assembly Node; USD: United States dollar; UWMAK: University of Wisconsin Magnetic Fusion Reactor Design Study.

| **Reactor** | **Confinement Type** | **Reactor Type** | **Capacity [MW]** | **Readiness** | **Value** | **Unit** | **Cost Reference Year** | **Comment** | **Reference** |
|---|---|---|---|---|---|---|---|---|---|
| STARFIRE | Magnetic | Tokamak | 1200 | $10^{th}$-OAK | 0.04 | mills/kWh | 1980 | | Baker et al. (1981) |
| Generomak - ESECOM - V-LI/TOK | Magnetic | Tokamak | 1200 | $10^{th}$-OAK | 0.35564 | USD/MWh | 1986 | | Delene et al. (1988) |
| Generomak - ESECOM - RAF-He/TOK | Magnetic | Tokamak | 1200 | $10^{th}$-OAK | 0.35564 | USD/MWh | 1986 | | Delene et al. (1988) |
| Generomak - ESECOM - RAF-LiPb/RFP | Magnetic | Reversed-Field Pinch | 1200 | $10^{th}$-OAK | 0.35564 | USD/MWh | 1986 | | Delene et al. (1988) |
| Generomak - ESECOM - V-Li/RFP | Magnetic | Reversed-Field Pinch | 1200 | $10^{th}$-OAK | 0.35564 | USD/MWh | 1986 | | Delene et al. (1988) |
| Generomak - ESECOM - SiC-He/TOK | Magnetic | Tokamak | 1200 | $10^{th}$-OAK | 0.35564 | USD/MWh | 1986 | | Delene et al. (1988) |
| Generomak - ESECOM - V-FLiBe/TOK | Magnetic | Tokamak | 1200 | $10^{th}$-OAK | 0.35564 | USD/MWh | 1986 | | Delene et al. (1988) |
| Generomak - ESECOM - V-MHD/TOK | Magnetic | Tokamak | 1200 | $10^{th}$-OAK | 0.30822 | USD/MWh | 1986 | | Delene et al. (1988) |
| Generomak - ESECOM - V-DHe3/TOK | Magnetic | Tokamak | 1200 | $10^{th}$-OAK | 2.00533 | USD/MWh | 1986 | D-$^{3}$He | Delene et al. (1988) |
| New ARIES | Magnetic | Tokamak | 1000 | $10^{th}$-OAK | 0.03 | mills/kWh | 1992 | | Dragojlovic et al. (2010) |
| DEMO2 | Magnetic | Tokamak | 953 | FOAK | 0.44 | mills/kWh | 2015 | | Entler et al. (2018) |
| Option 1-Reactor - restructured | Magnetic | Stellarator | 914 | FOAK | 0.20865 | USD/MWh | 2023 | | Famà et al. (2024) |
| Option 2-Reactor - restructured | Magnetic | Stellarator | 774 | FOAK | 0.24639 | USD/MWh | 2023 | | Famà et al. (2024) |
| Option 3-Reactor - restructured | Magnetic | Stellarator | 1945 | FOAK | 0.09805 | USD/MWh | 2023 | | Famà et al. (2024) |
| Option 4-Reactor - restructured | Magnetic | Stellarator | 1165 | FOAK | 0.1637 | USD/MWh | 2023 | | Famà et al. (2024) |
| Option 1-Reactor - Greenfield | Magnetic | Stellarator | 1204 | FOAK | 0.15839 | USD/MWh | 2023 | | Famà et al. (2024) |
| Option 2-Reactor - Greenfield | Magnetic | Stellarator | 1313 | FOAK | 0.14524 | USD/MWh | 2023 | | Famà et al. (2024) |
| Option 3-Reactor - Greenfield | Magnetic | Stellarator | 2272 | FOAK | 0.08394 | USD/MWh | 2023 | | Famà et al. (2024) |
| Option 4-Reactor - Greenfield | Magnetic | Stellarator | 1368 | FOAK | 0.13941 | USD/MWh | 2023 | | Famà et al. (2024) |
| Conventional R&D | Magnetic | Tokamak | 1000 | $10^{th}$-OAK | 2 | USD/MWh | 2000 | Fuel and back-end costs | Gi et al. (2020) |
| Advanced R&D | Magnetic | Tokamak | 1000 | $10^{th}$-OAK | 2 | USD/MWh | 2000 | Fuel and back-end costs | Gi et al. (2020) |
| FPP_low | Magneto-Inertial | SLC / PJMIF / Staged Z-Pinch / SFS Z-Pinch | 383.1 | Advanced NOAK | 0.03311 | USD/MWh | 2020 | | Hsu et al. (2020) |
| FPP_high | Magneto-Inertial | SLC / PJMIF / Staged Z-Pinch / SFS Z-Pinch | 814.4 | Advanced NOAK | 0.01557 | USD/MWh | 2020 | | Hsu et al. (2020) |
| Large Helical Device | Magnetic | Heliotron | 1604 | NOAK | 0.023 | mills/kWh | 2002 | | Kozaki et al. (2009) |
| Large Helical Device | Magnetic | Heliotron | 1601 | NOAK | 0.022 | mills/kWh | 2002 | | Kozaki et al. (2009) |
| Large Helical Device | Magnetic | Heliotron | 1598 | NOAK | 0.021 | mills/kWh | 2002 | | Kozaki et al. (2009) |
| Large Helical Device | Magnetic | Heliotron | 1598 | NOAK | 0.021 | mills/kWh | 2002 | | Kozaki et al. (2009) |
| ARIES-III | Magnetic | Tokamak | 1000 | 10$^{th}$-OAK | 17.5 | mills/kWh | 1992 | D-$^{3}$He | Krakowski (1995) |
| ARIES-II | Magnetic | Tokamak | 1000 | 10$^{th}$-OAK | 0.03 | mills/kWh | 1992 | | Krakowski (1995) |
| ARIES-IV | Magnetic | Tokamak | 1000 | 10$^{th}$-OAK | 0.03 | mills/kWh | 1992 | | Krakowski (1995) |
| ARIES-I | Magnetic | Tokamak | 1000 | 10$^{th}$-OAK | 0.03 | mills/kWh | 1992 | | Krakowski (1995) |
| Reactor based on ARIES Code | Magnetic | Tokamak | 1200 | 10$^{th}$-OAK | 0.06 | mills/kWh | 1990 | | Miller et al. (1991) |
| Reactor based on SCAN Code | Magnetic | Tokamak | 1200 | FOAK | 0.05 | mills/kWh | 1990 | | Miller et al. (1991) |
| HYLIFE-II | Inertia | Ion Beam Driver | 1083 | 10$^{th}$-OAK | 0.01 | mills/kWh | 1988 | | Moir (1992) |
| ARIES-I | Magnetic | Tokamak | 1000 | 10$^{th}$-OAK | 0.04 | mills/kWh | 1988 | | Najmabadi et al. (1991) |
| ARIES-AT | Magnetic | Tokamak | 1000 | 10$^{th}$-OAK | 0.03 | mills/kWh | 1992 | | Najmabadi et al. (2006) |
| Generic Toroidal Reactor | Magnetic | Tokamak | 1000 | 10$^{th}$-OAK | 0.05708 | USD/MWh | 2010 | | Sheffield and Milora (2016) |
| European Model C Reactor | Magnetic | Tokamak | 1450 | 10th-OAK | 0.04199 | USD/MWh | 2010 | | Sheffield and Milora (2016) |
| STARFIRE | Magnetic | Tokamak | 1200 | 10th-OAK | 0.05854 | USD/MWh | 1983 | | Sheffield et al. (1986) |
| Tokamak Reactor-1 | Magnetic | Tokamak | 1000 | NOAK | 0.04 | mills/kWh | 2003 | 30 years lifetime | Yamazaki et al. (2009) |
| Spherical Reactor-1 | Magnetic | Spherical Tokamak | 1000 | NOAK | 0.04 | mills/kWh | 2003 | 30 years lifetime | Yamazaki et al. (2009) |
| Helical Reactor-1 | Magnetic | Heliotron | 1000 | NOAK | 0.04 | mills/kWh | 2003 | 30 years lifetime | Yamazaki et al. (2009) |
| Tokamak (D-3He) | Magnetic | Tokamak | 1000 | NOAK | 5.54 | mills/kWh | 2011 | 30 years lifetime; D-$^{3}$He | Yamazaki et al. (2011) |
| Tokamak Reactor (DT) | Magnetic | Tokamak | 1000 | NOAK | 0.04 | mills/kWh | 2011 | 30 years lifetime | Yamazaki et al. (2011) |
| Spherical Tokamak (DT) | Magnetic | Spherical Tokamak | 1000 | NOAK | 0.04 | mills/kWh | 2008 | 30 years lifetime | Yamazaki et al. (2011) |
| Helical Reactor (DT) | Magnetic | Heliotron | 1000 | NOAK | 0.04 | mills/kWh | 2008 | 30 years lifetime | Yamazaki et al. (2011) |
| ARIES-RS | Magnetic | Tokamak | 1000 | 10th-OAK | 0.05372 | USD/MWh | 2010 | | Sheffield and Milora (2016) |
| ARIES-AT | Magnetic | Tokamak | 1000 | 10th-OAK | 0.05372 | USD/MWh | 2010 | | Sheffield and Milora (2016) |
| ARIES-AT Advanced | Magnetic | Tokamak | 1000 | 10$^{th}$-OAK | 0.05372 | USD/MWh | 2010 | | Sheffield and Milora (2016) |
| Prometheus-L | Inertia | Direct Drive | 972 | 10$^{th}$-OAK | 0.60349 | USD/MWh | 1991 | | Waganer et al. (1992) |
| Prometheus-H | Inertia | Indirect Drive | 999 | 10$^{th}$-OAK | 0.82832 | USD/MWh | 1991 | | Waganer et al. (1992) |

**Table 5: Literature-reported fuel cost assumptions for FPPs by confinement type and technological maturity.**

Note: In addition to fuel costs, supplementary techno-economic assumptions reported in the source studies – such as plant lifetime and additional cost components included in the fuel cost estimates – are extracted where available and documented in the table. Plant lifetime values reported here are used as input parameters for the LCOE simulations.

ARIES: Advanced Reactor Innovation and Evaluation Study; BOP: Balance of plant; DHe3: Deuterium–helium-3; DT: Deuterium-tritium; ESECOM: Energy Systems and Economic Modeling; FLiBe: Lithium fluoride–beryllium fluoride molten salt; FOAK: First-of-a-kind; FPP: Fusion power plant; GENEROMAK: Generic Reactor Model for Magnetic Fusion Power Plants; HYLIFE: High Yield Lithium Injection Fusion Energy; MHD: Magnetohydrodynamic; MIF: Magneto-inertial fusion; MW: Megawatt; MWh: Megawatt-hour; NOAK: N$^{th}$-of-a-kind; PJMIF: Plasma Jet Magneto-Inertial Fusion; RAF: Reduced activation ferritic; RFP: Reversed-field pinch; SFS: Sheared flow stabilized; SiC: Silicon carbide; SLC: Stabilized liner compressor; STARFIRE: STARFIRE tokamak reactor concept; 10$^{th}$-OAK: Tenth-of-a-kind; TOK: Tokamak; USD: United States dollar.

| **Reactor** | **Confinement type** | **Reactor type** | **Capacity [MW]** | **Readiness** | **Availability/ capacity factor** | **Reference** |
|---|---|---|---|---|---|---|
| LIFE1 | Inertia | Indirect Drive | 400 | FOAK | AV: 0.5 | Anklam et al. (2011) |
| LIFE2 | Inertia | Indirect Drive | 2200 | NOAK | AV: 0.92 | Anklam et al. (2011) |
| LIFE3 | Inertia | Indirect Drive | 2660 | 10$^{th}$-OAK | AV: 0.92 | Anklam et al. (2011) |
| UWMAK-III | Magnetic | Tokamak | 1985 | 10$^{th}$-OAK | AV: 0.95 (pulsed with TES) | (Badger et al. 1975) |
| n.a. | Magnetic | Tokamak | 1500 | NOAK | AV: 0.8 | Bednyagin and Gnansounou 2011) |
| n.a. | Magnetic | Tokamak | 1500 | FOAK | CF: 0.8 | Bednyagin and Gnansounou 2011) |
| n.a. | Magnetic | Tokamak | 1500 | FOAK | CF: 0.8 | Bednyagin and Gnansounou 2011) |
| n.a. | Magnetic | Tokamak | 1500 | 10th-OAK | CF: 0.8 | Bednyagin and Gnansounou 2011) |
| n.a. | Magnetic | Tokamak | 1500 | 10th-OAK | CF: 0.8 | Bednyagin and Gnansounou 2011) |
| n.a. | Magnetic | Tokamak | 1500 | FOAK | CF: 0.8 | Bednyagin and Gnansounou 2011) |
| n.a. | Magnetic | Tokamak | 1500 | 10th-OAK | CF: 0.8 | Bednyagin and Gnansounou 2011) |
| n.a. | Magnetic | Tokamak | 1500 | Advanced NOAK | CF: 0.8 | Bednyagin and Gnansounou 2011) |
| n.a. | Magnetic | Tokamak | 1500 | Highly Advanced NOAK | CF: 0.8 | Bednyagin and Gnansounou 2011) |
| n.a. | Magnetic | Tokamak | 1500 | Highly Advanced NOAK | CF: 0.8 | Bednyagin and Gnansounou 2011) |
| n.a. | Magnetic | Tokamak | 1500 | Highly Advanced NOAK | CF: 0.8 | Bednyagin and Gnansounou 2011) |
| n.a. | Magnetic | Tokamak | 1500 | FOAK | CF: 0.8 | Bednyagin and Gnansounou 2011) |
| MARS | Magnetic | Magnetic Mirror | 1198 | 10th-OAK | AV: 0.7 | Berwald et al. (1987) |
| HYLIFE-I (single unit) | Inertia | Direct Drive | 1010 | 10th-OAK | 0.7 | Blink et al. (1985) |
| HYLIFE-I (dual unit) | Inertia | Direct Drive | 2000 | 10th-OAK | 0.7 | Blink et al. (1985) |
| STARFIRE | Magnetic | Tokamak | 1200 | 10th-OAK | AV: 0.75 | Baker et al. (1981) |
| Modular Fusion Plant | Magnetic | Spherical Tokamak | 735.77 | FOAK | CF. 0.95 | Chuyanov and Gryaznevich (2017) |
| HYLIFE | Inertia | Direct Drive | 1010 | 10th-OAK | AV: 0.76 | Conn et al. (1990) |
| CASCADE | Inertia | Direct Drive | 815 | 10th-OAK | AV: 0.76 | Conn et al. (1990) |
| ESECOM GENEROMAK LI/LI/V | Magnetic | Reversed-Field Pinch | 1200 | 10th-OAK | AV: 0.76 | Conn et al. (1990) |
| TITAN-I | Magnetic | Reversed-Field Pinch | 970 | 10th-OAK | AV: 0.76 | Conn et al. (1990) |
| ARIES-II | Magnetic | Tokamak | 1000 | 10th-OAK | AV: 0.76 | Conn et al. (1990) |
| ARIES-I | Magnetic | Tokamak | 1000 | 10th-OAK | AV: 0.76 | Conn et al. (1990) |
| V-Li/TOK | Magnetic | Tokamak | 1200 | 10th-OAK | CF: 0.75 | Delene (1991) |
| V-Li/TOK | Magnetic | Tokamak | 1200 | 10th-OAK | CF: 0.75 | Delene (1991) |
| ARIES-I | Magnetic | Tokamak | 1200 | 10th-OAK | CF: 0.75 | Delene (1991) |
| Generomak - ESECOM - RAF-LiPb/RFP | Magnetic | Reversed-Field Pinch | 1200 | 10th-OAK | CF: 0.65 | Delene et al. (1988) |
| Generomak - ESECOM - V-Li/RFP | Magnetic | Reversed-Field Pinch | 1200 | 10th-OAK | CF: 0.65 | Delene et al. (1988) |
| Generomak - ESECOM - V-LI/TOK | Magnetic | Tokamak | 1200 | 10th-OAK | CF: 0.65 | Delene et al. (1988) |
| Generomak - ESECOM - RAF-He/TOK | Magnetic | Tokamak | 1200 | 10th-OAK | CF: 0.65 | Delene et al. (1988) |
| Generomak - ESECOM - SiC-He/TOK | Magnetic | Tokamak | 1200 | 10th-OAK | CF: 0.65 | Delene et al. (1988) |
| Generomak - ESECOM - V-FLiBe/TOK | Magnetic | Tokamak | 1200 | 10th-OAK | CF: 0.65 | Delene et al. (1988) |
| Generomak - ESECOM - V-MHD/TOK | Magnetic | Tokamak | 1200 | 10th-OAK | CF: 0.75 | Delene et al. (1988) |
| Generomak - ESECOM - V-DHe3/TOK | Magnetic | Tokamak | 1200 | 10th-OAK | CF: 0.75 | Delene et al. (1988) |
| ARIES-RS_max | Magnetic | Tokamak | 1000 | 10th-OAK | CF: 0.7 | Delene et al. (2001) |
| ARIES-ST_max | Magnetic | Tokamak | 1000 | 10th-OAK | CF: 0.7 | Delene et al. (2001) |
| ARIES-RS | Magnetic | Tokamak | 1300 | 10th-OAK | CF: 0.8 | Delene et al. (2001) |
| ARIES-RS | Magnetic | Tokamak | 1000 | 10th-OAK | CF: 0.8 | Delene et al. (2001) |
| ARIES-ST | Magnetic | Tokamak | 1300 | 10th-OAK | CF: 0.8 | Delene et al. (2001) |
| ARIES-ST | Magnetic | Tokamak | 1000 | 10th-OAK | CF: 0.8 | Delene et al. (2001) |
| ARIES-RS | Magnetic | Tokamak | 1000 | 10th-OAK | CF: 0.8 | Delene et al. (2001) |
| ARIES-ST | Magnetic | Tokamak | 1000 | 10th-OAK | CF: 0.8 | Delene et al. (2001) |
| ARIES-ST_min | Magnetic | Tokamak | 1300 | 10th-OAK | CF: 0.9 | Delene et al. (2001) |
| ARIES-RS_min | Magnetic | Tokamak | 1300 | 10th-OAK | CF: 0.9 | Delene et al. (2001) |
| NLHD-D1 Scaling Heliotron | Magnetic | Heliotron | 1000 | NOAK | AV: 0.75 | Dolan et al. (2005) |
| ARIES-SPSS | Magnetic | Stellarator | 1000 | 10th-OAK | AV: 0.75 | Dolan et al. (2005) |
| Base Case Heliotron | Magnetic | Heliotron | 1940 | NOAK | AV: 0.75 | Dolan et al. (2005) |
| New ARIES | Magnetic | Tokamak | 1000 | 10th-OAK | CF: 0.85 | Dragojlovic et al. (2010) |
| New ARIES | Magnetic | Tokamak | 1000 | 10th-OAK | CF: 0.85 | Dragojlovic et al. (2010) |
| DEMO2 | Magnetic | Tokamak | 953 | FOAK | AV: 0.75 | Entler et al. (2018) |
| Option 1-Reactor - Greenfield | Magnetic | Stellarator | 1204 | FOAK | CF: 0.85 | Famà et al. (2024) |
| Option 1-Reactor - Greenfield | Magnetic | Stellarator | 1204 | FOAK | CF: 0.85 | Famà et al. (2024) |
| Option 2-Reactor - Greenfield | Magnetic | Stellarator | 1313 | FOAK | CF: 0.85 | Famà et al. (2024) |
| Option 2-Reactor - Greenfield | Magnetic | Stellarator | 1313 | FOAK | CF: 0.85 | Famà et al. (2024) |
| Option 3-Reactor - Greenfield | Magnetic | Stellarator | 2272 | FOAK | CF: 0.85 | Famà et al. (2024) |
| Option 4-Reactor - Greenfield | Magnetic | Stellarator | 1368 | FOAK | CF: 0.85 | Famà et al. (2024) |
| Option 3-Reactor - Greenfield | Magnetic | Stellarator | 2272 | FOAK | CF: 0.85 | Famà et al. (2024) |
| Option 4-Reactor - Greenfield | Magnetic | Stellarator | 1368 | FOAK | CF: 0.85 | Famà et al. (2024) |
| Option 1-Reactor - restructured | Magnetic | Stellarator | 914 | FOAK | CF: 0.85 | Famà et al. (2024) |
| Option 1-Reactor - restructured | Magnetic | Stellarator | 914 | FOAK | CF: 0.85 | Famà et al. (2024) |
| Option 2-Reactor - restructured | Magnetic | Stellarator | 774 | FOAK | CF: 0.85 | Famà et al. (2024) |
| Option 2-Reactor - restructured | Magnetic | Stellarator | 774 | FOAK | CF: 0.85 | Famà et al. (2024) |
| Option 3-Reactor - restructured | Magnetic | Stellarator | 1945 | FOAK | CF: 0.85 | Famà et al. (2024) |
| Option 4-Reactor - restructured | Magnetic | Stellarator | 1165 | FOAK | CF: 0.85 | Famà et al. (2024) |
| Option 3-Reactor - restructured | Magnetic | Stellarator | 1945 | FOAK | CF: 0.85 | Famà et al. (2024) |
| Option 4-Reactor - restructured | Magnetic | Stellarator | 1165 | FOAK | CF: 0.85 | Famà et al. (2024) |
| Conventional R&D | Magnetic | Tokamak | 1000 | $10^{th}$-OAK | AV: 0.9 | Gi et al. (2020) |
| Advanced R&D | Magnetic | Tokamak | 1000 | $10^{th}$-OAK | AV: 0.9 | Gi et al. (2020) |
| Fusion Power Corporation Reactor (incl. Hydrogen production) | Inertia | Ion Beam Driver | 38000 | NOAK | CF: 0.8 | Helsley and Burke (2014) |
| n.a. | Magnetic | Tokamak | n.a. | 10th-OAK | AV: 0.75 | Hender et al. (1996) |
| n.a. | Magnetic | Stellarator | n.a. | 10th-OAK | AV: 0.75 | Hender et al. (1996) |
| n.a. | Magnetic | Spherical Tokamak | n.a. | 10th-OAK | AV: 0.75 | Hender et al. (1996) |
| CREST | Magnetic | Tokamak | 1160 | FOAK | AV: 0.9 | Hiwatari and Goto (2019) |
| CREST | Magnetic | Tokamak | 1160 | 10th-OAK | AV: 0.9 | Hiwatari and Goto (2019) |
| FPP_low | Magneto-Inertial | SLC / PJMIF / Staged Z-Pinch / SFS Z-Pinch | 383.1 | Advanced NOAK | AV: 0.9 | Hsu et al. (2020) |
| FPP_high | Magneto-Inertial | SLC / PJMIF / Staged Z-Pinch / SFS Z-Pinch | 814.4 | Advanced NOAK | AV: 0.9 | Hsu et al. (2020) |
| FPP_high | Magneto-Inertial | SLC / PJMIF / Staged Z-Pinch / SFS Z-Pinch | 814.4 | Advanced NOAK | AV: 0.9 | Hsu et al. (2020) |
| FPP_low | Magneto-Inertial | SLC / PJMIF / Staged Z-Pinch / SFS Z-Pinch | 383.1 | Advanced NOAK | AV: 0.9 | Hsu et al. (2020) |
| FPP_low | Magneto-Inertial | SLC / PJMIF / Staged Z-Pinch / SFS Z-Pinch | 383.1 | Advanced NOAK | AV: 0.9 | Hsu et al. (2020) |
| FPP_high | Magneto-Inertial | SLC / PJMIF / Staged Z-Pinch / SFS Z-Pinch | 814.4 | Advanced NOAK | AV: 0.9 | Hsu et al. (2020) |
| With Central Solenoid min | Magnetic | Tokamak | 1000 | NOAK | AV: 0.75 | Jo et al. (2021) |
| With Central Solenoid max | Magnetic | Tokamak | 1600 | NOAK | AV: 0.75 | Jo et al. (2021) |
| No Central Solenoid min | Magnetic | Tokamak | 1000 | NOAK | AV: 0.75 | Jo et al. (2021) |
| No Central Solenoid max | Magnetic | Tokamak | 1420 | NOAK | AV: 0.75 | Jo et al. (2021) |
| Large Helical Device | Magnetic | Heliotron | 1604 | NOAK | AV: 0.68 | Kozaki et al. (2009) |
| Large Helical Device | Magnetic | Heliotron | 1601 | NOAK | AV: 0.706 | Kozaki et al. (2009) |
| Large Helical Device | Magnetic | Heliotron | 1598 | NOAK | AV: 0.726 | Kozaki et al. (2009) |
| Large Helical Device | Magnetic | Heliotron | 1598 | NOAK | AV: 0.727 | Kozaki et al. (2009) |
| PULSAR-I | Magnetic | Tokamak | 1000 | $10^{th}$-OAK | AV: 0.75 | Krakowski (1995) |
| Compact Spheromak Reactor | Magnetic | Spheromak | 1000 | $10^{th}$-OAK | AV: 0.75 | Krakowski (1995) |
| HELIAC | Magnetic | Stellarator | 1000 | $10^{th}$-OAK | AV: 0.75 | Krakowski (1995) |
| High-field Stellarator Reactor | Magnetic | Stellarator | 1000 | $10^{th}$-OAK | AV: 0.75 | Krakowski (1995) |
| TITAN-I | Magnetic | Reversed-Field Pinch | 1000 | $10^{th}$-OAK | AV: 0.75 | Krakowski (1995) |
| TITAN-II | Magnetic | Reversed-Field Pinch | 1000 | $10^{th}$-OAK | AV: 0.75 | Krakowski (1995) |
| ARIES-I | Magnetic | Tokamak | 1000 | $10^{th}$-OAK | AV: 0.75 | Krakowski (1995) |
| ARIES-II | Magnetic | Tokamak | 1000 | $10^{th}$-OAK | AV: 0.75 | Krakowski (1995) |
| ARIES-III | Magnetic | Tokamak | 1000 | $10^{th}$-OAK | AV: 0.75 | Krakowski (1995) |
| ARIES-IV | Magnetic | Tokamak | 1000 | $10^{th}$-OAK | AV: 0.75 | Krakowski (1995) |
| n.a. | Magnetic | Tokamak | 1000 | FOAK | CF: 0.675 | Lako & Seebregts (1998) |
| Commercial breeder | Magnetic | Tandem Mirror | 1100 | 10th-OAK | CF: 0.7 | Lee (1987) |
| ARC-2035 | Magnetic | Tokamak | 190 | FOAK | CF: 0.685 | Lerede et al. (2023) |
| ARC-2050 | Magnetic | Tokamak | 250 | NOAK | CF: 0.685 | Lerede et al. (2023) |
| EU-DEMO | Magnetic | Tokamak | 500 | FOAK | CF: 0.685 | Lerede et al. (2023) |
| EU-DEMO Advanced | Magnetic | Tokamak | 500 | NOAK | CF: 0.685 | Lerede et al. (2023) |
| Asian-DEMO | Magnetic | Tokamak | 500 | FOAK | CF: 0.685 | Lerede et al. (2023) |
| Early large | Magnetic | Tokamak | 1500 | NOAK | CF: 0.56 (pulsed) | Lindley et al. (2023) |
| Early large | Magnetic | Tokamak | 1500 | 10th-OAK | CF: 0.56 (pulsed) | Lindley et al. (2023) |
| Advanced Large | Magnetic | Tokamak | 1500 | NOAK | CF: 0.75 | Lindley et al. (2023) |
| Advanced Large | Magnetic | Tokamak | 1500 | 10th-OAK | CF: 0.75 | Lindley et al. (2023) |
| PPCS Model A Min | Magnetic | Tokamak | 1546 | 10th-OAK | AV: 0.75 | Maisonnier et al. (2007) |
| PPCS Model A Max | Magnetic | Tokamak | 1546 | FOAK | AV: 0.75 | Maisonnier et al. (2007) |
| PPCS Model AB Min | Magnetic | Tokamak | 1500 | 10th-OAK | AV: 0.75 | Maisonnier et al. (2007) |
| PPCS Model AB Max | Magnetic | Tokamak | 1500 | FOAK | AV: 0.75 | Maisonnier et al. (2007) |
| PPCS Model B Min | Magnetic | Tokamak | 1332 | 10th-OAK | AV: 0.75 | Maisonnier et al. (2007) |
| PPCS Model B Max | Magnetic | Tokamak | 1332 | FOAK | AV: 0.75 | Maisonnier et al. (2007) |
| PPCS Model C Min | Magnetic | Tokamak | 1449 | 10th-OAK | AV: 0.75 | Maisonnier et al. (2007) |
| PPCS Model C Max | Magnetic | Tokamak | 1449 | FOAK | AV: 0.75 | Maisonnier et al. (2007) |
| PPCS Model D Min | Magnetic | Tokamak | 1527 | 10th-OAK | AV: 0.75 | Maisonnier et al. (2007) |
| PPCS Model D Max | Magnetic | Tokamak | 1527 | FOAK | AV: 0.75 | Maisonnier et al. (2007) |
| CASCADE (all Conventional ) | Inertia | Direct Drive | 815 | 10$^{th}$-OAK | AV: 0.75 | Maya et al. (1985) |
| CASCADE | Inertia | Direct Drive | 815 | 10$^{th}$-OAK | AV: 0.75 | Maya et al. (1985) |
| CASCADE (all Conventional ) | Inertia | Direct Drive | 815 | 10$^{th}$-OAK | AV: 0.75 | Maya et al. (1985) |
| CASCADE | Inertia | Direct Drive | 815 | 10$^{th}$-OAK | AV: 0.75 | Maya et al. (1985) |
| Reactor based on SCAN Code | Magnetic | Tokamak | 1200 | FOAK | AV: 0.75 | Miller et al. (1991) |
| Reactor based on ARIES Code | Magnetic | Tokamak | 1200 | 10$^{th}$-OAK | AV: 0.75 | Miller et al. (1991) |
| ARIES-I | Magnetic | Tokamak | 1000 | 10$^{th}$-OAK | AV: 0.76 | Najmabadi et al. (1991) |
| ARIES-I | Magnetic | Tokamak | 1000 | 10$^{th}$-OAK | AV: 0.76 | Najmabadi et al. (1991) |
| ARIES-AT | Magnetic | Advanced Tokamak | 1000 | 10$^{th}$-OAK | AV: 0.85 | Najmabadi et al. (2006) |
| ARIES-AT | Magnetic | Tokamak | 1000 | 10$^{th}$-OAK | AV: 0.9 | Najmabadi et al. (2006) |
| ARIES-CS | Magnetic | Stellarator | 1000 | 10$^{th}$-OAK | AV: 0.85 | Najmabadi et al. (2008) |
| SSTR-like DEMO | Magnetic | Tokamak | 1080 | FOAK | AV: 0.75 | Oishi et al. (2012) |
| European Model C Reactor | Magnetic | Tokamak | 1450 | 10$^{th}$-OAK | AV: 0.75 | Sheffield and Milora (2016) |
| ARIES-RS | Magnetic | Tokamak | 1000 | 10$^{th}$-OAK | AV: 0.85 | Sheffield and Milora (2016) |
| ARIES-AT | Magnetic | Tokamak | 1000 | 10$^{th}$-OAK | AV: 0.85 | Sheffield and Milora (2016) |
| ARIES-AT Advanced | Magnetic | Tokamak | 1000 | 10th-OAK | AV: 0.85 | Sheffield and Milora (2016) |
| Generic Toroidal Reactor | Magnetic | Tokamak | 1000 | 10th-OAK | AV: 0.85 | Sheffield and Milora (2016) |
| ARIES-RS | Magnetic | Tokamak | 1000 | 10th-OAK | AV: 0.85 | Sheffield and Milora (2016) |
| ARIES-AT | Magnetic | Tokamak | 1000 | 10th-OAK | AV: 0.85 | Sheffield and Milora (2016) |
| ARIES-AT Advanced | Magnetic | Tokamak | 1000 | 10th-OAK | AV: 0.85 | Sheffield and Milora (2016) |
| STARFIRE | Magnetic | Tokamak | 1200 | 10th-OAK | AV: 0.65 | Sheffield et al. (1986) |
| PCSR-E | Magnetic | Tokamak | 1247 | FOAK | AV: 0.73 (pulsed with 5000 seconds) | Spears (1990) |
| Reactor-1 | Magnetic | Tokamak | 1203 | FOAK | AV: 0.75 | Spears (1990) |
| Reactor-2 | Magnetic | Tokamak | 1209 | FOAK | AV: 0.75 | Spears (1990) |
| 1 GW class (t=0) | Magnetic | Tokamak | 1000 | FOAK | AV: 0.6 | Tokimatsu et al. (2002) |
| 1 GW class (t=25) | Magnetic | Tokamak | 1000 | NOAK | AV: 0.78 | Tokimatsu et al. (2002) |
| 1 GW class (t=50) | Magnetic | Tokamak | 1000 | NOAK | AV: 0.83 | Tokimatsu et al. (2002) |
| UWMAK-III | Magnetic | Tokamak | 1185 | 10th-OAK | CF: 0.65 | Tsoulfanidis (1981) |
| UWMAK-III | Magnetic | Tokamak | 1185 | 10th-OAK | CF: 0.65 | Tsoulfanidis (1981) |
| Prometheus-L | Inertia | Direct Drive | 972 | 10th-OAK | AV: 0.794 | Waganer et al. (1992) |
| Prometheus-L | Inertia | Direct Drive | 972 | 10th-OAK | AV: 0.794 | Waganer et al. (1992) |
| Prometheus-H | Inertia | Indirect Drive | 999 | 10th-OAK | AV: 0.808 | Waganer et al. (1992) |
| Prometheus-H | Inertia | Indirect Drive | 999 | 10th-OAK | AV: 0.808 | Waganer et al. (1992) |
| PPCS-A | Magnetic | Tokamak | 1546 | 10th-OAK | AV: 0.75 | Maisonnier et al. (2005) |
| PPCS-B | Magnetic | Tokamak | 1332 | $10^{th}$-OAK | AV: 0.75 | Maisonnier et al. (2005) |
| PPCS-C | Magnetic | Tokamak | 1449 | $10^{th}$-OAK | AV: 0.75 | Maisonnier et al. (2005) |
| PPCS-D | Magnetic | Tokamak | 1527 | $10^{th}$-OAK | AV: 0.75 | Maisonnier et al. (2005) |
| Tokamak Reactor-1 | Magnetic | Tokamak | 1000 | NOAK | AV: 0.75 | Yamazaki and Dolan (2006) |
| Tokamak Reactor-2 | Magnetic | Tokamak | 2000 | NOAK | AV: 0.75 | Yamazaki and Dolan (2006) |
| Helical Reactor-1 | Magnetic | Heliotron | 1000 | NOAK | AV: 0.75 | Yamazaki and Dolan (2006) |
| Helical Reactor-2 | Magnetic | Heliotron | 2000 | NOAK | AV: 0.75 | Yamazaki and Dolan (2006) |
| Tokamak Reactor-1 | Magnetic | Tokamak | 1000 | NOAK | AV: 0.75 | Yamazaki et al. (2009) |
| Spherical Reactor-1 | Magnetic | Spherical Tokamak | 1000 | NOAK | AV: 0.75 | Yamazaki et al. (2009) |
| Helical Reactor-1 | Magnetic | Heliotron | 1000 | NOAK | AV: 0.75 | Yamazaki et al. (2009) |
| Tokamak (Fusion-Fission Hybrid) | Magnetic | Tokamak | 1000 | NOAK | AV: 0.75 | Yamazaki et al. (2011) |
| Tokamak (D-3He) | Magnetic | Tokamak | 1000 | NOAK | AV: 0.75 | Yamazaki et al. (2011) |
| Tokamak Reactor (DT) | Magnetic | Tokamak | 1000 | NOAK | AV: 0.75 | Yamazaki et al. (2011) |
| Tokamak (Fusion-Fission Hybrid) | Magnetic | Tokamak | 1000 | NOAK | AV: 0.75 | Yamazaki et al. (2011) |
| Tokamak (D-3He) | Magnetic | Tokamak | 1000 | NOAK | AV: 0.75 | Yamazaki et al. (2011) |
| Tokamak Reactor (DT) | Magnetic | Tokamak | 1000 | NOAK | AV: 0.75 | Yamazaki et al. (2011) |
| Spherical Tokamak (DT) | Magnetic | Spherical Tokamak | 1000 | NOAK | AV: 0.75 | Yamazaki et al. (2011) |
| Helical Reactor (DT) | Magnetic | Heliotron | 1000 | NOAK | AV: 0.75 | Yamazaki et al. (2011) |
| IFE (with FLIBE Breeder) | Inertia | Laser Driver | 1000 | NOAK | AV: 0.65-0.85 | Yamazaki et al. (2011) |

**Table 6: Literature-reported availability and capacity factor assumptions for FPPs by confinement type and technological maturity.**

Abbreviations used: ARIES: Advanced Reactor Innovation and Evaluation Study; ARC: Advanced Reactor Concept; AV: Availability; CASCADE: Conceptual Advanced Stellarator/Direct-Drive Energy; CF: Capacity factor; CREST: Commercial Reactor based on Equilibrium and Steady-State Tokamak; DEMO: Demonstration Fusion Power Plant; DHe3: Deuterium-helium-3; DT: Deuterium-tritium; ESECOM: Energy Systems and Economic Modeling; FLiBe: Lithium fluoride-beryllium fluoride molten salt; FOAK: First-of-a-kind; FPP: Fusion power plant; GENEROMAK: Generic Reactor Model for Magnetic Fusion Power Plants; HYLIFE: High Yield Lithium Injection Fusion Energy; IFE: Inertial fusion energy; LIFE: Laser Inertial Fusion Energy; MHD: Magnetohydrodynamic; MIF: Magneto-inertial fusion; MW: Megawatt; NLHD: National Large Helical Device; NOAK: N$^{th}$-of-a-kind; PCSR: Power Core Support Reactor; PPCS: Power Plant Conceptual Study; PJMIF: Plasma Jet Magneto-Inertial Fusion; RAF: Reduced activation ferritic; RFP: Reversed-field pinch; RS: Reversed shear; SFS: Sheared flow stabilized; SLC: Stabilized liner compressor; SPSS: Stellarator Power Plant Study; SSTR: Steady-State Tokamak Reactor; 10$^{th}$-OAK: Tenth-of-a-kind; TOK: Tokamak; UWMAK: University of Wisconsin Magnetic Fusion Reactor Design Study.

| Reactor | Confinement Type | Reactor Type | Capacity [MW] | Readiness | Comment | Reference |
|---|---|---|---|---|---|---|
| LIFE1 | Inertia | Indirect Drive | 400 | FOAK | Learning rates based on experiences from NIF but not further specified | Anklam et al. (2011) |
| LIFE2 | Inertia | Indirect Drive | 2200 | NOAK | Average value between 4 and 6 billion USD; Learning rates based on experiences from NIF but not further specified | Anklam et al. (2011) |
| LIFE3 | Inertia | Indirect Drive | 2660 | 10$^{th}$-OAK | Learning rates based on experiences from NIF but not further specified | Anklam et al. (2011) |
| n.a. | Magnetic | Tokamak | 1500 | NOAK | Pessimistic scenario A; 40 years lifetime; 2060-2070; 10 years construction time | Bednyagin and Gnansounou (2011) |
| n.a. | Magnetic | Tokamak | 1500 | FOAK | Pessimistic scenario A; 40 years lifetime; 2060-2070; 10 years construction time | Bednyagin and Gnansounou (2011) |
| n.a. | Magnetic | Tokamak | 1500 | FOAK | Scenario B; 40 years lifetime; 2050-2060; 10 years construction time | Bednyagin and Gnansounou (2011) |
| n.a. | Magnetic | Tokamak | 1500 | $10^{th}$-OAK | Pessimistic scenario A; 40 years lifetime; 2060-2070; 10 years construction time | Bednyagin and Gnansounou (2011) |
| n.a. | Magnetic | Tokamak | 1500 | $10^{th}$-OAK | Scenario B; 40 years lifetime; 2050-2060; 10 years construction time | Bednyagin and Gnansounou (2011) |
| n.a. | Magnetic | Tokamak | 1500 | FOAK | Optimistic scenario C; 40 years lifetime; 2050-2060; 10 years construction time | Bednyagin and Gnansounou (2011) |
| n.a. | Magnetic | Tokamak | 1500 | $10^{th}$-OAK | Optimistic scenario C; 40 years lifetime; 2050-2060; 10 years construction time | Bednyagin and Gnansounou (2011) |
| n.a. | Magnetic | Tokamak | 1500 | Advanced NOAK | Optimistic scenario C; 40 years lifetime; 2050-2060; 10 years construction time | Bednyagin and Gnansounou (2011) |
| n.a. | Magnetic | Tokamak | 1500 | Highly Advanced NOAK | Pessimistic scenario A; 40 years lifetime; 2060-2070; 10 years construction time | Bednyagin and Gnansounou (2011) |
| n.a. | Magnetic | Tokamak | 1500 | Highly Advanced NOAK | Scenario B; 40 years lifetime; 2050-2060; 10 years construction time | Bednyagin and Gnansounou (2011) |
| n.a. | Magnetic | Tokamak | 1500 | Highly Advanced NOAK | Optimistic scenario C; 40 years lifetime; 2050-2060; 10 years construction time | Bednyagin and Gnansounou (2011) |
| n.a. | Magnetic | Tokamak | 1500 | NOAK | Scenario B; 40 years lifetime; 2050-2060; 10 years construction time | Bednyagin and Gnansounou (2011) |
| n.a. | Magnetic | Tokamak | 1500 | NOAK | Optimistic scenario C; 40 years lifetime; 2050-2060; 10 years construction time | Bednyagin and Gnansounou (2011) |
| n.a. | Magnetic | Tokamak | 1500 | Advanced NOAK | Scenario B; 40 years lifetime; 2050-2060; 10 years construction time | Bednyagin and Gnansounou (2011) |
| n.a. | Magnetic | Tokamak | 1500 | Advanced NOAK | Optimistic scenario C; 40 years lifetime; 2050-2060; 10 years construction time | Bednyagin and Gnansounou (2011) |
| Modular Fusion Plant | Magnetic | Spherical Tokamak | 735.77 | FOAK | Modular plant; learning effects due to mass prodction for modular parts | Chuyanov and Gryaznevich (2017) |
| NLHD-D1 Scaling Heliotron | Magnetic | Heliotron | 1000 | NOAK | 6 years construction time | Dolan et al. (2005) |
| ARIES-SPSS | Magnetic | Stellarator | 1000 | 10th-OAK | Modular; 6 years construction time | Dolan et al. (2005) |
| New ARIES | Magnetic | Tokamak | 1000 | 10th-OAK | 6 years construction time | Dragojlovic et al. (2010) |
| Option 1-Reactor - Greenfield | Magnetic | Stellarator | 1204 | FOAK | Greenfield reference plant where everything has been built from zero/without restructuring | Famà et al. (2024) |
| Option 1-Reactor - Greenfield | Magnetic | Stellarator | 1204 | FOAK | Greenfield reference plant where everything has been built from zero/without restructuring | Famà et al. (2024) |
| Option 2-Reactor - Greenfield | Magnetic | Stellarator | 1313 | FOAK | Greenfield reference plant where everything has been built from zero/without restructuring | Famà et al. (2024) |
| Option 2-Reactor - Greenfield | Magnetic | Stellarator | 1313 | FOAK | Greenfield reference plant where everything has been built from zero/without restructuring | Famà et al. (2024) |
| Option 3-Reactor - Greenfield | Magnetic | Stellarator | 2272 | FOAK | Greenfield reference plant where everything has been built from zero/without restructuring | Famà et al. (2024) |
| Option 4-Reactor - Greenfield | Magnetic | Stellarator | 1368 | FOAK | Greenfield reference plant where everything has been built from zero/without restructuring | Famà et al. (2024) |
| Option 3-Reactor - Greenfield | Magnetic | Stellarator | 2272 | FOAK | Greenfield reference plant where everything has been built from zero/without restructuring | Famà et al. (2024) |
| Option 4-Reactor - Greenfield | Magnetic | Stellarator | 1368 | FOAK | Greenfield reference plant where everything has been built from zero/without restructuring | Famà et al. (2024) |
| Option 1-Reactor - restructured | Magnetic | Stellarator | 914 | FOAK | Based on ARIES calculations; 30 years lifetime; applies only to the retrofit of PWR plant and completely reuses the existing systems | Famà et al. (2024) |
| Option 1-Reactor - restructured | Magnetic | Stellarator | 914 | FOAK | Based on ARIES calculations; 30 years lifetime; applies only to the retrofit of PWR plant and completely reuses the existing systems | Famà et al. (2024) |
| Option 2-Reactor - restructured | Magnetic | Stellarator | 774 | FOAK | Based on ARIES calculations; 30 years lifetime; updates the existing steam turbines of the PWR plant by adding higher pressure stages | Famà et al. (2024) |
| Option 2-Reactor - restructured | Magnetic | Stellarator | 774 | FOAK | Based on ARIES calculations; 30 years lifetime; updates the existing steam turbines of the PWR plant by adding higher pressure stages | Famà et al. (2024) |
| Option 3-Reactor - restructured | Magnetic | Stellarator | 1945 | FOAK | Based on ARIES calculations; 30 years lifetime; Option 3 updates the PWR plant | Famà et al. (2024) |
| Option 4-Reactor - restructured | Magnetic | Stellarator | 1165 | FOAK | Based on ARIES calculations; 30 years lifetime; Retrofit of a supercritical coal fired power plant | Famà et al. (2024) |
| Option 3-Reactor - restructured | Magnetic | Stellarator | 1945 | FOAK | Based on ARIES calculations; 30 years lifetime; Option 3 updates the PWR plant | Famà et al. (2024) |
| Option 4-Reactor - restructured | Magnetic | Stellarator | 1165 | FOAK | Based on ARIES calculations; 30 years lifetime; Retrofit of a supercritical coal fired power plant | Famà et al. (2024) |
| Conventional R&D | Magnetic | Tokamak | 1000 | $10^{th}$-OAK | 40 years lifetime | Gi et al. (2020) |
| Advanced R&D | Magnetic | Tokamak | 1000 | $10^{th}$-OAK | 40 years lifetime | Gi et al. (2020) |
| With Central Solenoid | Magnetic | Tokamak | 1000 | NOAK | 6 years costruction time; 40 years lifetime | Jo et al. (2021) |
| With Central Solenoid | Magnetic | Tokamak | 1600 | NOAK | 6 years costruction time; 40 years lifetime | Jo et al. (2021) |
| No Central Solenoid | Magnetic | Tokamak | 1000 | NOAK | 6 years costruction time; 40 years lifetime | Jo et al. (2021) |
| No Central Solenoid | Magnetic | Tokamak | 1420 | NOAK | 6 years costruction time; 40 years lifetime | Jo et al. (2021) |
| ACT1 | Magnetic | Tokamak | 1013.2 | $10^{th}$-OAK | 40 years lifetime | Kessel et al. (2015) |
| ACT2 | Magnetic | Tokamak | 999.7 | $10^{th}$-OAK | 40 years lifetime | Kessel et al. (2015) |
| ACT3 | Magnetic | Tokamak | 1010 | $10^{th}$-OAK | 40 years lifetime | Kessel et al. (2015) |
| ACT4 | Magnetic | Tokamak | 999.2 | $10^{th}$-OAK | 40 years lifetime | Kessel et al. (2015) |
| PULSAR-I | Magnetic | Tokamak | 1000 | $10^{th}$-OAK | 40 years lifetime | Krakowski (1995) |
| Compact Spheromak | Magnetic | Spheromak | 1000 | 10th-OAK | 40 years lifetime | Krakowski (1995) |
| HELIAC | Magnetic | Stellarator | 1000 | 10th-OAK | 40 years lifetime | Krakowski (1995) |
| High-field Stellarat | Magnetic | Stellarator | 1000 | 10th-OAK | 40 years lifetime | Krakowski (1995) |
| TITAN-I | Magnetic | Reversed-Field Pinch | 1000 | 10th-OAK | 40 years lifetime | Krakowski (1995) |
| TITAN-II | Magnetic | Reversed-Field Pinch | 1000 | 10th-OAK | 40 years lifetime | Krakowski (1995) |
| 1 GW class (t=0) | Magnetic | Tokamak | 1000 | FOAK | 5% discount rate | Tokimatsu et al. (2002) |
| 1 GW class (t=25) | Magnetic | Tokamak | 1000 | NOAK | 5% discount rate; after 25 years | Tokimatsu et al. (2002) |
| 1 GW class (t=50) | Magnetic | Tokamak | 1000 | NOAK | 5% discount rate; after 50 years | Tokimatsu et al. (2002) |
| Tokamak Reactor- | Magnetic | Tokamak | 1000 | NOAK | 30 years lifetime | Yamazaki et al. (2009) |
| Spherical Reactor- | Magnetic | Spherical Tokamak | 1000 | NOAK | 30 years lifetime | Yamazaki et al. (2009) |
| Helical Reactor-1 | Magnetic | Heliotron | 1000 | NOAK | 30 years lifetime | Yamazaki et al. (2009) |
| IFE (with FLIBE Breeder) | Inertia | Laser Driver | 1000 | NOAK | 30 years lifetime; high repetition rate 12Hz | Yamazaki et al. (2011) |

**Table 7: Literature-reported techno-economic assumptions for fusion power plants by confinement type and maturity level used as inputs for the LCOE analysis.**

Note: This table reports non-cost techno-economic parameters used as inputs for the LCOE simulations, including plant lifetime, construction time, availability, and discount rate. Additional reactor-specific assumptions are included for context but are not explicitly modeled.

Abbreviations used: ACT: Advanced tokamak concept; ARIES: Advanced Reactor Innovation and Evaluation Study; AV: Availability; CF: Capacity factor; DEMO: Demonstration Fusion Power Plant; ESECOM: Energy Systems and Economic Modeling; FLiBe: Lithium fluoride-beryllium fluoride molten salt; FOAK: First-of-a-kind; GW: Gigawatt; HELIAC: Helically symmetric advanced stellarator concept; Hz: Hertz; IFE: Inertial fusion energy; LIFE: Laser Inertial Fusion Energy; MW: Megawatt; NLHD: National Large Helical Device; NOAK: Nth-of-a-kind; PPCS: Power Plant Conceptual Study; PULSAR: Pulsed tokamak reactor concept; PWR: Pressurized water reactor; R&D: Research and development; RFP: Reversed-field pinch; SFS: Sheared flow stabilized; SLC: Stabilized liner compressor; SSTR: Steady-State Tokamak Reactor; 10th-OAK: Tenth-of-a-kind; TOK: Tokamak; UWMAK: University of Wisconsin Magnetic Fusion Reactor Design Study.

| Year | US CPI | USD/EUR exchange rate | DEM/USD exchange rate | USD/EUR combined exchange rate | Inflation factor to $USD_{2018}$ | EUR to $USD_{2018}$ conversion factor |
|---|---|---|---|---|---|---|
| **1965** | 31.5 | | 3.994 | 0.48965526 | 7.97165079 | 3.90336073 |
| **1966** | 32.4 | | 3.998 | 0.48917763 | 7.75021605 | 3.79123232 |
| **1967** | 33.4 | | 3.987 | 0.49060101 | 7.51817365 | 3.68842361 |
| **1968** | 34.8 | | 3.992 | 0.48990056 | 7.21571839 | 3.53498447 |
| **1969** | 36.7 | | 3.924 | 0.49837682 | 6.84215259 | 3.40997026 |
| **1970** | 38.8 | | 3.646 | 0.53638757 | 6.4718299 | 3.47140912 |
| **1971** | 40.5 | | 3.480 | 0.56210088 | 6.20017284 | 3.48512259 |
| **1972** | 41.8 | | 3.189 | 0.61332434 | 6.0073445 | 3.68445062 |
| **1973** | 44.4 | | 2.659 | 0.73555096 | 5.65556306 | 4.15995483 |
| **1974** | 49.3 | | 2.590 | 0.7552342 | 5.09344828 | 3.84674632 |
| **1975** | 53.8 | | 2.463 | 0.79405221 | 4.66741636 | 3.70617228 |
| **1976** | 56.9 | | 2.517 | 0.77695547 | 4.4131283 | 3.42880416 |
| **1977** | 60.6 | | 2.322 | 0.84241289 | 4.14367987 | 3.49068932 |
| **1978** | 65.2 | | 2.008 | 0.97382494 | 3.85133436 | 3.75052543 |
| **1979** | 72.6 | | 1.833 | 1.06701037 | 3.4587741 | 3.69054782 |
| **1980** | 82.4 | | 1.816 | 1.07711752 | 3.04741505 | 3.28242415 |
| **1981** | 90.9 | | 2.261 | 0.86502875 | 2.76245325 | 2.38960147 |
| **1982** | 96.5 | | 2.429 | 0.80529913 | 2.60214508 | 2.09550517 |
| **1983** | 99.6 | | 2.555 | 0.76543128 | 2.52115462 | 1.9297706 |
| **1984** | 103.9 | | 2.846 | 0.68731726 | 2.41681424 | 1.66111815 |
| **1985** | 107.6 | | 2.942 | 0.66470568 | 2.33370818 | 1.55122909 |
| **1986** | 109.6 | | 2.171 | 0.90097199 | 2.29112226 | 2.06423699 |
| **1987** | 113.6 | | 1.798 | 1.08765988 | 2.21044894 | 2.40421664 |
| **1988** | 118.3 | | 1.758 | 1.11227821 | 2.12262891 | 2.36095388 |
| **1989** | 124.0 | | 1.881 | 1.03961622 | 2.02505645 | 2.10528154 |
| **1990** | 130.7 | | 1.616 | 1.21021595 | 1.92124713 | 2.32512393 |
| **1991** | 136.2 | | 1.661 | 1.17735974 | 1.84366373 | 2.17065545 |
| **1992** | 140.3 | | 1.560 | 1.25413915 | 1.78978617 | 2.2446409 |
| **1993** | 144.5 | | 1.654 | 1.18219898 | 1.73776471 | 2.05438367 |
| **1994** | 148.2 | | 1.622 | 1.20596251 | 1.69437922 | 2.04335782 |
| **1995** | 152.4 | | 1.434 | 1.36408844 | 1.64768373 | 2.24758632 |
| **1996** | 156.9 | | 1.504 | 1.30067833 | 1.60042702 | 2.08164074 |
| **1997** | 160.5 | | 1.735 | 1.1274095 | 1.5645296 | 1.76386553 |
| **1998** | 163.0 | | 1.759 | 1.1117724 | 1.54053374 | 1.71272289 |
| **1999** | 166.6 | 1.0653 | | 1.0653 | 1.5072449 | 1.60566799 |
| **2000** | 172.2 | 0.9232 | | 0.9232 | 1.4582288 | 1.34623683 |
| **2001** | 177.1 | 0.8952 | | 0.8952 | 1.41788255 | 1.26928846 |
| **2002** | 179.9 | 0.9454 | | 0.9454 | 1.39581434 | 1.31960288 |
| **2003** | 184.0 | 1.1321 | | 1.1321 | 1.36471196 | 1.54499041 |
| **2004** | 188.9 | 1.2438 | | 1.2438 | 1.32931181 | 1.65339802 |
| **2005** | 195.3 | 1.2449 | | 1.2449 | 1.28575013 | 1.60063033 |
| **2006** | 201.6 | 1.2563 | | 1.2563 | 1.24557044 | 1.56481014 |
| **2007** | 207.342 | 1.3711 | | 1.3711 | 1.21107639 | 1.66050683 |
| **2008** | 215.303 | 1.4726 | | 1.4726 | 1.16629587 | 1.7174873 |
| **2009** | 214.537 | 1.3935 | | 1.3935 | 1.17046011 | 1.63103616 |
| **2010** | 218.056 | 1.3261 | | 1.3261 | 1.15157116 | 1.52709851 |
| **2011** | 224.939 | 1.3931 | | 1.3931 | 1.11633376 | 1.55516456 |
| **2012** | 229.594 | 1.2859 | | 1.2859 | 1.09370018 | 1.40638907 |
| **2013** | 232.957 | 1.3281 | | 1.3281 | 1.07791137 | 1.4315741 |
| **2014** | 236.736 | 1.3297 | | 1.3297 | 1.06070475 | 1.41041911 |
| **2015** | 237.017 | 1.1096 | | 1.1096 | 1.05944721 | 1.17556263 |
| **2016** | 240.007 | 1.1072 | | 1.1072 | 1.04624865 | 1.15840651 |
| **2017** | 245.120 | 1.1301 | | 1.1301 | 1.02442477 | 1.15770243 |
| **2018** | 251.107 | 1.1817 | | 1.1817 | 1 | 1.1817 |
| **2019** | 255.657 | 1.1194 | | 1.1194 | 0.98220272 | 1.09947772 |
| **2020** | 258.811 | 1.1410 | | 1.141 | 0.9702331 | 1.10703597 |
| **2021** | 270.970 | 1.1830 | | 1.183 | 0.92669668 | 1.09628218 |
| **2022** | 292.655 | 1.0534 | | 1.0534 | 0.85803079 | 0.90384963 |
| **2023** | 304.702 | 1.0817 | | 1.0817 | 0.82410683 | 0.89143636 |
| **2024** | 313.689 | 1.0820 | | 1.082 | 0.80049667 | 0.8661374 |

**Table 8: Inflation indices and historical exchange rates used to convert literature-reported fusion cost data to constant 2018 USD.**

Source: U.S. Consumer Price Index (CPI) data are taken from (BLS undated), USD/EUR exchange rates are taken from (FRED 2025), DEM/USD exchange rates are taken from (BBk undated).

Abbreviations used: CPI: Consumer price index; DEM: Deutsche Mark; EUR: Euro; USD: United States dollar.